\documentclass[twocolumn,tighten]{aastex62}

\newcommand{\ud}{\ensuremath{\mathrm{d}}}

\newcommand{\cb}{\color{blue}}

\newcommand{\nulls}{{$\varnothing$}}
\hypersetup{linkcolor=red,citecolor=blue,filecolor=cyan,urlcolor=magenta}

\usepackage{natbib}
\usepackage{amsmath}
\usepackage{multirow}
\usepackage{ulem}
\usepackage{graphicx}
\usepackage[maxfloats=500]{morefloats}
\usepackage{bold-extra}
\received{00, 00, 2020}
\revised{00, 00, 2020}
\accepted{00, 00, 2020}
\submitjournal{ApJ}
\shorttitle{3D Supernova Models: From $-$7 Minutes to $+$7 Seconds}
\shortauthors{R.~Bollig, et al.}

\begin{document}

\title{\textsc{Self-consistent 3D Supernova Models From $-$7 Minutes to $+$7 Seconds:\\ 
	       A 1-bethe Explosion of a $\sim$19\,M$_\odot$ Progenitor}}

\correspondingauthor{Hans-Thomas Janka}
\email{thj@MPA-Garching.MPG.DE}

\author[0000-0003-2354-2454]{Robert Bollig}
\affil{Max Planck Institute for Astrophysics, 
       Karl-Schwarzschild-Str.~1, 85748 Garching, Germany}

\author[0000-0002-4107-9443]{Naveen Yadav}
\affil{Max Planck Institute for Astrophysics, 
       Karl-Schwarzschild-Str.~1, 85748 Garching, Germany}
\affil{Excellence Cluster ORIGINS,
       Boltzmannstr.~2, 85748 Garching, Germany}

\author[0000-0003-1120-2559]{Daniel Kresse}
\affil{Max Planck Institute for Astrophysics, 
       Karl-Schwarzschild-Str.~1, 85748 Garching, Germany}
\affil{Physik-Department, Technische Universit\"at M\"unchen,
       James-Franck-Str.~1, 85748 Garching, Germany}

\author[0000-0002-0831-3330]{Hans-Thomas Janka}
\affil{Max Planck Institute for Astrophysics, 
       Karl-Schwarzschild-Str.~1, 85748 Garching, Germany}

\author[0000-0002-4470-1277]{Bernhard~M\"uller}
\affil{School of Physics and Astronomy,
       Monash University, Victoria 3800, Australia}
\affil{Australian Research Council Centre of Excellence for
       Gravitational Wave Discovery (OzGrav), Clayton, VIC 3800, Australia}
\affil{Astrophysics Research Centre, School of Mathematics and Physics,
       Queen's University Belfast, BT7 1NN, Belfast, Northern Ireland}

\author[0000-0002-3684-1325]{Alexander~Heger}
\affil{School of Physics and Astronomy,
       Monash University, Victoria 3800, Australia}
\affil{Australian Research Council Centre of Excellence for
       Gravitational Wave Discovery (OzGrav), Clayton, VIC 3800, Australia}
\affil{Center of Excellence for Astrophysics in Three Dimensions
       (ASTRO-3D), Australia}
\affil{Joint Institute for Nuclear Astrophysics, 1 Cyclotron Laboratory,
       National Superconducting Cyclotron Laboratory, Michigan State
       University, East Lansing, MI 48824-1321, USA}


\begin{abstract}
To date, modern three-dimensional (3D) supernova (SN) simulations have not
demonstrated that explosion energies of $10^{51}$\,erg ($=1\,\mathrm{bethe} = 1$\,B)
or more are possible for neutrino-driven SNe of 
non/slow-rotating $M<20$\,M$_\odot$ progenitors. We present the first such model,
considering a non-rotating, solar-metallicity 18.88$\,$M$_\odot$ progenitor, whose
final 7\,minutes of convective oxygen-shell burning were simulated in 3D and showed
a violent oxygen-neon shell merger prior to collapse. A large set of 3D SN-models
was computed with the \textsc{Prometheus-Vertex} code, whose improved convergence 
of the two-moment equations with Boltzmann closure allows
now to fully exploit the implicit neutrino-transport treatment. Nuclear 
burning is treated with a 23-species network. We vary the angular
grid resolution and consider different nuclear equations of state and muon formation
in the proto-neutron star (PNS), which requires six-species transport
with coupling of all neutrino flavors across all energy-momentum groups.
Elaborate neutrino transport was applied until $\sim$2\,s after bounce. In one
case the simulation was continued to $>$7\,s with an approximate treatment of
neutrino effects that allows for seamless continuation without transients.  A
spherically symmetric neutrino-driven wind does not develop.  Instead, accretion downflows
to the PNS and outflows of neutrino-heated matter establish a monotonic rise of the
explosion energy until $\sim$7\,s post bounce, when the outgoing shock reaches
$\sim$50,000\,km and enters the He-layer. The converged value of the explosion energy
at infinity (with overburden subtracted) is $\sim$\,1\,B and the ejected $^{56}$Ni mass
$\lesssim$\,0.087\,M$_\odot$, both within a few 10\% of the SN~1987A values. The
final NS mass and kick are $\sim$1.65\,M$_\odot$ and $>$450\,km\,s$^{-1}$, respectively.
\end{abstract}

\section{Introduction} 
\label{sec:Introduction}

Self-consistent 3D simulations with energy-dependent neutrino transport
\citep[e.g.,][]{Takiwaki+2014,Melson+2015a,Melson+2015b,Lentz+2015,Ott+2018,Janka+2016,Summa+2018,Mueller+2017,Mueller+2019,Burrows+2020,Stockinger+2020} 
have recently demonstrated the viability of the neutrino-driven
mechanism~\citep{Colgate+1966,Arnett1966,Bethe+1985} for core-collapse supernovae
(CCSNe), supporting earlier 3D results with
gray neutrino transport and smoothed-particle hydrodynamics~\citep{Fryer+2002}. 
Besides a detailed treatment of the crucial neutrino physics, modern grid-based 
hydrodynamics schemes are necessary for a reliable description of explosion-assisting
hydrodynamic instabilities in the form of convective overturn 
\citep{Herant+1994,Burrows+1995,Janka+1996}, the standing accretion shock instability
\citep[SASI;][]{Blondin+2003,Blondin+2007}, and turbulent flows 
\citep[e.g.,][]{Murphy+2013,Couch+2015a,Radice+2016} in the postshock layer.
Pre-collapse asymmetries in the convectively burning oxygen and silicon shells
\citep{Arnett+2011,Couch+2015b,Mueller+2016,Yadav+2020,Yoshida+2019} 
turned out to facilitate the onset of the 
explosion by strengthening the growth and power of postshock instabilities
\citep{Couch+2013,Mueller+2015,Mueller+2017}.
Moreover, moderate or rapid rotation
\citep{Fryer+2004,Summa+2018,Kuroda+2020,Obergaulinger+2021}, magnetic
field effects also in non-rotating collapsing stellar cores
\citep{Mueller+2020}, or enhanced neutrino luminosities
or harder neutrino spectra due to a
faster contraction of the proto-neutron star (PNS) connected to 
microphysics \citep[e.g.,][]{Melson+2015b,Bollig+2017,Yasin+2020} 
can play a role in triggering an (earlier) beginning of the explosion.

An open problem still to be solved is the question about SN energies that can
be reached by the neutrino-driven mechanism. In view of current self-consistent
multi-D simulations that follow the post-bounce evolution for typically 
$\lesssim$1\,s, concerns were expressed that the explosions might be under-energetic
compared to observations \citep[e.g.,][]{Papish+2015,Murphy+2019}. But only 
for low-mass ($\lesssim$10\,M$_\odot$) progenitors with very low core compactness
\citep{Kitaura+2006,Melson+2015a,Stockinger+2020} 
and some ultra-stripped progenitors \citep[e.g.,][]{Mueller+2019}
this time scale may be sufficient to determine the final value of the explosion
energy. For more massive progenitors with high mass-accretion rates of 
the PNS, however, it can take much longer for the energy to saturate. 
So far, only in axi-symmetry (2D) the corresponding explosions
could be tracked sufficiently long to reach energy saturation
\citep{Mueller2015,Bruenn+2016,Nakamura+2019,Burrows2021}. 
In 3D, the final values were still not saturated and
well below 1\,B \citep{Mueller+2017} and thus below the margin
considered to be typical of common Type-II SN progenitors.
Exceptions were only found for special, 
extremely rapidly rotating, strongly
magnetized, or very massive black-hole forming stars 
\citep{Kuroda+2020,Obergaulinger+2021,Chan+2020,Powell+2020,Fujibayashi+2021}.

In this paper, we present the first self-consistent 3D simulations 
of a non-rotating progenitor in the typical mass range of CCSNe that yield
an explosion with the canonical value near 1\,B and an ejecta mass of $^{56}$Ni 
around the average amount \citep[around 0.05\,M$_\odot$;][]{MuellerT+2017}. 
We consider a solar-metallicity star of initially 
18.88\,M$_\odot$, whose evolution was followed continuously in 3D through the 
final 7~minutes of convective oxygen-shell burning, core collapse and bounce,
until $\sim$7\,s after core bounce, at which stage the shock passes the C/He
interface and the explosion saturates at its final energy. 

In Section~\ref{sec:setup} we briefly describe the considered progenitor model,
numerical methods, and set of 3D simulations. In Section~\ref{sec:results} we
discuss our results with a focus on the influence of progenitor perturbations
and resolution (Section~\ref{sec:dimresphys}), the evolution of the explosion asymmetry 
(Section~\ref{sec:explosionasymmetry}), the explosion energy (Section~\ref{sec:explenergy}), 
explosive nucleosynthesis (Section~\ref{sec:nucleosynthesis}),
and PNS properties (Section~\ref{sec:PNSproperties}) in our longest simulation.
In Section~\ref{sec:conclusions} we finish with our conclusions.
Appendix~\ref{app:addinfos} provides additional information on a subset of
our larger sample of models for comparison, Appendix~\ref{app:angmom}
reports on our investigation of the angular momentum evolution,
and Appendix~\ref{app:LESA} contains additional details on the neutrino 
emission with respect to the electron-neutrino lepton-number emission asymmetry 
termed LESA).

\section{Modeling set-up and simulations} 
\label{sec:setup}

\subsection{Progenitor model}
\label{sec:progenitor}

The 18.88\,M$_\odot$ progenitor model was evolved with the implicit
hydrodynamics code KEPLER 
in spherical symmetry (1D) until 7~minutes before iron-core collapse. 
At this stage it was mapped to the \textsc{Prometheus} hydrodynamics code to 
follow the convective oxygen-shell burning for the final 7~minutes in 3D until
the gravitational instability of the iron core set in \citep{Yadav+2020}.
The star was chosen because it lies in a mass regime where semi-analytic
models \citep{Mueller+2016a} and calibrated 1D simulations \citep{Sukhbold+2016}
suggest high explosion energies by the neutrino-driven
mechanism, and where a radially extended oxygen-burning 
shell is expected to favor low-mode (i.e., low spherical harmonics modes, $\ell = 1,\,2$) 
convection \citep{Collins+2018}, which is most supportive for the onset
of SN explosions \citep{Mueller+2015}. Moreover, the progenitor's initial mass 
and He-core mass (5.78\,M$_\odot$)
are near those discussed for the progenitors of SN~1987A and Cassiopeia~A
(see, e.g., \citealt{Utrobin+2019} and \citealt{Utrobin+2021} for SN~1987A and 
\citealt{Orlando+2016} for Cassiopeia~A).
During the 3D simulation a vigorous O-Ne shell merger occurs in the convectively 
burning oxygen shell, leading to large-scale, large-amplitude 
asymmetries of density, velocity, and composition in the entire oxygen layer
\citep[for details, see][]{Yadav+2020}.
This perturbed state constitutes self-consistent initial conditions for the 
3D core-collapse simulations reported in the present paper.

Ne-O shell mergers may not be an ubiquitous phenomenon in the mass range between
15\,M$_\odot$ and 20\,M$_\odot$, but they can still occur in a significant 
fraction of progenitors (\citealt{Rauscher+2002}; \citealt{Sukhbold+2018};
\citealt{Collins+2018}, who found them
in about 40\% of the cases in the mentioned mass range).
In particular, around 18\,M$_\odot$ there are large variations in the 
stellar structure, almost chaotic, because of the transition of convective
central C burning to radiative central C burning 
\citep{Sukhbold+2018,Sukhbold+2020}.
The fact that our model displays this phenomenon is interesting
but not crucial for the success of explosion simulations. The relevant aspect
in this context is the presence of large-scale convective perturbations of 
considerable amplitude in the inner burning shells prior to core collapse
(see the discussions by \citealt{Couch+2013} and \citealt{Mueller+2015}). Such
perturbations are predicted to be widespread in the convective 
O-shells of progenitors in the mentioned mass interval \citep{Collins+2018}
and are not limited to cases with O-Ne shell mergers.

In the following we discuss results from a large set of new
3D core-collapse calculations, based on this 18.88\,M$_\odot$ progenitor and
conducted with the \textsc{Prometheus-Vertex} code. If started from the 
3D initial data, these simulations reveal relatively weak sensitivity of the onset
of the explosion to the high-density equation-of-state (EoS) of the PNS, the
inclusion of muons, and also to
the angular resolution, because the pre-collapse perturbations of the progenitor
foster the development of violent postshock instabilities when the Si/O
interface falls through the stagnant shock.

\begin{deluxetable*}{lcccccccccc}[!htbp]
    \tablecolumns{11}
    \tablecaption{Summary of 3D CCSN simulations. \label{tab:simulations}}
    \tablehead{
        \colhead{Model Name\tablenotemark{{\it a}}}
      & \colhead{$t_{\rm bounce}$\tablenotemark{{\it b}}}
      & \colhead{$t^{\rm exp}_{\rm pb}$}
      & \colhead{$t^{\rm f}_{\rm pb}$}
      & \colhead{${M^{\rm f}_{\rm PNS,b}}$}
      & \colhead{${M^{\rm f}_{\rm PNS,g}}$}
      & \colhead{${R^{\rm f}_{\rm PNS}}$}
      & \colhead{${E^{\rm diag}_{\rm exp}}$}
      & \colhead{${E^{\rm OB-}_{\rm exp}}$}
      & \colhead{${R^{\rm 270ms}_{\rm s}}$}
      & \colhead{${R^{\rm f}_{\rm s}}$} \tabularnewline
      & \colhead{[ms]}
      & \colhead{[ms]}
      & \colhead{[ms]}
      & \colhead{[M$_{\odot}$]}
      & \colhead{[M$_{\odot}$]}
      & \colhead{[km]}
      & \colhead{[B]}
      & \colhead{[B]}
      & \colhead{[km]}
      & \colhead{[km]}}
\startdata
$\texttt{H\_P1D\_LS220\_m-~~}$    & 357 &        & 288  &        &        &       &         & & $107^{120}_{96}$  & $98^{107}_{89}$\tabularnewline
$\texttt{H\_P3D\_LS220\_m-~~}$    & 357 &        & 285  &        &        &       &         & & $158^{213}_{114}$ & $168^{245}_{120}$\tabularnewline
\hline
$\texttt{M\_P1D\_LS220\_m-~~}$    & 357 & \nulls & 579  & 1.8788 & 1.8115 & 26.00 & \nulls  & \nulls  & $142^{170}_{122}$ & $82^{95}_{64}$ \tabularnewline
$\texttt{\textbf{M\_P3D\_LS220\_m-~~}}$    & 357 & 418    & 1675 & 1.8655 & 1.7548 & 17.89 & 0.5071  & $\phantom{-}$0.2024  & $165^{213}_{126}$ & $9704^{12203}_{7852}$ \tabularnewline
$\texttt{\cb \textbf{M\_P3D\_LS220\_m-HC}}$    & {\cb \nulls}  & {\cb \nulls} & \cb 7035 & \cb1.8654 & \cb 1.6749 & \cb 13.57   & \cb 0.9779  & \cb $\phantom{-}$0.9411 & {\cb \nulls} & $\cb 49470^{66024}_{38333}$ \tabularnewline 
$\texttt{\textbf{M\_P3D\_SFHo\_m-~~}}$     & 362 & 426    & 545  & 1.8635 & 1.8025 & 28.97 & 0.0184  & $-$0.3978 & $156^{206}_{122}$ & $549^{948}_{251}$ \tabularnewline
\hline
$\texttt{L\_P1D\_LS220\_m-~~}$    & 357 & \nulls & 489  & 1.8503 & 1.7910 & 27.96 & \nulls  & \nulls  & $173^{213}_{141}$ & $81^{96}_{70}$ \tabularnewline
$\texttt{\textbf{L\_P3D\_LS220\_m-~~}}$    & 357 & 400    & 1884 & 1.8530 & 1.7359 & 17.41 & 0.6314  & $\phantom{-}$0.3728  & $159^{190}_{136}$ & $11996^{15332}_{9425}$ \tabularnewline
$\texttt{L\_P1D\_SFHo\_m-~~}$     & 362 & \nulls & 486  & 1.8302 & 1.7798 & 30.12 & \nulls  & \nulls  & $148^{169}_{128}$ & $87^{108}_{67}$ \tabularnewline
$\texttt{\textbf{L\_P3D\_SFHo\_m-~~}}$     & 362 & 602    & 742  & 1.9154 & 1.8399 & 25.27 & 0.1001  & $-$0.2994 & $162^{184}_{141}$ & $1254^{1933}_{545}$ \tabularnewline
\hline
\hline
$\texttt{M\_P1D\_LS220\_m+~~}$    & 357 & \nulls & 321  & 1.8100 & 1.7646 & 31.32 & \nulls  & \nulls  & $144^{170}_{126}$ & $109^{124}_{98}$ \tabularnewline
$\texttt{\textbf{M\_P3D\_LS220\_m+~~}}$    & 357 & 338    & 395  & 1.8235 & 1.7722 & 29.45 & 0.0137  & $-$0.4140 & $170^{213}_{133}$ & $398^{719}_{201}$ \tabularnewline
$\texttt{\textbf{M\_P3D\_SFHo\_m+~~}}$     & 362 & 379    & 430  & 1.8416 & 1.7875 & 30.46 & 0.0063  & $-$0.4220 & $154^{195}_{121}$ & $319^{590}_{179}$ \tabularnewline
\hline
$\texttt{L\_P1D\_LS220\_m+~~}$    & 357 & \nulls & 391  & 1.8250 & 1.7734 & 29.33 & \nulls  & \nulls  & $151^{185}_{120}$ & $80^{98}_{65}$ \tabularnewline
$\texttt{\textbf{L\_P3D\_LS220\_m+~~}}$    & 357 & 376    & 694  & 1.8971 & 1.8212 & 22.91 & 0.2316  & $-$0.1613 & $172^{201}_{136}$ & $2726^{3590}_{1767}$ \tabularnewline
$\texttt{L\_P1D\_SFHo\_m+~~}$     & 362 & \nulls & 270  & 1.7889 & 1.7487 & 36.23 & \nulls  & \nulls  & $139^{165}_{117}$ & $137^{156}_{119}$ \tabularnewline
$\texttt{\textbf{L\_P3D\_SFHo\_m+~~}}$     & 362 & 426    & 515  & 1.8668 & 1.8053 & 28.14 & 0.0306  & $-$0.3899 & $135^{174}_{115}$ & $621^{910}_{298}$ \tabularnewline
\enddata
	\tablenotetext{a}{Naming convention: \texttt{H} ($1^{\circ}$), \texttt{M} ($2^{\circ}$), \texttt{L} ($4^{\circ}$), \texttt{P1D} (1D progenitor), \texttt{P3D} (3D progenitor), \texttt{LS220} \citep[EoS of][]{Lattimer_1991}, \texttt{SFHo} \citep[EoS of][]{Hempel_2010,Steiner_2013}, $\tt m-$ (without muons), $\tt m+$ (with muons), and the suffix \texttt{HC} and blue color indicate the simulation that continues model \texttt{M\_P3D\_LS220\_m$-$} after 1.675\,s and uses the heating and cooling scheme instead of the \textsc{Vertex} neutrino transport. Names of models with determined explosion times are boldfaced.}
\tablenotetext{b}{$t_{\rm bounce}$ is the time from the onset of core collapse until bounce;
$t^{\rm exp}_{\rm pb}$ is the post-bounce time when the explosion sets in, defined as the instant when the maximum shock radius reaches $400~\rm km$;
$t^{\rm f}_{\rm pb}$ is the post-bounce time when the simulation was terminated;
$M^{\rm f}_{\rm PNS,b}$ is the baryonic PNS mass at that time,
$M^{\rm f}_{\rm PNS,g}$ the corresponding gravitational mass,
$R^{\rm f}_{\rm PNS}$ the corresponding PNS radius (all of these PNS quantities are defined where the average density is $10^{11}$\,g\,cm$^{-3}$);
$E^{\rm diag}_{\rm exp}$ is the diagnostic explosion energy of all ejected postshock matter with positive total energy at $t^{\rm f}_{\rm pb}$,
$E^{\rm OB-}_{\rm exp}$ the corresponding net explosion energy, where the ``overburden'' of the gravitational binding energy of stellar matter ahead of the SN shock is subtracted, and
$R^{\rm 270ms}_{\rm s}$ and $R^{\rm f}_{\rm s}$ are the shock radii (average$^\mathrm{maximum}_\mathrm{minimum}$ values) at $t_\mathrm{pb} = 270$\,ms and $t^{\rm f}_{\rm pb}$, respectively. Blank spaces indicate values that could not be determined yet, whereas \nulls\ symbols correspond to values that do not exist for the given model. }
\vspace{-6mm}
\end{deluxetable*}

\subsection{Numerical methods}
\label{sec:numerics}

The neutrino-hydrodynamics simulations presented here were performed with the
\textsc{Prometheus-Vertex} code \citep{Rampp+2002,Buras+2006}. 
\textsc{Prometheus-Vertex} is a pseudo-Newtonian neutrino-hydrodynamics code
for 1D, 2D, and 3D CCSN simulations in polar coordinates.
It employs an axis-free polar Yin-Yang grid \citep{Wongwathanarat+2010} and
a Poisson solver for the components of a spherical harmonics expansion of the
gravitational potential up to chosen order \citep{Mueller+1995}. The
monopole is corrected for general relativistic effects according to Case~A
of \citet{Marek_2006}.

The \textsc{Prometheus} module solves the equations of nonrelativistic hydrodynamics
on an Eulerian grid (either fixed or moving). It is a higher-order, conservative, 
Godunov-type finite-volume scheme with an exact Riemann solver 
\citep{Fryxell1991,Mueller1991}, which is based on the piecewise parabolic method 
(PPM) of \citet{Colella+1984}. It is supplemented by the consistent multifluid
advection (CMA) scheme of \citet{Plewa+1999} to treat the advection of 
nuclear species in regions where nuclear statistical equilibrium (NSE) does
not hold.

The \textsc{Vertex}
transport module solves the two-moment set of neutrino energy and momentum
equations with a Boltzmann closure for the variable Eddington factor in the comoving
frame of the stellar medium including all ${\cal{O}}(v/c)$ terms ($v$ is the
fluid velocity, $c$ the speed of light) as well as gravitational
time dilation and redshifting \citep{Rampp+2002}. Multi-D aspects are treated
by the ray-by-ray-plus (RbR+) approximation \citep{Buras+2006}, and the full set of neutrino
interactions presented in these papers \citep[including neutrino-neutrino scattering
and pair conversion;][]{Buras+2003} is taken into account. The energy-momentum 
bins are coupled across the entire momentum space to treat inelastic effects
in neutrino-lepton scatterings and in both neutral-current and charged-current
neutrino-nucleon interactions. The last two are implemented by high-resolution 
rate tables including a random-phase approximation for nucleon correlations 
\citep{Burrows+1998,Burrows_1999}, weak-magnetism corrections \citep{Horowitz2002}, and 
virial corrections for neutrino-nucleon scattering \citep{Horowitz+2017},
smoothly connecting virial expansion results at low densities and random phase approximation
results at high densities (C.~Horowitz, private communication, 2016). Nucleon 
correlations and mean-field potentials \citep[e.g.,][]{Reddy+1998,Martinez-Pinedo+2012} 
are implemented consistently with the employed EoS.
Some of our models include muon physics in the EoS and in the neutrino transport
\citep{Bollig+2017}, in which case six-species transport of 
$\nu_e$, $\bar\nu_e$, $\nu_\mu$, $\bar\nu_\mu$, $\nu_\tau$, $\bar\nu_\tau$
instead of our standard four-species treatment (not discriminating $\nu_\mu$ 
and $\nu_\tau$) is applied.

Another new feature of the \textsc{Prometheus-Vertex} code is the fact that 
it contains an optimized version of the nuclear burning network 
\texttt{XNet} of \cite{Hix_1999} now, which replaces the previous flashing 
treatment of \citet{Rampp+2002}. Consistent with the tabulated NSE
composition applied at temperatures $T \ge 0.689$\,MeV and
densities $\rho \le 10^{11}$\,g\,cm$^{-3}$ after bounce,
it includes 23 nuclear species (neutrons,
protons, deuterium, tritium, $^3$He, $^4$He, the 12 $\alpha$-nuclei from 
$^{12}$C to $^{56}$Ni, $^{14}$N, $^{54}$Mn, $^{56}$Fe, $^{60}$Fe, $^{70}$Ni),
coupled by $\alpha$-reactions.

Network and \textsc{Vertex} transport were used for evolution periods of
$\lesssim$2\,s after bounce. One 3D model was continued beyond that time
by replacing \textsc{Vertex} with a cheaper heating and cooling (HC) scheme
described in detail in \citet{Stockinger+2020} and, after a transition period
of another $\sim$0.1\,s, by applying NSE down to
$T = 0.343$\,MeV instead of using the nuclear network.\footnote{This
lower limit for the NSE description was only applied for high-entropy material 
with $s/k_{\mathrm B} > 6.5$, whereas in lower-entropy regions of the infalling
O+Ne layer and parts of the O+C layer the lower temperature limit for NSE was
set to 0.5\,MeV \citep[see Section~3.1 of][]{Stockinger+2020}.}
This permitted a nearly
transient-free and computationally much cheaper prolongation of the simulated evolution. 
Free parameters in the HC treatment were determined from a 1D PNS cooling 
simulation of the same 3D SN model, using \textsc{Prometheus-Vertex}. 

The large set of 3D calculations that partly cover long evolution periods
has became possible by a considerable computational acceleration
of the original \textsc{Prometheus-Vertex} code implemented by \cite{Rampp+2002}
and \citet{Buras+2006} through a rigorous optimization of vectorization
and parallelization for the new SuperMUC-NG
architecture.\footnote{{\tt https://doku.lrz.de/display/PUBLIC/SuperMUC-NG}}
Moreover,
improved convergence of the transport module permits now full exploitation
of the implicit nature of the transport solver in computing the transport
with time steps up to $\sim$10$^{-5}$\,s, which is 100 times bigger than the
typical hydrodynamics steps. This was achieved by improved accuracy of the
Jacobian matrix through a higher precision in the calculation of the
partial derivatives. Moreover, the numerical stability in solving the
model-Boltzmann equation was improved at steep discontinuities in energy space
that are caused by Doppler shifting at shocks. These occur mainly in narrowly
localized regions in space and time when fast downflows hit the PNS surface.
The enhanced stability and thus reduced number of iterations could be accomplished
by spreading velocity jumps seen by the transport solver over a few more zones
than computed for the stellar medium by the hydrodynamics module.
In addition, a predictor-corrector
scheme was implemented for treating the neutrino source terms in the hydrodynamics
solver at 2$^\mathrm{nd}$ order accuracy, thus increasing accuracy and stability
for long transport time steps.

The use of large transport time steps in combination with much smaller
time steps for the hydrodynamics has been tested extensively in 1D. Even
beyond the factor of 100 mentioned above, this did not lead to any visible
differences in the evolution. There are several reasons why the results are
not sensitive to increased transport times: First, the conditions
for neutrino transport are close to steady-state during most of the post-bounce
evolution after shock stagnation. Second, the
time scales for changes in the stellar medium connected to neutrino-transport
source terms are usually much longer than the typical length of the time steps
for the hydrodynamics solver, which are constrained by the short sound travel 
time (determining the Courant-Friedrich-Lewy (CFL) condition). Both aspects in 
combination also imply that changes of the background medium during each
hydrodynamics time step are usually very small and have little effect on the
exact solution of the transport problem. Third, we apply the neutrino source
terms not at once but in fractions during each hydrodynamics step. In multi-D
we do not expect any fundamental differences and more serious intricacies,
although short periods of time occur where fast downflows penetrate to the PNS
and the neutrino radiation field can become non-stationary. However, during
these periods also the transport time steps tend to become smaller and the
transport is therefore well able to follow the faster changes of the neutrino
distributions in space and momentum space. 
Inside the PNS, where neutrinos are strongly coupled to the fluid, the
flow is no longer as close to hydrostatic equilibrium as in 1D. The far
subsonic flow implies, however, that the neutrino radiation field will
still change on time scales that are ${\cal O}(10^2)$ longer than the CFL 
time step. We stress that detailed and
conclusive tests are much more difficult in multi-D than in 1D because
of the nature of turbulent plasma flows, whose stochastic and chaotic behavior
can be influenced by any small(er) changes in the coupling between hydrodynamics
and neutrinos as well as any other changes of the physics inputs, grid structure
and resolution, besides the time stepping procedure used in the numerical codes.

\begin{figure*}[!]
        \centering
        \includegraphics[width=0.475\textwidth]{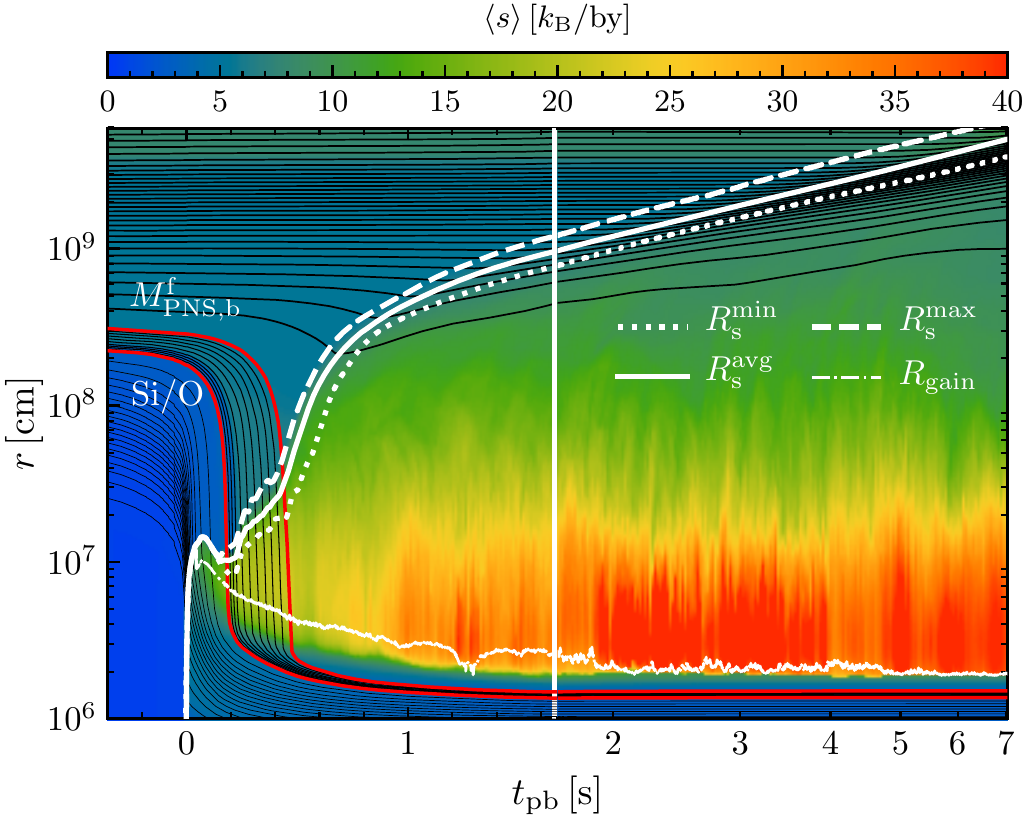}\hskip 15pt
        \includegraphics[width=0.495\textwidth]{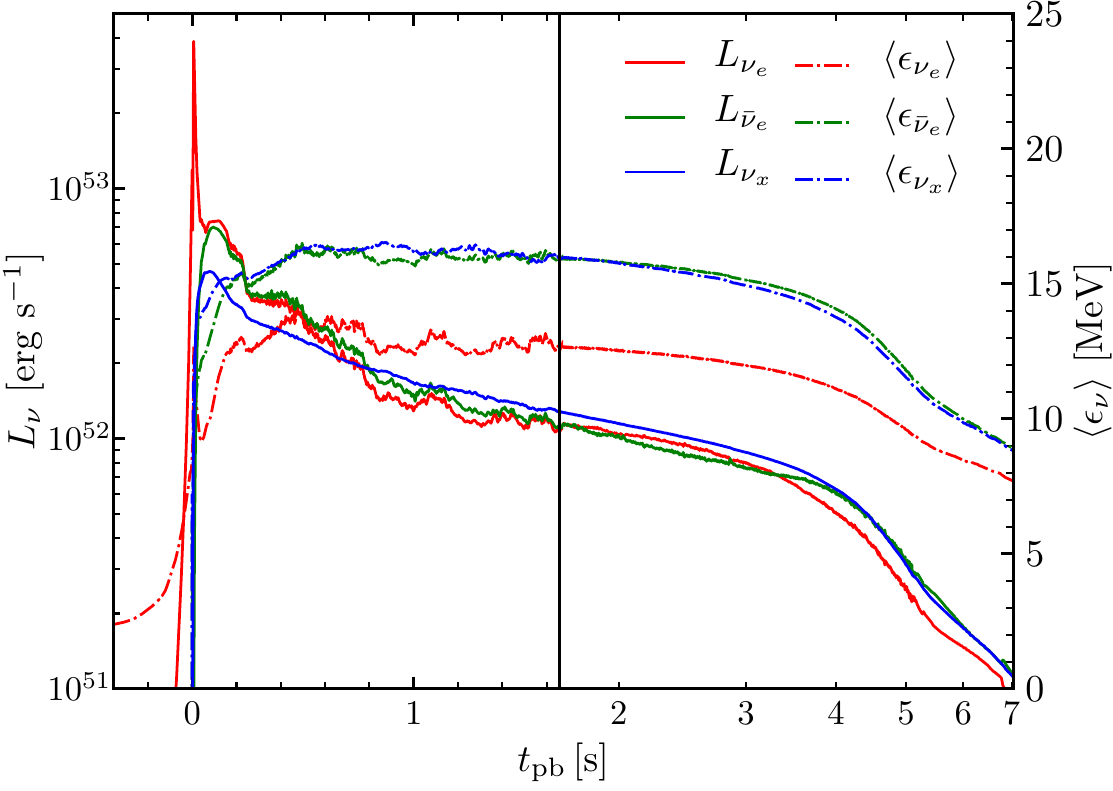}
        \caption{Explosion dynamics and neutrino emission of model \texttt{M\_P3D\_LS220\_m$-$} 
	and its extension \texttt{M\_P3D\_LS220\_m$-$HC}.
        The time axes are chosen for optimal visibility.
        {\em Left:} Mass shells with entropy per nucleon color-coded. Maximum, minimum, and average
        shock radii, gain radius, and the mass shells of Si/O shell interface and final NS mass are marked.
        The vertical white line separates \textsc{Vertex} transport (left, time linear) and HC neutrino
        approximation (right, time logarithmic). 
        {\em Right:} Emitted luminosities and mean energies of $\nu_e$, $\bar\nu_e$, and a single
        species of heavy-lepton neutrinos. The time axis is split as in the left panel.
	Right of the vertical solid line we show neutrino data from the artificially exploded 1D 
	simulation.
        }
\label{fig:explosionprops1}
\end{figure*}

\begin{figure*}[!]
        \centering
        \includegraphics[width=0.48\textwidth]{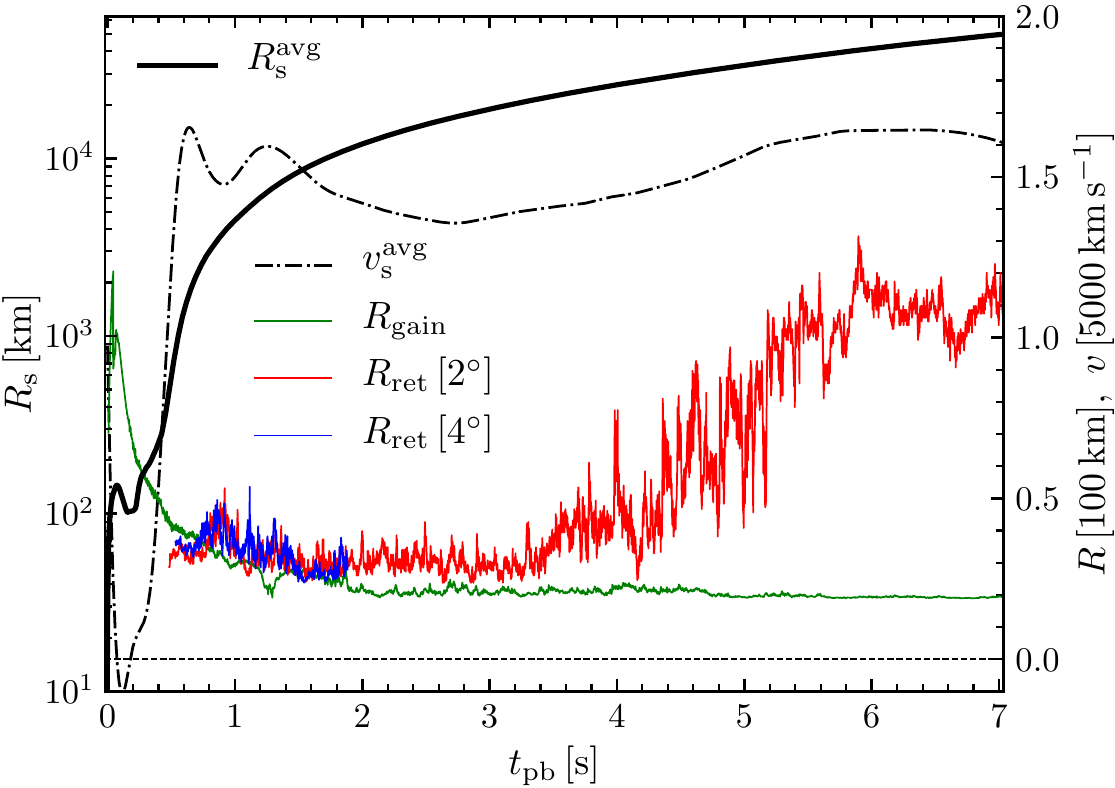}\hskip 15pt
        \includegraphics[width=0.48\textwidth]{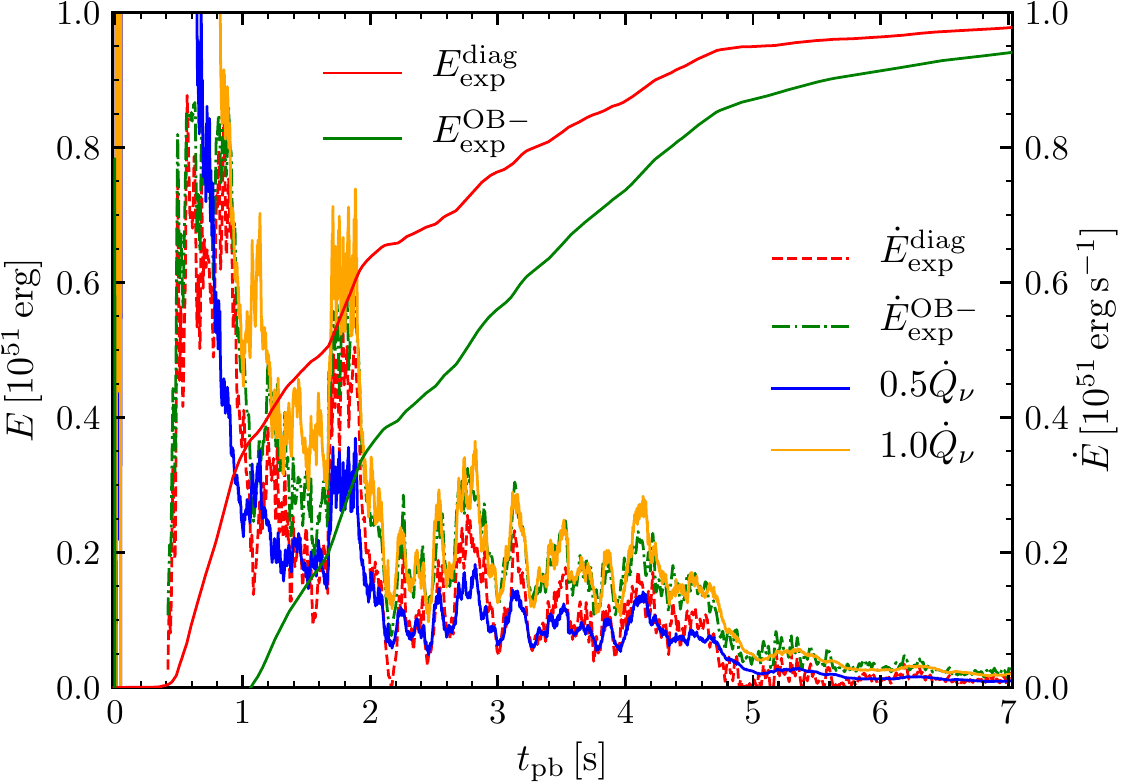}\\\vskip 0pt
        \includegraphics[width=0.48\textwidth]{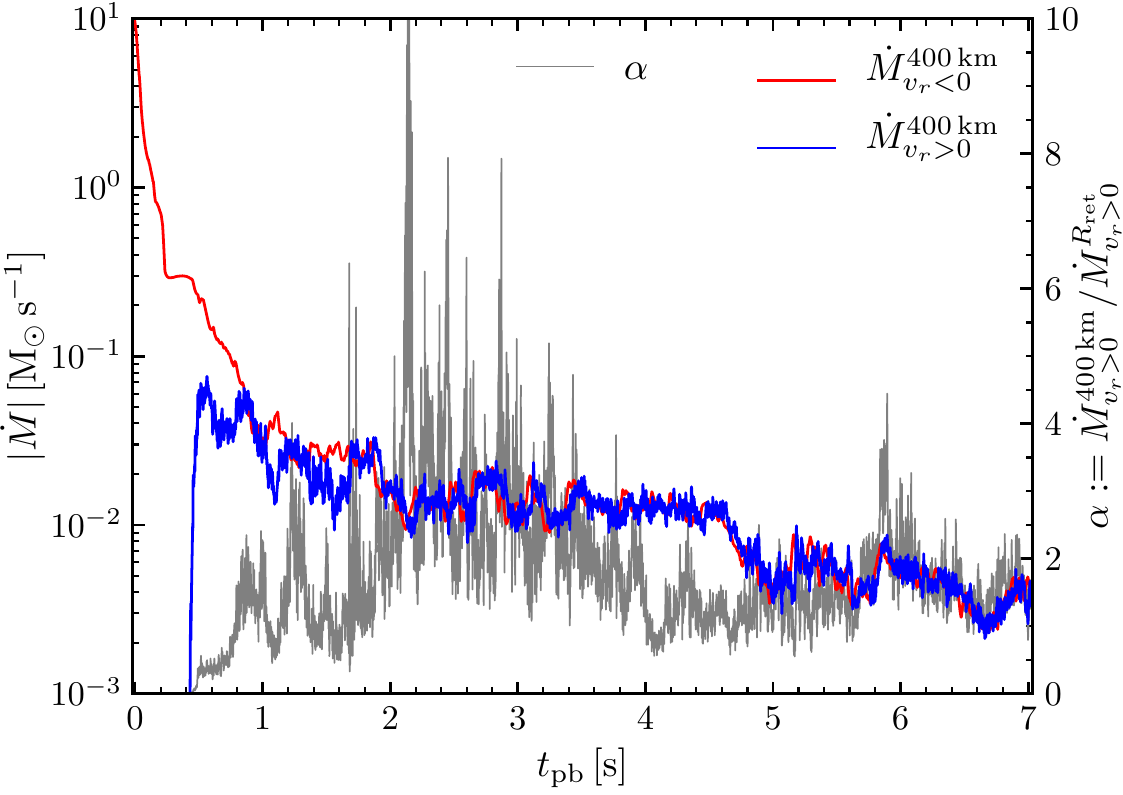}\hskip 15pt
        \includegraphics[width=0.48\textwidth]{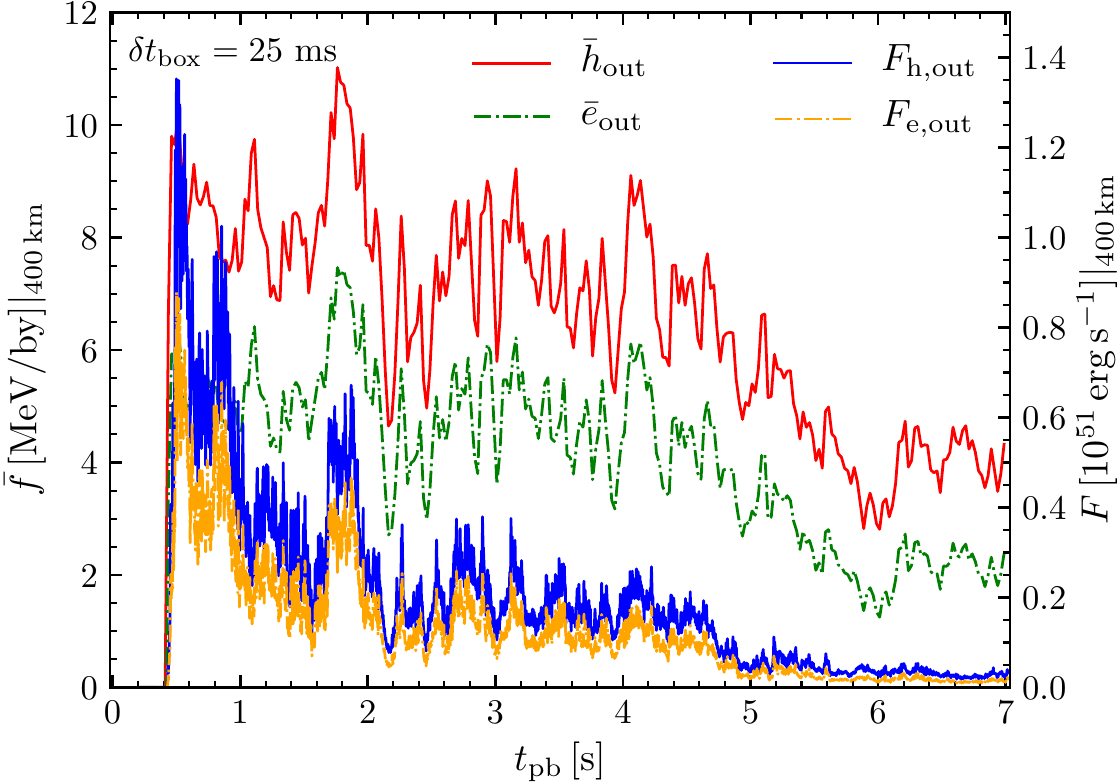}
	\caption{Evolution of explosion dynamics and explosion energy of model 
        \texttt{M\_P3D\_LS220\_m$-$}
	and its extension \texttt{M\_P3D\_LS220\_m$-$HC}.
	{\em Top left:} Average values of shock radius and velocity, gain radius, and turnaround radius
	$R_\mathrm{ret}$ for model \texttt{M\_P3D\_LS220\_m$-$} and its \texttt{HC} extension
	(both of which possess 2$^\circ$ angular resolution).
	For comparison, we also plot $R_\mathrm{ret}$ for model \texttt{L\_P3D\_LS220\_m$-$}
        (having 4$^\circ$ angular resolution).
	{\em Top right:} Explosion energy, diagnostic and without overburden (OB$-$), and corresponding 
        time derivatives compared to 0.5 and 1.0 of the net neutrino heating rate in the gain layer.
	At 7\,s $E_\mathrm{exp}^\mathrm{OB-}$ is still growing with a rate of 
	$\sim$\,$0.02$\,B\,s$^{-1}$.
	{\em Bottom left:} Mass accretion rate in downflows and ejection rate in outflows at 400\,km,
	and ratio $\alpha$ of the mass outflow rates at 400\,km and at the average turnaround radius 
        $R_\mathrm{ret}$. 
	{\em Bottom right:} Total enthalpy and energy fluxes, $F_\mathrm{h,out}$ and $F_\mathrm{e,out}$,
	in outflows at 400\,km and corresponding mean enthalpy and energy per baryon, 
	averaged over a running window of 25\,ms to reduce fluctuation amplitudes.
        }
\label{fig:explosionprops2}
\end{figure*}

\begin{figure*}[!]
	\begin{center}
\includegraphics[width=0.32\textwidth]{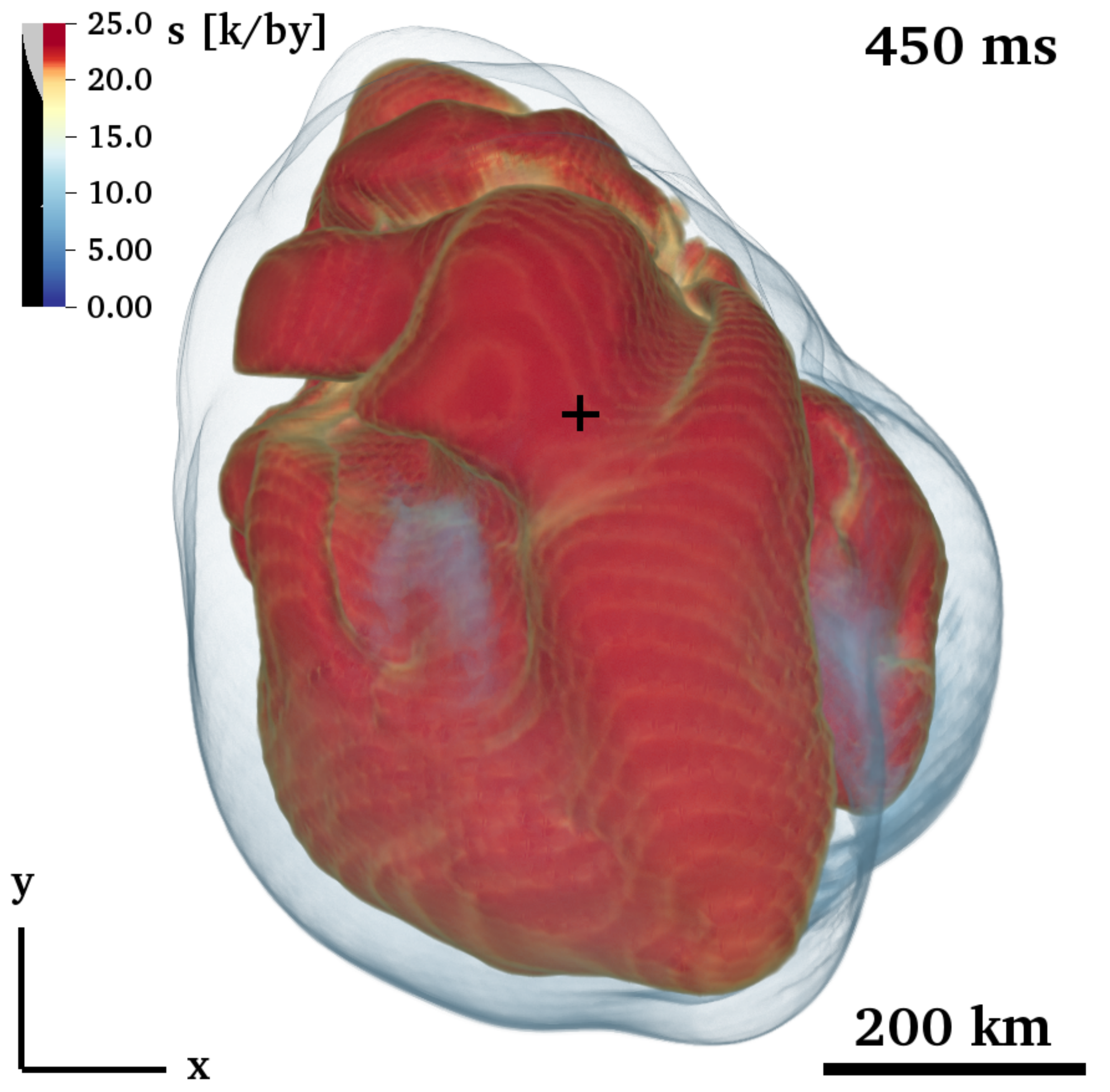}
\includegraphics[width=0.32\textwidth]{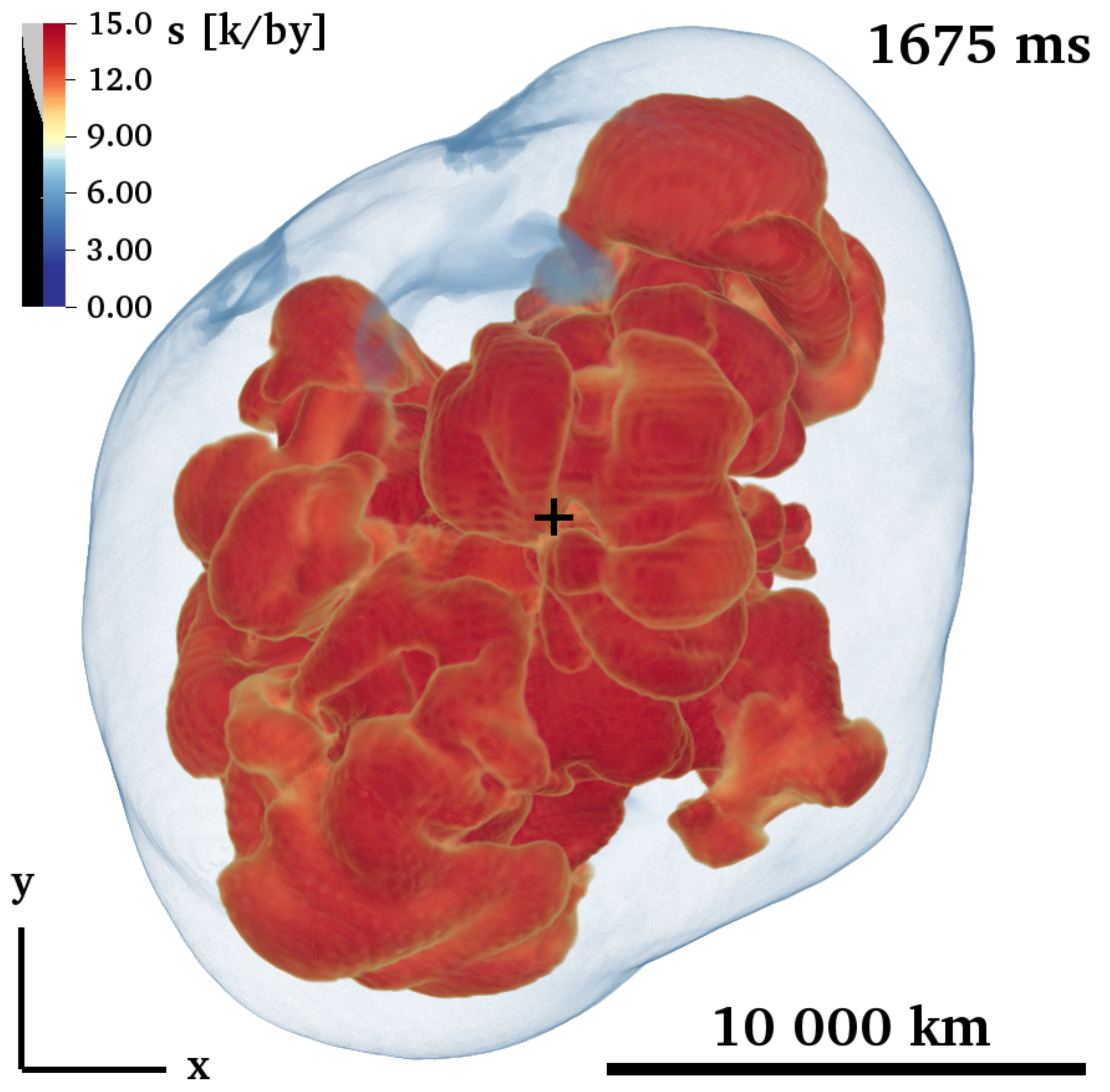}
\includegraphics[width=0.32\textwidth]{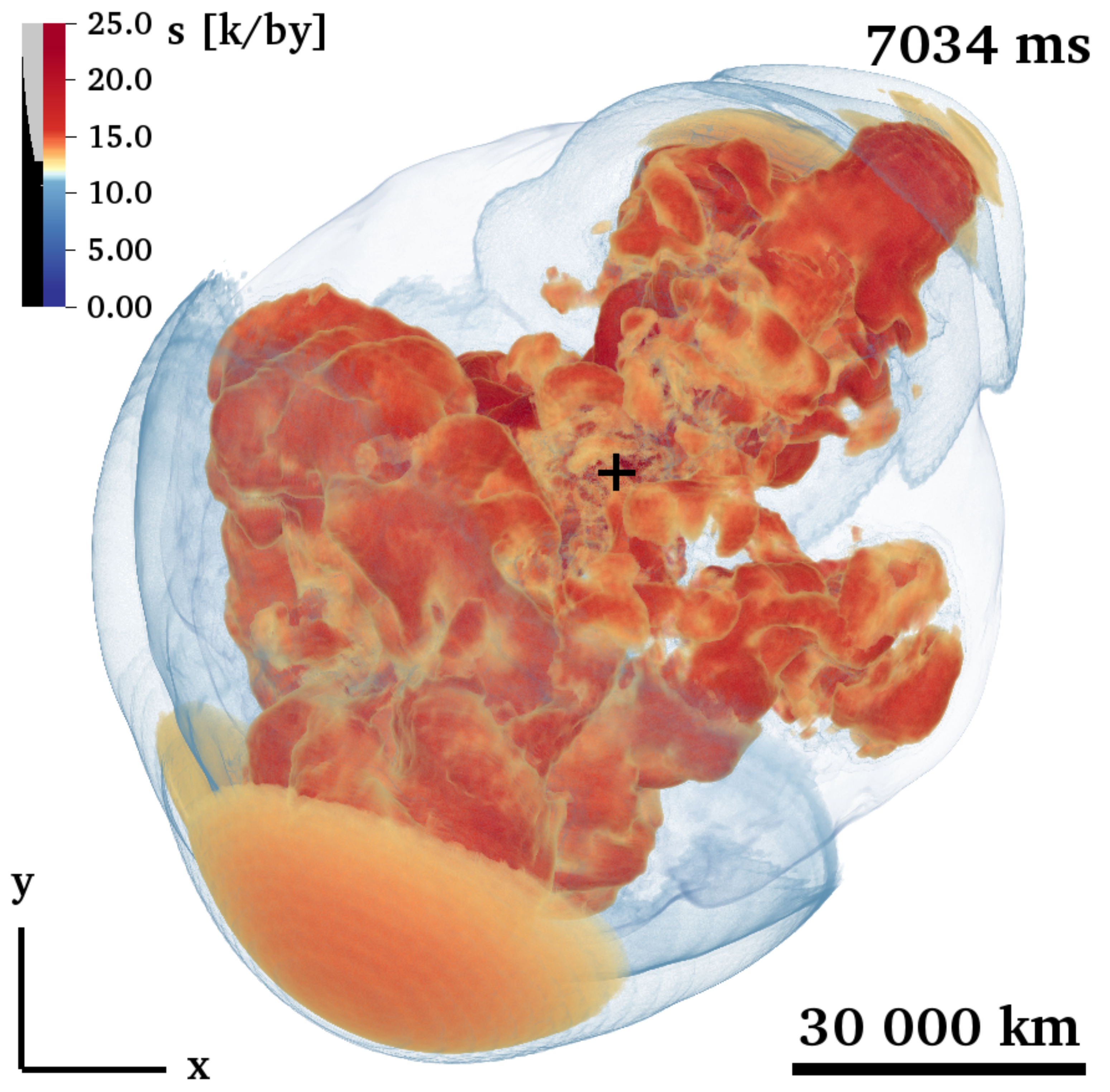}\\\vskip 15pt
\includegraphics[width=\textwidth]{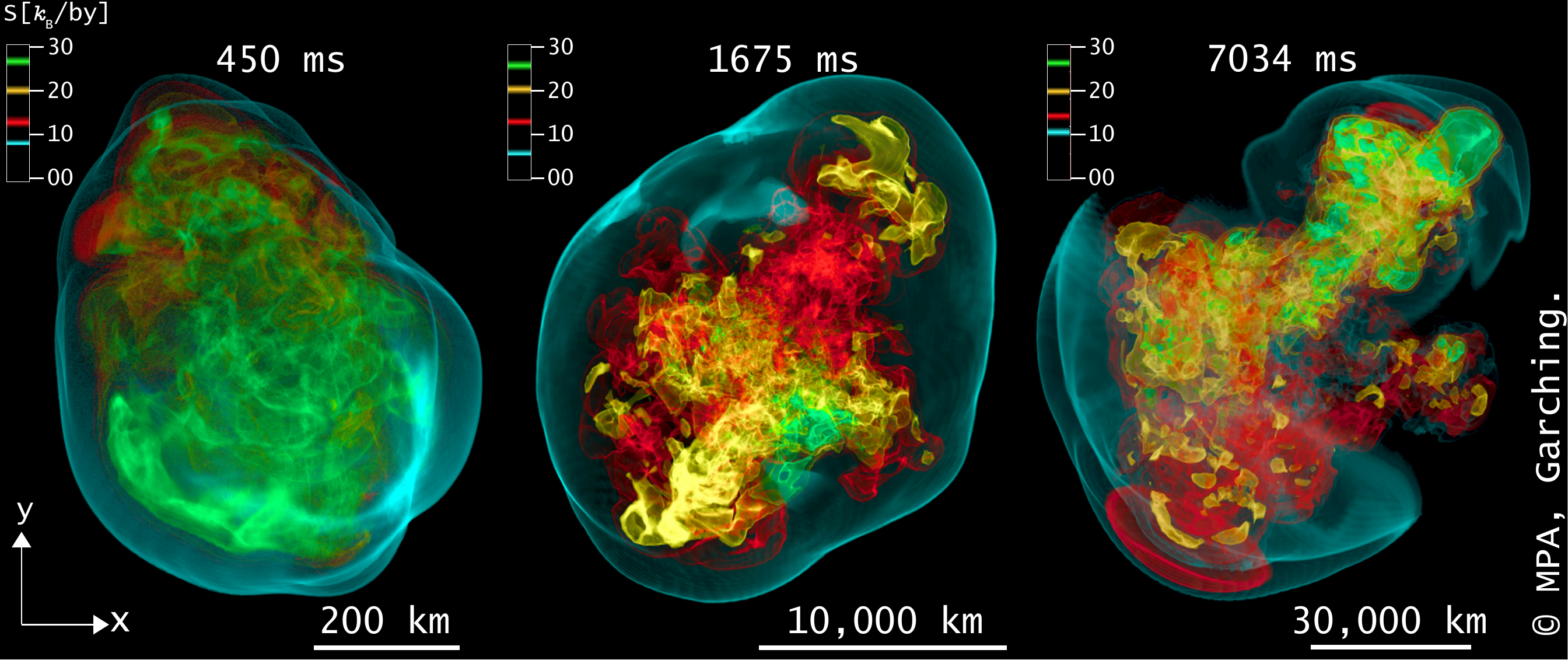}
	\end{center}
\caption{3D explosion geometry of model \texttt{M\_P3D\_LS220\_m$-$} 
	 and its extension \texttt{M\_P3D\_LS220\_m$-$HC}.
         {\em Top row:} 3D volume-renderings of plumes of high-entropy neutrino-heated
	 matter viewed from the $+z$-direction at 0.450\,s (just after the onset of 
         the explosion), 1.675\,s (at the end of the simulation with \textsc{Vertex} 
	 neutrino transport), and 7.034\,s (end of the simulation with HC neutrino
	 approximation) after core bounce. The enveloping, transparent, bluish 
         surface is the SN shock. The black cross marks the position of the PNS.
	 At 7.034\,s the shock has entered the He-layer and the entropy per baryon
	 of postshock matter and neutrino-heated gas become similar.
	 {\em Bottom row:} Volume renderings with higher transparency to
	 allow for the better visibility of turbulent structures. The times, 
	 viewing directions, and scales are the same as in the panels of the upper
	 row.
         } 
\label{fig:3D}
\end{figure*}

\begin{figure*}[!]
        \begin{center}
\includegraphics[width=1.0\textwidth]{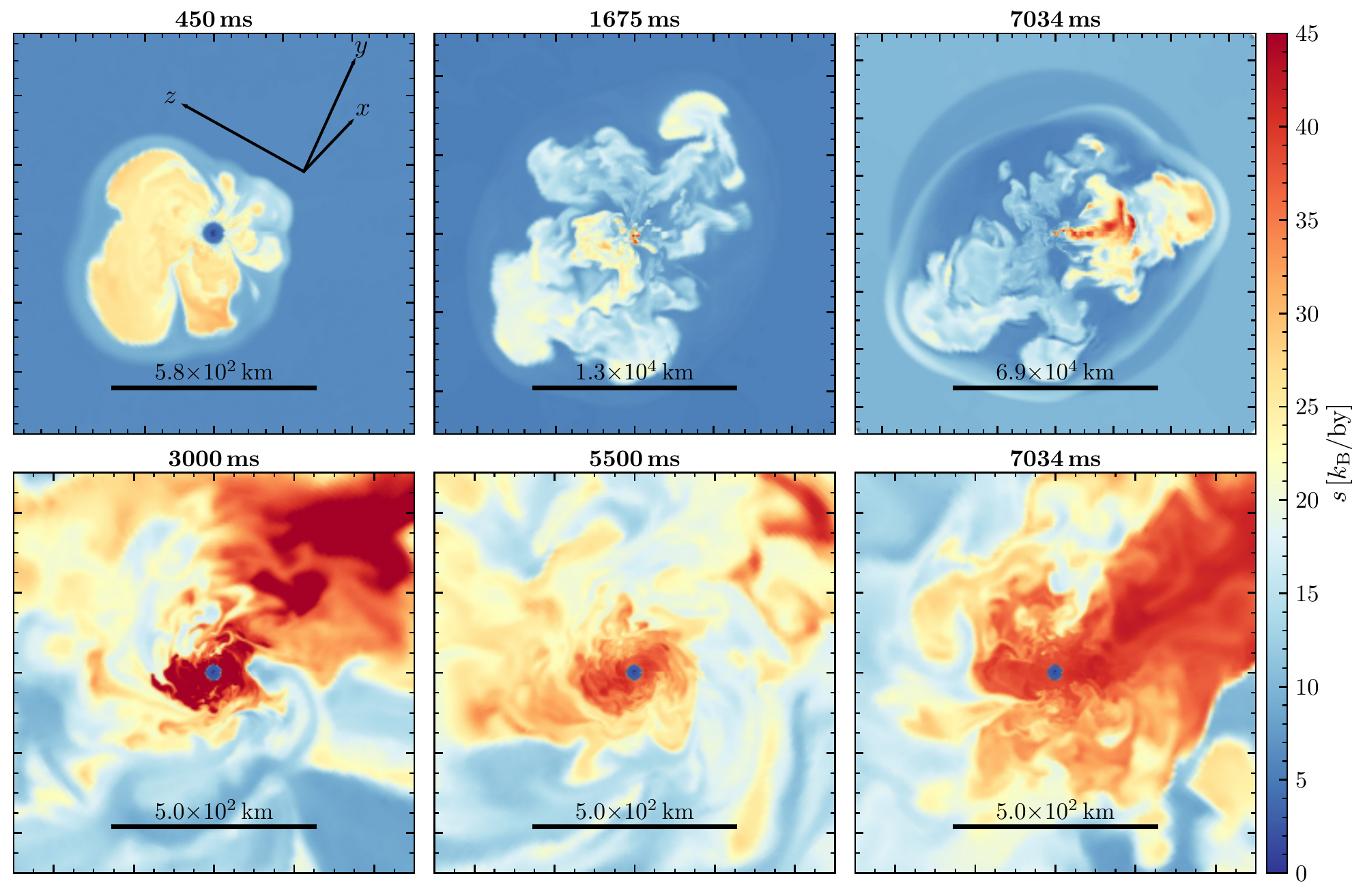}
        \end{center}
\caption{Explosion and accretion geometry of model \texttt{M\_P3D\_LS220\_m$-$}
         and its extension \texttt{M\_P3D\_LS220\_m$-$HC}.
         {\em Top row:} Entropy per baryon color-coded in a cross-sectional plane
	 at 0.45\,s, 1.675\,s, and 7.034\,s after bounce,
         rotated by roughly 90$^\circ$ relative to the orientation of the images in
	 Figure~\ref{fig:3D}.
         {\em Bottom row:} Close-ups of the turbulent vicinity of the PNS at 3.00\,s,
         5.50\,s, and 7.034\,s after bounce, showing lower-entropy downflows and
         high-entropy outflows of neutrino-heated matter. A spherical neutrino-driven
         wind does not develop until $>$7\,s after bounce.
         }
\label{fig:2D}
\end{figure*}

\subsection{Model simulations}
\label{sec:simulations}

We performed CCSN simulations with the nuclear EoS of \citet{Lattimer_1991} 
with an incompressibility modulus of $K = 220$\,MeV (LS220) and the SHFo EoS
of \citet{Hempel_2010} and \citet{Steiner_2013} with $K = 245$\,MeV, 
applied before core bounce in the NSE regime and after core bounce
at densities above $10^{11}$\,g\,cm$^{-3}$. We consider models with
angular resolutions of 4$^\circ$ (``low''; \texttt{L}-models), 2$^\circ$ (``medium'';
\texttt{M}-models), and 1$^\circ$ (``high''; \texttt{H}-models). We used a moving 
radial grid during the pre-bounce infall, and a Eulerian grid of initially 550 radial
zones with geometric spacing (resolution $\Delta r/r<0.028$ everywhere) out to 
60,000\,km after bounce (open outer boundary), covering the
entire perturbed O-shell, where large convective asymmetries were present
(up to a mass coordinate of $\sim$4.2\,M$_\odot$ and radius of 45,000\,km).
To resolve the steepening density gradient at the PNS surface, the radial grid
was successively refined until it had 755 grid points with a minimum of $\Delta r/r
= 0.002$. The central 3\,km (1.6\,km until $\sim$1\,s post bounce)
were calculated in spherical symmetry to
improve stability and reduce the time-step constraint for the explicit hydrodynamics
module. The neutrino spectrum was discretized with 15 geometrically-spaced energy
bins for $0\le\epsilon_\nu\le 380$\,MeV.

For the long-time run with the HC scheme we extended the radial grid to
approximately 100,000\,km with $\Delta r/r \sim 0.01$. To regain computational
efficiency we slightly coarsened the radial grid in the PNS interior and also
increased the 1D core from the final value of 3\,km in the \textsc{Vertex} runs 
to roughly 15\,km in the continuation calculation with the HC treatment
\citep[see Appendix~E of][]{Stockinger+2020}.

We provide an overview of all models in Table~\ref{tab:simulations}.  
The models are sorted and named
by their \texttt{H}, \texttt{M}, or \texttt{L} resolution; \texttt{P1D} or 
\texttt{P3D} denote 1D or 3D pre-collapse conditions; \texttt{LS220}
and \texttt{SFHo} correspond to the employed high-density EoS; 
and \texttt{m}$-$ and \texttt{m}$+$ mean computations without and with muons, 
respectively. Starting our simulations from 1D progenitors requires the 
use of artificial perturbations in order to seed the growth of nonradial hydrodynamic
instabilities (otherwise the models preserve spherical symmetry during their
evolution). Following our previous works for reasons of comparison, we applied
the standard procedure used in \textsc{Prometheus-Vertex} simulations for this
purpose, namely we imposed random cell-by-cell density perturbations imposed 
at bounce in the entire computational domain with an amplitude of 0.1\% of
the local average density.

We emphasize that in the current paper we will confine our discussion
mainly to the two models with the longest simulated periods of evolution,
namely \texttt{L\_P3D\_LS220\_m}$-$ with 4$^\circ$ angular resolution and 
\texttt{M\_P3D\_LS220\_m}$-$ (and its extension \texttt{M\_P3D\_LS220\_m$-$HC})
with 2$^\circ$ angular resolution, because these models
permit the prediction of astronomically relevant explosion and remnant properties. 
As for most of the other
models listed in Table~\ref{tab:simulations}, we will refer only to a few basic
aspects connected to a subset of the cases or to some relevant entries in the table. 
These aspects shall allow the reader to judge the two highlighted models in the broader 
context of our entire sample of 3D simulations. More information on the models not 
further considered here will be presented in a follow-up paper.

\begin{figure*}[tb!]
        \centering
        \includegraphics[width=0.48\textwidth]{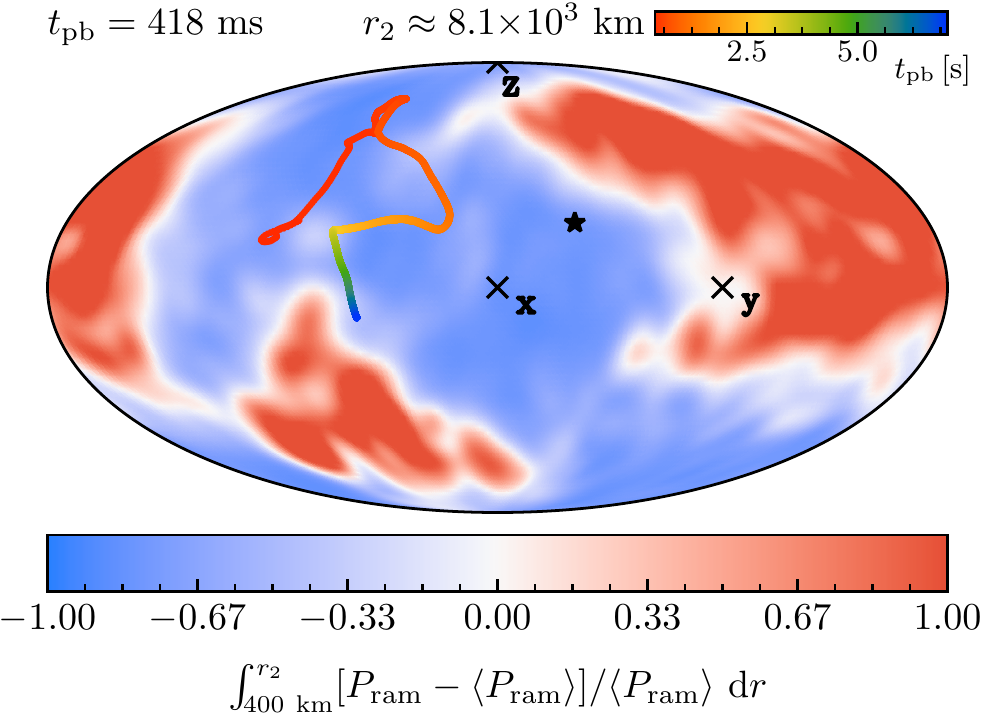}\hskip 15pt
        \includegraphics[width=0.48\textwidth]{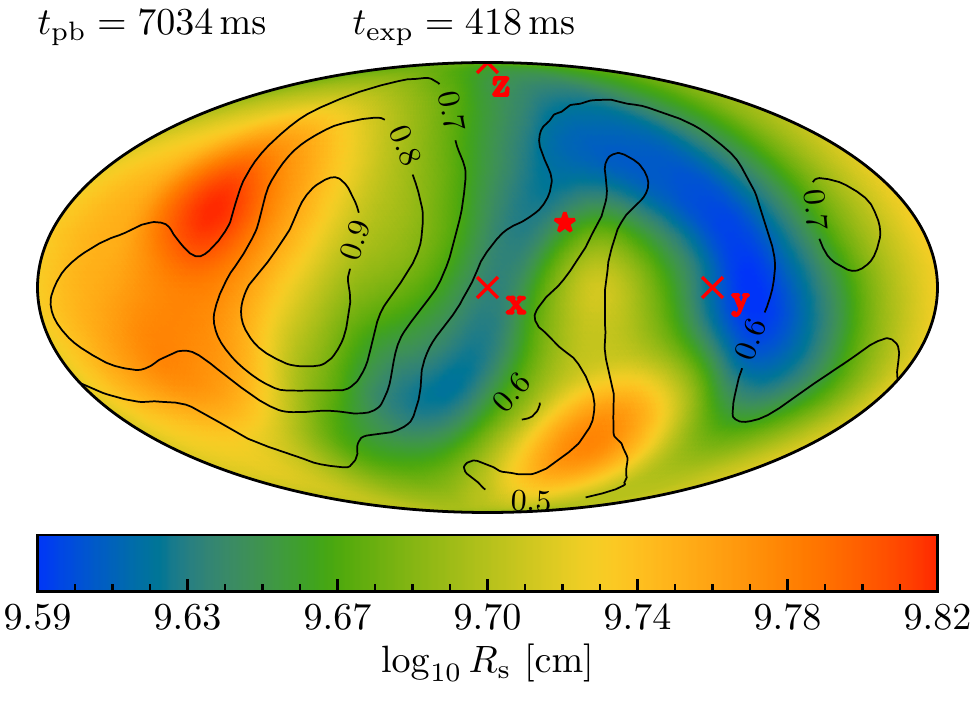}\\\vskip 20pt
        \includegraphics[width=0.48\textwidth]{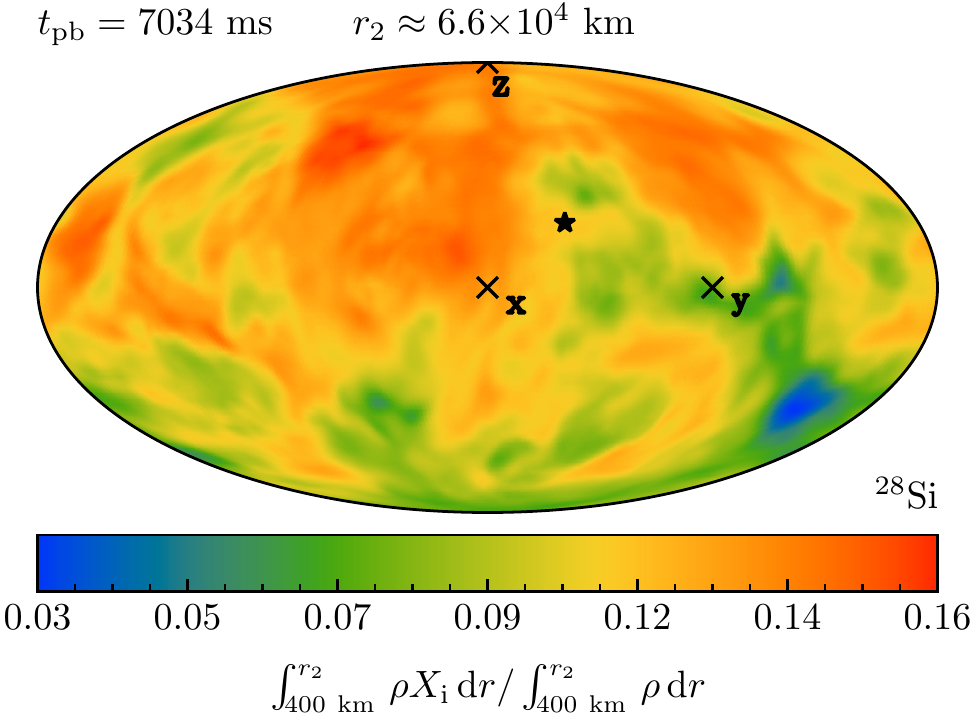}\hskip 15pt
        \includegraphics[width=0.48\textwidth]{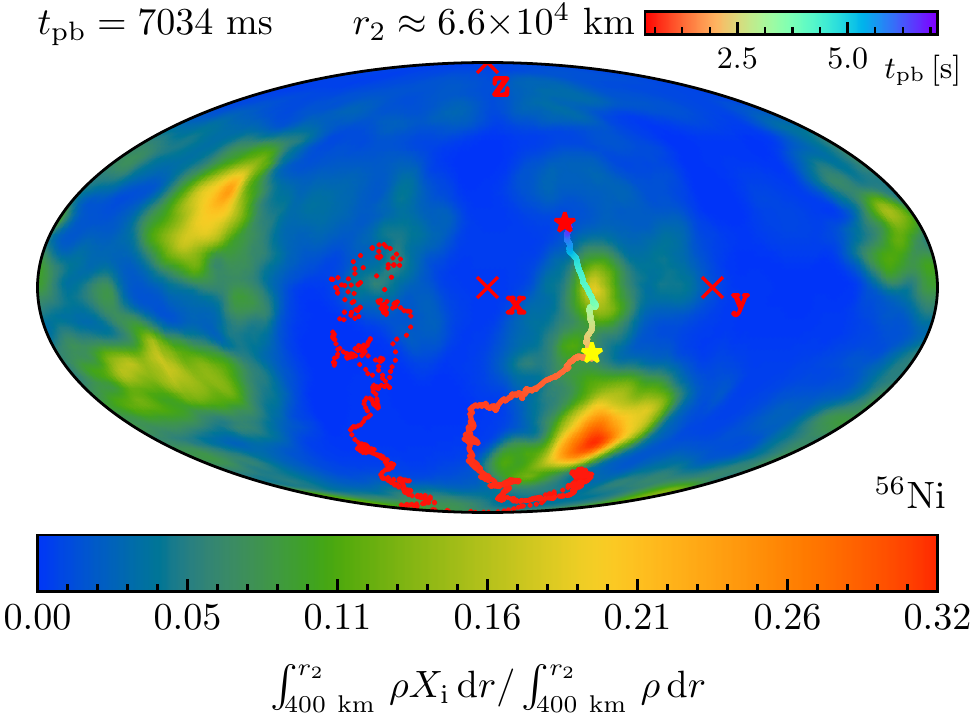}
\caption{Mollweide projections of different quantities of model \texttt{M\_P3D\_LS220\_m$-$}
        and its extension \texttt{M\_P3D\_LS220\_m$-$HC}. In all panels the $x$, $y$, and
	$z$ directions of the coordinate axes shown in Figures~\ref{fig:3D}, \ref{fig:2D},
	and \ref{fig:Ye} are indicated for enabling 3D orientation.
        {\em Top left:} Ram-pressure variations associated with large-scale asymmetries from
	convective O-shell burning in the 3D progenitor model, radially integrated from 400\,km to 8,100\,km
	at 0.418\,s after bounce. These variations determine the development of the anisotropic SN
	explosion because the shock expansion is faster in low-resistance ``cavities''.
	The marked path describes the movement of the dipole direction of the deformed shock with
	time (color coded along the path). It stays within the cavity in the left hemisphere.
	{\em Top right:} Shock-deformation contours around the onset of the explosion at 0.418\,s 
	after bounce (black lines corresponding to 0.6, 0.7, 0.8, and 0.9 times the maximum shock radius)
	overlaid on the shock surface at 7.034\,s (color-coded). At the latter time the shock has a bipolar
	shape with a considerable dipole asymmetry (see also Figures~\ref{fig:3D} and \ref{fig:2D}).
        {\em Bottom:} Line-of-sight averages of the mass fractions of $^{28}$Si ({\em left}) and
	$^{56}$Ni ({\em right}) at 7.034\,s after bounce, averaged between 400\,km and 6,600\,km.
	Black or red asterisks in the panels mark the direction of the total (hydrodynamic plus
	neutrino-induced) PNS kick at 7.034\,s (different colors of the asterisks were chosen for
	better visibility against the colored backgrounds). The path indicated by colored dots describes the 
        migration of the PNS kick direction with time, with the yellow asterisk marking its direction
	at 1.675\,s, when the \textsc{Vertex} neutrino transport is switched off and only the further
	acceleration of the PNS by the gravitational tug-boat effect is tracked later on. 
	}
\label{fig:explosionasymmetry}
\end{figure*}

\section{Results}
\label{sec:results}

\subsection{Dependencies on progenitor perturbations, resolution, and microphysics}
\label{sec:dimresphys}

Our simulations starting from 1D pre-collapse conditions do not develop
favorable conditions for an explosion, because the shock radius contracts quickly
after reaching a maximum at $\sim$200\,ms post bounce, when the infalling 
Si/O interface has crossed the shock. Strong SASI motions of the postshock
layer lead to large-amplitude, quasi-periodic shock expansion and contraction
episodes in all of these cases (see Figure~\ref{figapp:shockradii} in 
Appendix~\ref{app:addinfos} and 
\citealt{Glas+2019} for a detailed discussion of similar behavior in
non-exploding 20\,M$_\odot$ simulations).

In contrast, all models with 3D initial conditions exhibit
a phase of quasi-continuous shock expansion at $t_\mathrm{pb}\gtrsim 200$\,ms
and explode, if followed long enough (see models with boldfaced names in
Table~\ref{tab:simulations}). 
In these cases, the infall of the large-amplitude density and velocity 
perturbations in the convective O-shell fosters vigorous convection as the 
dominant hydrodynamic instability in the gain layer behind the shock
\citep[discussed in detail by, e.g.,][]{Mueller+2017,Mueller+2015,Couch+2013}. 
The typically larger shock radii 
in models started from 3D progenitor conditions can be verified for
270\,ms post bounce (the latest common time of all models) in 
Table~\ref{tab:simulations} and for a subset of models in 
Figure~\ref{figapp:shockradii}.

For \texttt{L}-models the shock-radius difference at 270\,ms after bounce
is reduced or even reversed 
(in \texttt{L\_P1D\_LS220\_m}$-$ compared to \texttt{L\_P3D\_LS220\_m}$-$, and 
\texttt{L\_P1D\_SFHo\_m}$+$ compared to \texttt{L\_P3D\_SFHo\_m}$+$) because 
here the models started from the 1D progenitor 
develop stronger and longer-lasting SASI activity, aided by a lack 
of resolution to follow in detail the growth of 
Rayleigh-Taylor and Kelvin-Helmholtz (``parasitic'') instabilities.
Such secondary instabilities tap energy from the coherent large-amplitude,
large-scale SASI motions and
thus limit the nonlinear saturation amplitude of the SASI 
\citep{Guilet+2010}. The latter authors argued that a resolution of about
2$^\circ$ or better may be needed to resolve the relevant scales of the 
flow structures at the onset of their growth, 
and 4$^\circ$ or coarser may not suffice. Such quantitative 
statements, however, depend on the quality of the hydrodynamics solver, 
in particular on the order of the employed discretization scheme, and therefore
have to be investigated for each numerical code specifically. 

Based on the resolution study performed with the \textsc{Prometheus} code by 
\citet{Melson+2020}, one may expect that our 2$^\circ$ simulations are close 
to convergence in their global behavior (although local effects and short-time 
variations are always subject to changes when turbulent flows play a role),
whereas the 4$^\circ$ models might hardly provide sufficient resolution to
achieve convergence. In that study, however, 1D initial conditions were 
employed and our default small-amplitude random perturbations were applied
to seed the convection, which in those models was the dominant postshock
instability instead of SASI. Employing 3D progenitor data changes the growth
conditions for postshock convection fundamentally. 
In Figure~\ref{figapp:shockradii} all models started from the 3D
progenitor (\texttt{L\_P3D\_LS220\_m}$-$, \texttt{M\_P3D\_LS220\_m}$-$, and
\texttt{H\_P3D\_LS220\_m}$-$) exhibit a similar evolution of the average 
shock radius, independent of their angular resolution.
In these models convection dominates in the postshock layer, too, but
convergence (in the sense of overall similarity in the evolution)
is fostered by the 3D progenitor asymmetries, which act as 
strong drivers of the hydrodynamic instabilities in the postshock volume.
In contrast, models started from the 1D progenitor, where SASI becomes the 
dominant nonradial instability in the postshock layer, exhibit the 
resolution dependence discussed by \citet{Guilet+2010}. These models still
display differences between the 1$^\circ$ and 2$^\circ$ simulations
with the tendency of smaller shock radii and less vigorous SASI activity
with better resolution (Figure~\ref{figapp:shockradii}).
Therefore  1$^\circ$ models would be preferable for these conditions, 
but are practically not feasible with the computing resources available to us.

The onset of the explosion is measured by the time when the maximum shock
radius reaches 400\,km. 
Lower resolution (in the \texttt{L}-cases) may tend to delay the explosion compared 
to the \texttt{M}-cases (for as much as 176\,ms in \texttt{L\_P3D\_SFHo\_m}$-$),
although \texttt{M\_P3D\_LS220\_m}$-$ and \texttt{L\_P3D\_\-LS220\_m}$-$ are
exceptions as mentioned above. 
The tendency of earlier and faster explosions in 3D simulations 
with higher resolution has been demonstrated before with the \textsc{Prometheus}
code by \citet{Melson+2020} and with \textsc{Fornax}, another SN code using
spherical polar coordinates, by \citet{Nagakura+2019} and \citet{Burrows+2020}.
This trend is typical of models where convective mass motions instead of SASI
are the dominant nonradial hydrodynamic instability in the postshock layer.
The trend is opposite when SASI dominates in the postshock layer. Under such
circumstances the average shock radii are found to be smaller with higher 
resolution (see Figure~\ref{figapp:shockradii} for an example), because the
aforementioned influence of the parasitic instabilities reduces the saturation
amplitude of the SASI by tapping energy from the coherent SASI mass motions.
Smaller shock radii and delayed
explosions with higher spatial resolution are also witnessed in 3D simulations 
of models with turbulent convection as the leading postshock instability when
Cartesian grids are employed \citep[see][]{Radice+2016}. The reason for this
finding could be smaller grid-induced perturbations with finer Cartesian grids,
an effect that inhibits the instigation of postshock convection by such 
perturbations
(for a detailed discussion of such resolution dependent aspects, see also
\citealt{Janka+2016} and \citealt{Melson+2020}).

In our set of models the shock runaway signalling the beginning of
the explosion occurs somewhat later ($\sim$10--200\,ms)
when using the SFHo EoS instead of the LS220 EoS, because with the latter
nuclear EoS the hot PNS contracts faster during the first several 100\,ms
after bounce. The correspondingly hotter neutrinosphere leads to
higher neutrino luminosities and harder neutrino spectra, both of
which enhance the postshock neutrino heating and thus support earlier explosions
\citep[in line with results presented in, e.g.,][]{Janka2012,Suwa+2013}.
For the same reason explosions set in earlier ($\sim$25--175\,ms)
when muons are included \citep[as first demonstrated by][]{Bollig+2017},
where the biggest shifts are witnessed in \texttt{L}-models
(Table~\ref{tab:simulations}).

The rest of our report is intended to be focused on the long-time evolution
towards energy saturation and the corresponding changes of the nucleosynthetic
composition in the shock-heated and neutrino-heated ejecta, both of which are
observationally relevant. This aim naturally centers our entire discussion on 
model \texttt{M\_P3D\_LS220\_m}$-$, its extension \texttt{M\_P3D\_LS220\_m$-$HC},
and their lower-resolution counterpart \texttt{L\_P3D\_LS220\_m}$-$. The available
computing resources did not permit us to continue all 3D runs to similarly
late evolution stages, for which reason our high-resolution (\texttt{H}) models
offer information limited to only $\sim$290\,ms after bounce, i.e., to a period
before the onset of the explosion. For the same
reason we could not conduct long-time evolution simulations for the SFHo EoS
and for models including muons. This is unfortunate because SFHo is a more
modern nuclear EoS, and the models including muons have the more complete
microphysics, but they are also computationally significantly more expensive.
We therefore decided to report additional results only of simulations with the 
LS220 EoS and without muon physics in the present paper.
We refer the reader to a follow-up paper for details on 
the results of our simulations with muons and the two versions of the nuclear
EoS until about 750\,ms after bounce.

\subsection{O-shell perturbations, shock revival, and explosion asymmetry}
\label{sec:explosionasymmetry}

We mainly concentrate on the 2$^\circ$ model \texttt{M\_P3D\_LS220\_m}$-$, 
which was followed with \textsc{Prometheus-Vertex} until 1.675\,s after bounce,
and compare some aspects with the 4$^\circ$ model \texttt{L\_P3D\_LS220\_m}$-$, 
whose evolution was tracked until 1.884\,s, also using the 
\textsc{Vertex} neutrino transport. By applying the neutrino-HC scheme mentioned
in Section~\ref{sec:numerics}, we continued model \texttt{M\_P3D\_LS220\_m}$-$ 
(maintaining 2$^\circ$ angular resolution) from 1.675\,s until 7.035\,s,
when the shock had entered the He-layer at $r = 52,000$\,km and an
enclosed mass of 4.45\,M$_\odot$, with a velocity of $\sim$8,000\,km\,s$^{-1}$
(Figures~\ref{fig:explosionprops1} and \ref{fig:explosionprops2}). 
The run extended from 1.675\,s to 7.035\,s is named 
\texttt{M\_P3D\_LS220\_m$-$HC} in Table~\ref{tab:simulations}.

\texttt{M\_P3D\_LS220\_m}$-$ and \texttt{L\_P3D\_LS220\_m}$-$
employ the same microphysics (LS220 EoS and no muons),
are both based on the 3D progenitor model, and differ only in their angular
resolutions. In Appendix~\ref{app:addinfos} we compare the evolution of their
average shock radii as functions of time. We also show these results for the
corresponding 1D SN runs, without and with artificial explosion,
and for the 3D SN runs of 
\texttt{L\_P1D\_LS220\_m}$-$, and \texttt{M\_P1D\_LS220\_m}$-$, all of which
were started from the 1D progenitor data. Moreover, the two high-resolution
cases of \texttt{H\_P1D\_LS220\_m}$-$ and \texttt{H\_P3D\_LS220\_m}$-$, 
which are based on the 1D and 3D progenitor data, respectively, are added
for comparion of the evolution prior to the onset of the SN explosion.

External forcing by infalling O-shell perturbations acts as an additional
driver of large-scale, nonradial fluid motions (convective overturn or SASI)
in the postshock layer besides neutrino heating and thus supports shock revival.
This can be quantified by an 
increase of efficiency factors for the conversion of neutrino heating to turbulent
kinetic energy, defined as
\begin{equation}
        \eta_{\mathrm{conv},i} = \frac{E_{\mathrm{kin},i}/M_\mathrm{g}}
        {[(R_\mathrm{s}-R_\mathrm{g})(\dot Q_\nu/M_\mathrm{g})]^{2/3}}
\end{equation}
\citep{Mueller+2015,Mueller+2017},
where $i$ denotes radial ($r$) or nonradial ($\theta$ plus $\phi$) motions.
The $E_{\mathrm{kin},i}$ are the corresponding turbulent kinetic energies (equations
(10) and (11) in \citealt{Mueller+2017}), $M_\mathrm{g}$ is the mass in the gain
layer, $R_\mathrm{s}$ and $R_\mathrm{g}$ angular averages of shock radius and gain radius,
respectively, and $\dot Q_\nu$ the integrated neutrino-heating rate in the gain layer.
Between $t_\mathrm{pb}\approx 200$\,ms until shortly after the explosion begins
at $t_\mathrm{pb}\approx 400$\,ms, we find efficiency factors between 0.3 and 0.4, 
in rough agreement with values obtained in SN simulations with a 
3D progenitor by \citet{Mueller+2017}.

The large-scale density variations in the infalling O-shell trigger a highly
asymmetric explosion (Figures~\ref{fig:3D} and \ref{fig:2D}) with the shock 
expanding faster in
directions of lower ram pressure (Figure~\ref{fig:explosionasymmetry}). The 
biggest expanding bubble is located close to the negative $y$-direction, and although
the shock dipole vector drifts considerably during the first second and
finds a stable position only after a few seconds, the deformation of the 
shock remains stable during the whole simulation, characterized by a huge
outward bulge in the negative $y$-hemisphere (close to the dipole direction) and 
a second big plume between the positive $y$-axis and negative $z$-axis 
(Figure~\ref{fig:explosionasymmetry}). Since the low-resolution model
\texttt{L\_P3D\_LS220\_m}$-$ was started from the same asymmetric 3D progenitor 
conditions, the most prominent plume driving the shock expansion develops also
in the negative $y$-direction of this model (see Figure~\ref{figapp:3Ds-cutsL}
in Appendix~\ref{app:addinfos}). However, a second, 
smaller plume grows between the positive $y$-axis and positive $z$-axis, 
which lies in the periphery of the wide ram-pressure ``depression'' 
extending around the positive $x$-direction in the upper left panel of
Figure~\ref{fig:explosionasymmetry}, just as the secondary plume does in 
model \texttt{M\_P3D\_LS220\_m}$-$.

\subsection{Explosion energy}
\label{sec:explenergy}

The blast-wave energy increases continuously from the onset of the explosion
until several seconds later. The diagnostic energy, 
$E_\mathrm{exp}^\mathrm{diag}$ (which is the integrated
internal plus gravitational plus kinetic energy of all postshock matter with
a positive value of this total energy), effectively saturates at $\sim$5\,s, whereas the 
explosion energy that accounts for the negative binding energy of overlying
stellar layers (``overburden'', OB), $E_\mathrm{exp}^{\mathrm{OB}-}$,
rises further to nearly converge to the 
diagnostic energy at a value around 0.98\,B at 7.035\,s 
(Figure~\ref{fig:explosionprops2}). During all this time, a ``classical'' spherically
symmetric neutrino-driven wind does not develop, but the PNS environment displays
turbulent mass flows, because accretion downdrafts carry matter to locations near
the gain radius ($R_\mathrm{g}$), where the gas absorbs energy from neutrino heating 
before it returns back outward to enhance the power of the explosion 
(Figure~\ref{fig:2D}). The entropy per baryon in the outflows stays
moderately high ($s\sim 20$--45\,k$_\mathrm{B}$ per nucleon) and $Y_e$ very 
close to 0.5.

As described by \citet{Marek+2009} and quantified by \citet{Mueller2015} and
\citet{Mueller+2017}, the mass outflow rate, $\dot M_\mathrm{out}$ (for
$v_r > 0$), is determined by neutrino
energy deposition, which provides the energy to gravitationally unbind the
matter carried to the vicinity of the PNS in downflows, whereas the long-time 
growth of the explosion energy is mainly fueled by the subsequent recombination 
energy of free nucleons re-assembling into $\alpha$-particles and heavy nuclei
when the downflow material expands back outward. The net effect is captured
by relations for the outgoing energy and enthalpy fluxes 
(Figure~\ref{fig:explosionprops2}):
\begin{equation}
F_\mathrm{e,out} = \dot M_\mathrm{out}\,\bar{e}_\mathrm{out} = 
\dot E_\mathrm{exp}^\mathrm{diag} \approx 0.5\,\dot Q_\nu \,,
\label{eq:feout}
\end{equation}
confirming results in \citet{Mueller+2017}, and
\begin{equation}
F_\mathrm{h,out} = \dot M_\mathrm{out}\,\bar{h}_\mathrm{out} =
\dot E_\mathrm{exp}^{\mathrm{OB}-} \approx \dot Q_\nu \,.
\label{eq:fhout}
\end{equation}
Here, $\dot M_\mathrm{out}$, $F_\mathrm{e,out}$, and $F_\mathrm{h,out}$ are
defined as in equations~(5), (6), and (7) of \cite{Mueller2015}, $\bar{e}_\mathrm{out}$ 
and $\bar{h}_\mathrm{out}$ are the total energy and enthalpy per baryon (including
gravitational energy) in outflows, and all of these quantities are evaluated at 400\,km.
These relations express energy conservation during the buildup of the explosion
energy. The latter is provided by neutrino energy deposition, which does not only
account for the growth of the positive energy transported by neutrino-heated ejecta 
($\dot E_\mathrm{exp}^\mathrm{diag}$ in Equation~(\ref{eq:feout})) but also for the 
energy that is needed to gravitationally unbind the stellar layers that are
swept up by the propagating shock, an effect that is included in 
$\dot E_\mathrm{exp}^{\mathrm{OB}-}$ in Equation~(\ref{eq:fhout}).
At $t_\mathrm{pb}\lesssim 2$\,s one can find time intervals when
$\dot Q_\nu$ exceeds $F_\mathrm{h,out} \sim \dot E_\mathrm{exp}^{\mathrm{OB}-}$.
This indicates that some neutrino energy is deposited in downflows instead of outflows,
and subsequently this energy gets lost again by the re-emission of neutrinos around the
(non-spherical) gain radius. Therefore the heating mechanism is not 
perfectly efficient to feed energy into the explosion. 

\begin{figure}[!]
        \begin{center}
\includegraphics[width=1.0\columnwidth]{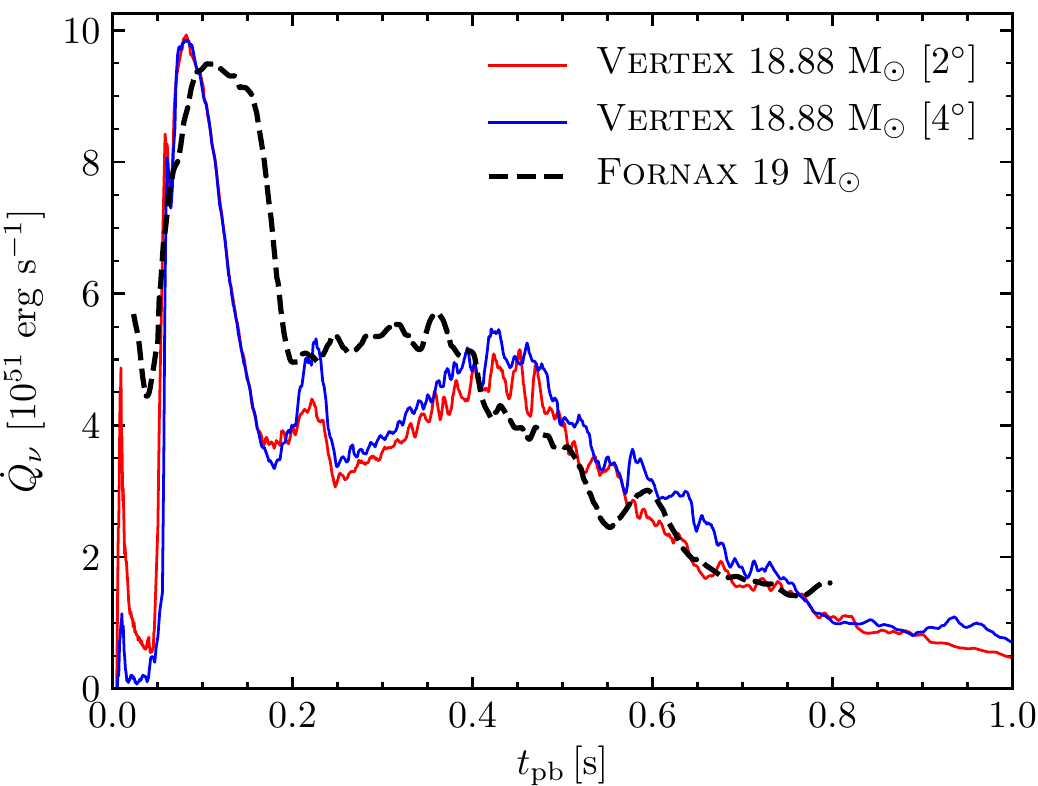}
        \end{center}
        \caption{Net neutrino energy deposition rate in the gain layer 
	 as function of time, $\dot Q_\nu(t)$,
	 for our models \texttt{M\_P3D\_LS220\_m$-$} (with 2$^\circ$ angular resolution)
	 and \texttt{L\_P3D\_LS220\_m$-$} (4$^\circ$ angular resolution) compared to 
	 the 19\,M$_\odot$ SN calculation with the \textsc{Fornax} code by
	 \citet{Burrows+2020}, see their Figure~5. The data from our
	 \textsc{Prometheus-Vertex} simulations are smoothed with a 
	 running window of 5\,ms and are plotted only until
	 1\,s post bounce for good comparison; the later evolution for model
	 \texttt{M\_P3D\_LS220\_m$-$} and its extension \texttt{M\_P3D\_LS220\_m$-$HC}
	 is depicted in the upper right panel of Figure~\ref{fig:explosionprops2}.
	 The considerable differences between \textsc{Prometheus-Vertex} and 
	 \textsc{Fornax} results between $\sim$0.1\,s and $\sim$0.4\,s are 
	 probably connected to the fact that the SN shock contracts after a local
	 maximum at $\sim$0.07\,s in our models (see Figure~\ref{figapp:shockradii}),
	 whereas it expands continuously in the \textsc{Fornax} simulation
	 \citep[see Figure~2 in][]{Burrows+2020}, which leads to a larger mass in
	 the gain layer and therefore a higher net neutrino energy deposition rate.}
\label{fig:fornax}
\end{figure}

Neutrinos thus power the explosion and there is no major net gain
of energy from nuclear dissociation in downflows and later recombination in
outflows as pointed out by \citet{Marek+2009}: 
Matter from the Si-, O-, and Ne-shells gets disintegrated during infall,
stores the nuclear binding energy transiently in free nucleons, which release it
again when they recombine to iron-group nuclei and $\alpha$-particles 
during re-ejection. The net gain of energy this way is $\lesssim$(0.5--1)\,MeV/nucleon,
and therefore the net energy carried on average per outflowing nucleon, which 
is $\bar{h}_\mathrm{out}\sim (3$--11)\,MeV
or $\bar{e}_\mathrm{out}\sim (1.5$--7.5)\,MeV (Figure~\ref{fig:explosionprops2}),
is provided by neutrino absorption.

These energies exhibit decreasing trends with time because at early times the 
downflows reach or penetrate $R_\mathrm{g}$, whereas at later times they return
outward from larger radii because of turbulent and shear interaction with 
outflows. \citet{Mueller+2017} demonstrated this by tracer particles.
Instead, we compute an average turnaround radius, $R_\mathrm{ret}$, from the
condition that neutino-heating drives the outflows, i.e.:
\begin{equation}
\eta_\mathrm{out} = 
\frac{\dot M_\mathrm{out}(R_\mathrm{ret})\,|e_\mathrm{tot}(R_\mathrm{ret})|}
{\dot Q_\nu} = 1 \,.
\label{eq:etaout}
\end{equation}
With equally good motivation this equation could be evaluated with the total 
enthalpy per nucleon, $|h_\mathrm{tot}(R_\mathrm{ret})|$, instead of 
$|e_\mathrm{tot}(R_\mathrm{ret})|$, but this would only lead to a small 
(outward) shift of $R_\mathrm{ret}$, because the gravitational binding 
energy typically dominates. Figure~\ref{fig:explosionprops2} shows that
$R_\mathrm{ret}$ is initially around 30\,km but at $t_\mathrm{pb}\gtrsim 3$\,s
grows to $\sim$100\,km, with no clear dependence on 2$^\circ$ or 4$^\circ$
resolution until $\sim$2\,s. The ratio $\alpha = \dot M_\mathrm{out}(400\,\mathrm{km})/
\dot M_\mathrm{out}(R_\mathrm{ret})$ is typically between 1 and 2, but
fluctuates stochastically to reach $\sim$10 in peaks, signalling that a large
fraction of the infalling matter returns outward from 
$r > R_\mathrm{ret} \gg R_\mathrm{g} \lesssim 30$\,km.
At 400\,km and $t_\mathrm{pb} \gtrsim 0.8$\,s we measure mass-inflow ($v_r < 0$) and
outflow ($v_r > 0$) rates roughly in balance, i.e., 
$\dot M_\mathrm{in} \approx \dot M_\mathrm{out}$
(Figure~\ref{fig:explosionprops2}), and consequently the PNS baryonic mass 
of 1.865\,M$_\odot$ becomes effectively constant (Figure~\ref{fig:PNSproperties};
Table~\ref{tab:simulations}), corresponding to a gravitational mass of 
1.675\,M$_\odot$ at 7\,s after bounce.

\citet{Burrows+2020} have recently published a large set of 3D models, exploding
and non-exploding, which might partly also develop sizable explosion
energies. However, all of these simulations were terminated well before 1\,s
after bounce at a stage when the diagnostic explosion energy had just started to grow.
Their most energetic model, a 19\,M$_\odot$ case, has reached between 0.2\,B
and 0.3\,B at the end of their computations at 0.8--0.87\,s (depending on the 
employed resolution). This model is also the most suitable one for a comparison
with \texttt{M\_P3D\_LS220\_m}$-$, although a detailed assessment is hampered 
by the fact that the progenitors are not the same. Despite a slightly smaller 
PNS mass and the correspondingly lower neutrino luminosities in the simulation
by \citet{Burrows+2020}, the neutrino heating rate as plotted in their Figure~5
is fairly similar to our model at $t_\mathrm{pb}\gtrsim 0.4$\,s 
(see Figure~\ref{fig:fornax}).\footnote{We
point out that the net neutrino heating rate is considerably higher than the
growth rate of the diagnostic explosion energy, $\dot E_\mathrm{exp}^\mathrm{diag}$,
plotted in Figure~\ref{fig:explosionprops2}, because most of the deposited energy
is used up to gravitationally unbind the expelled matter, i.e., to overcome its
gravitational binding energy. Moreover, we remark that the neutrino energy 
deposition rate during the first second after bounce dwarfs by far estimates of
the heating in the postshock layer due to the energy flux carried by gravity waves
and outgoing acoustic waves. Such waves can be excited by PNS 
convection and their energy and its fraction potentially transmitted upwards to
the postshock region have recently been estimated on grounds of 1D simulations by
\citet{Gossan+2020}.}
In line with this finding, the diagnostic explosion
energy and its growth rate at 0.87\,s are also quite close to our 
model.\footnote{It should be noted, however, that the diagnostic explosion
energies plotted in Figure~6 of \citet{Burrows+2020} differ in their definition
from our use of $E_\mathrm{exp}^\mathrm{diag}$. The diagnostic explosion energies 
given by \citet{Burrows+2020} include `reassociation energy' and gravitational
binding energy of matter exterior to the shock, whereas our diagnostic 
explosion energies do not.} Similarly, although the shock expansion sets in
earlier and without a transient phase of contraction in the simulation of
\citet{Burrows+2020}, and although the diagnostic explosion energy rises correspondingly 
earlier, the shock radius and shock velocity are again compatible with those 
in our model \texttt{M\_P3D\_LS220\_m}$-$ at the time when \citet{Burrows+2020}
stopped their calculation.

Based on a large set of 2D simulations, \citet{Nakamura+2015} explored
systematic features of axi-symmetric neutrino-driven SN models. They
concluded that the accretion luminosity becomes higher for progenitor stars with high
core-compactness parameter \citep{OConnor+2011}, because the latter correlates with
the density of the shells surrounding the degenerate iron core. \citet{Nakamura+2015}
further argued that the higher accretion luminosity leads to a higher growth rate of 
the diagnostic energy and of the synthesized nickel mass.
While we agree that there is a close correlation between the rate of mass-infall
towards the PNS and the core compactness of the progenitor, we emphasize that
the long-time growth of the explosion energy and nickel yield in our 3D model is
{\em not} connected to a higher accretion luminosity. As mentioned above, the PNS 
stops accreting at about 0.8\,s after bounce. Instead, the long-time growth of the 
explosion energy is fueled by the efficient neutrino heating of the infalling
flows of gas, which absorb energy from neutrinos and then return outward. At 
$t_\mathrm{pb} \gtrsim 0.8$\,s the mass inflow and outflow rates have become effectively 
equal, and the subsequent steep rise of the explosion energy is a consequence of the 
long-lasting high mass-infall rate (which is correlated with the core compactness)
on the one hand, and the efficient neutrino heating due to high PNS cooling luminosities,
which blows the infalling matter out again, on the other hand.
The process of reversing inflow to outflow is significantly more efficient in 3D models 
than in the case of axi-symmetric geometry 
\citep[as discussed in detail by][]{Mueller2015,Mueller+2017,Mueller+2019}.

\begin{figure}[!]
        \begin{center}
\includegraphics[width=1.0\columnwidth]{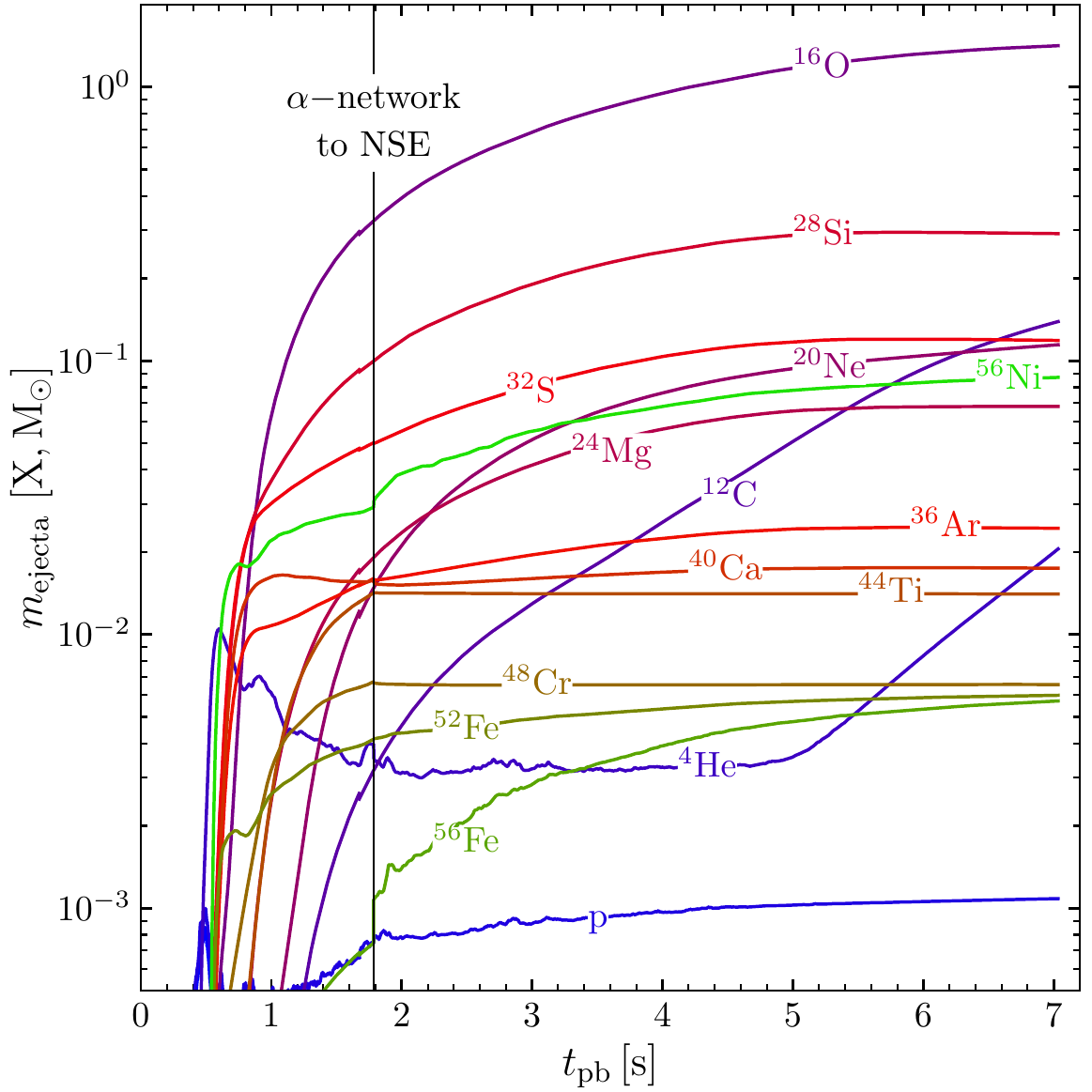}
        \end{center}
\caption{Ejecta masses of chemical species in the postshock volume of model
	 \texttt{M\_P3D\_LS220\_m$-$} and its extension 
	 \texttt{M\_P3D\_LS220\_m$-$HC}.
         The vertical line marks the moment when the $\alpha$-network was replaced
         by an NSE treatment applied above $T = 0.343$\,MeV.}
\label{fig:elementmasses}
\end{figure}

\begin{figure}[!]
        \begin{center}
\includegraphics[width=1.0\columnwidth]{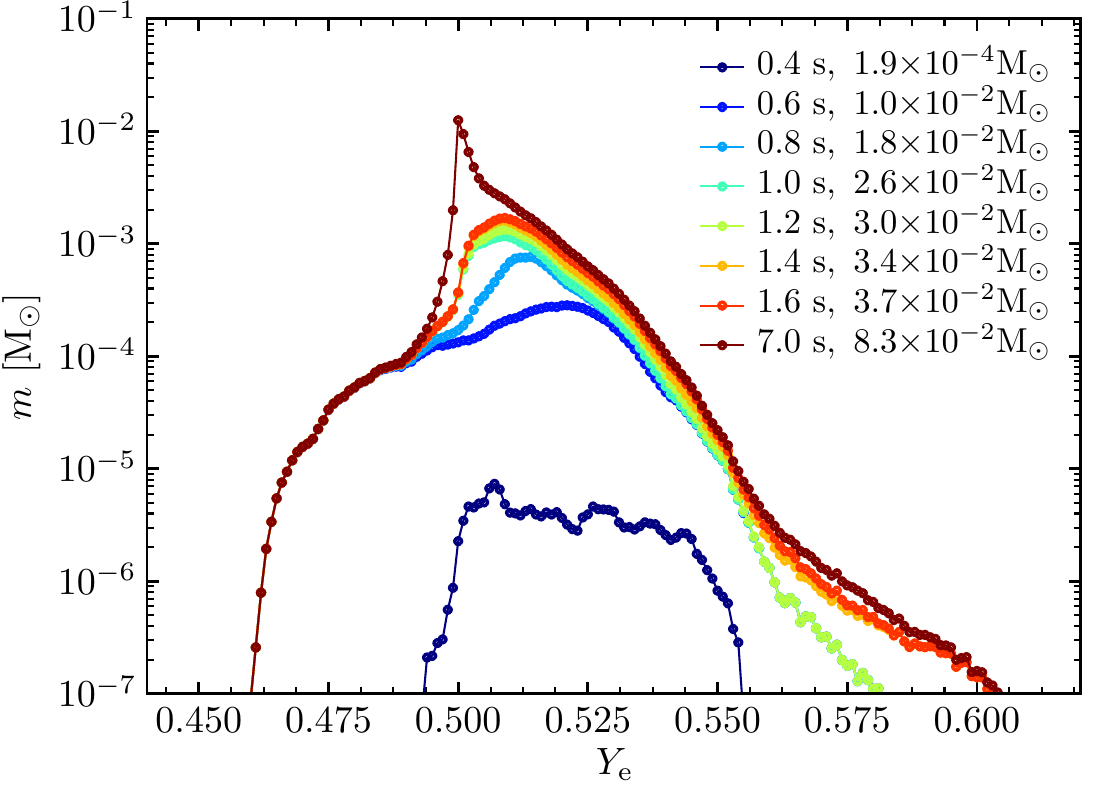}
        \end{center}
	\caption{Mass distributions versus electron fraction $Y_e$ of 
	 neutrino-heated ejecta at different post-bounce times for model
         \texttt{M\_P3D\_LS220\_m$-$} and its extension \texttt{M\_P3D\_LS220\_m$-$HC}.
         The distributions were constructed from all ejecta flowing out through a sphere
	 of 250\,km radius.} 
\label{fig:Yemass}
\end{figure}

\begin{figure*}[!]
        \begin{center}
\includegraphics[width=1.0\textwidth]{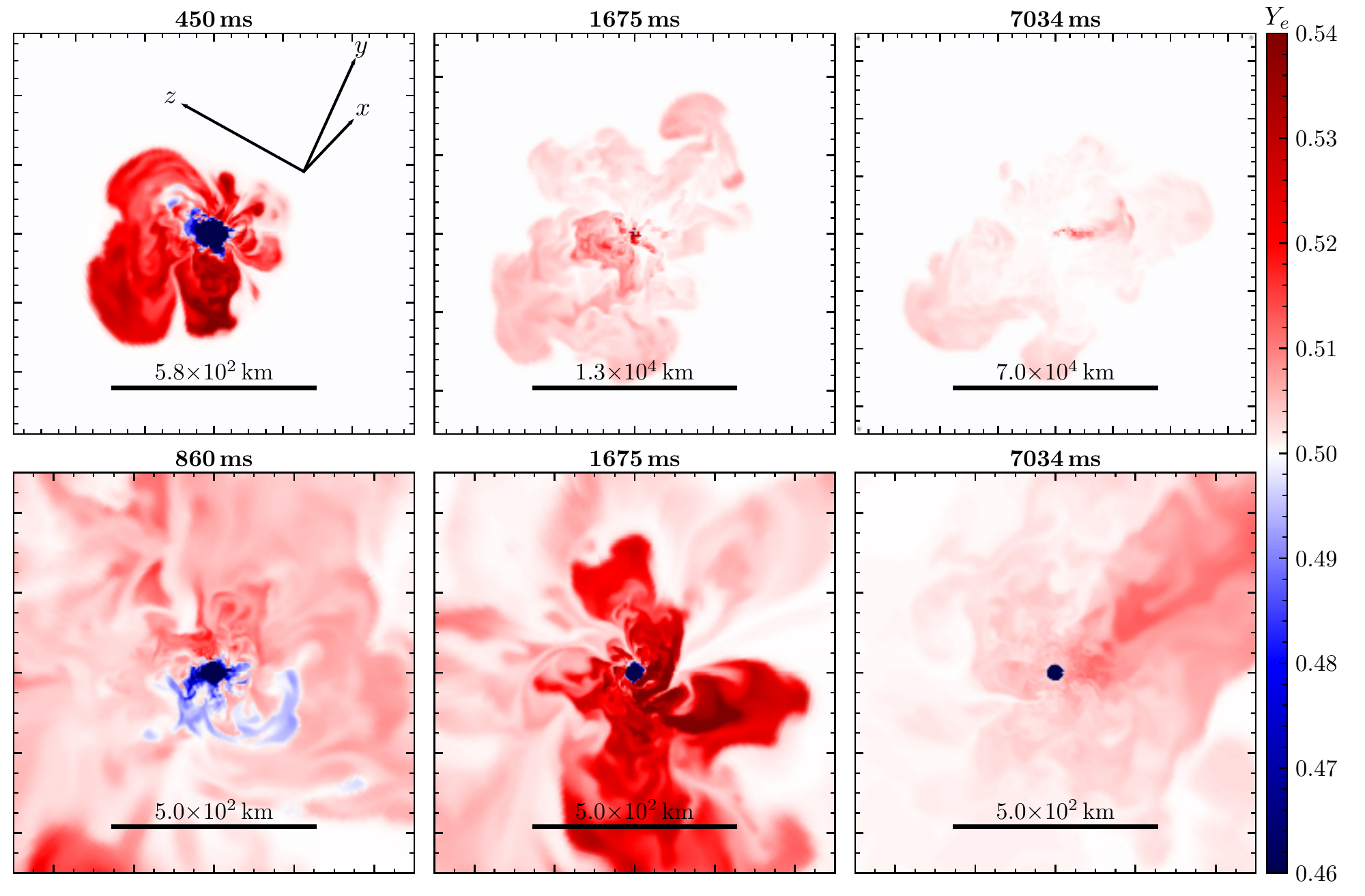}
        \end{center}
\caption{Electron fraction in the outflows of model \texttt{M\_P3D\_LS220\_m$-$}
         and its extension \texttt{M\_P3D\_LS220\_m$-$HC}.
         The cross-sectional plane is the same as the one chosen in
         Figure~\ref{fig:2D}.
         {\em Top:} $Y_e$ color-coded at 0.450\,s, 1.675\,s, and 7.034\,s matching
         the times in Figure~\ref{fig:2D}.
         {\em Bottom:} Close-ups of the turbulent vicinity of the PNS at 0.86\,s,
         1.675\,s, and 7.034\,s after bounce.
         $Y_e$ in the ejecta is close to or slightly larger than 0.50.
         }
\label{fig:Ye}
\end{figure*}

\subsection{Nucleosynthesis yields}
\label{sec:nucleosynthesis}

Radioactive $^{56}$Ni and other elements heavier than $^{28}$Si are
produced either by explosive burning or normal and alpha-rich freeze-out
\citep[see, e.g.,][]{Hix+1999a,Magkotsios+2010}. They are
ejected in the highest concentrations in regions where the shock is strongest
(Figure~\ref{fig:explosionasymmetry}), in agreement with findings of
\citet{Wongwathanarat+2013}, whereas such a correlation is effectively lost for
$^{28}$Si and lighter elements (Figure~\ref{fig:explosionasymmetry}).
Using the small network we obtain ejecta
masses of 0.29\,M$_\odot$ of silicon and 1.42\,M$_\odot$ of oxygen.
A mass of 0.028\,M$_\odot$ of $^{56}$Ni is nucleosynthesized during the
\textsc{Prometheus-Vertex} run of model \texttt{M\_P3D\_LS220\_m}$-$ until $\sim$1.7\,s,
increasing to an upper estimate of $\lesssim$0.087\,M$_\odot$ until $\sim$7\,s 
during the continuation by model \texttt{M\_P3D\_LS220\_m$-$HC}
(Figure~\ref{fig:elementmasses}). Efficient formation of $^{56}$Ni under
conditions of alpha-rich freeze-out is enabled because the electron fraction,
$Y_e$, of most of the neutrino-processed ejecta varies between $\sim$0.495 and 
$\sim$0.52 (Figures~\ref{fig:Yemass} and \ref{fig:Ye}).
Since the extended run was done with an NSE solver applied down to lower temperatures
instead of the network, the $^{56}$Ni production may be overestimated, as suggested
by the discontinuous growth rate of the $^{56}$Ni mass after the network-NSE
switch (Figure~\ref{fig:elementmasses}). Our final $^{56}$Ni mass is therefore an
upper limit, and we expect the actual mass to be around 0.05\,M$_\odot$ (see below). 
Nevertheless, values in this range demonstrate that $^{56}$Ni masses close to those
of typical CCSNe can be ejected in 3D neutrino-driven explosions.
The continuous growth of the ejected masses of intermediate-mass elements displayed in
Figure~\ref{fig:elementmasses} mostly reflects the composition of the stellar shells
swept up by the outgoing shock.

Interestingly, the mass-versus-$Y_e$-distribution of the neutrino heated-ejecta
in our simulation of \texttt{M\_P3D\_LS220\_m$-$} and its extension
\texttt{M\_P3D\_LS220\_m$-$HC} (Figure~\ref{fig:Yemass};
see also Figure~\ref{figapp:Yemass4d} for the corresponding result of our 
4$^\circ$-model \texttt{L\_P3D\_LS220\_m$-$}), is perfectly compatible with 
important nucleosynthetic constraints discussed by \citet{Hoffman+1996}. 
According to these authors, typical Type-II SNe should eject 
$\lesssim$$10^{-4}$\,M$_\odot$ of neutrino-heated matter with $Y_e\lesssim 0.47$
in order not to overproduce $N = 50$ closed neutron shell nuclei.
This approximate limit is fulfilled by the $2^\circ$ model as well as the $4^\circ$
explosion model (see also Appendix~\ref{app:addinfos}).

The use of a small $\alpha$-network in the \textsc{Prometheus-Vertex} run does not
permit accurate predictions of the nucleosynthetic yields. For example, it overestimates
the production of $^{44}$Ti by up to a factor of 100 \citep[see also][]{Wongwathanarat+2013}.
Large-network calculations in a post-processing analysis will have to clarify whether
most of this $^{44}$Ti as well as some fraction of $^{48}$Cr might rather end up as iron
and $^{56}$Ni, as it happens when we replace the network by the NSE solver.

\begin{figure*}[!]
        \centering
        \includegraphics[width=0.48\textwidth]{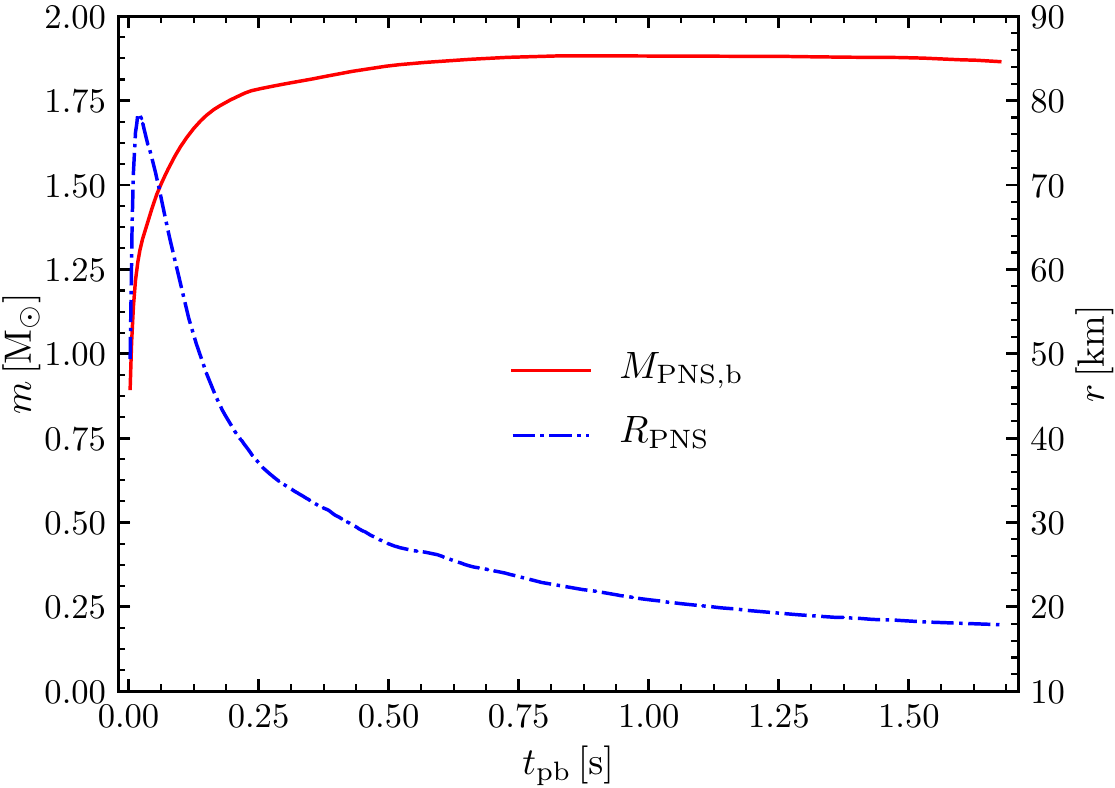}\hskip 15pt
        \includegraphics[width=0.48\textwidth]{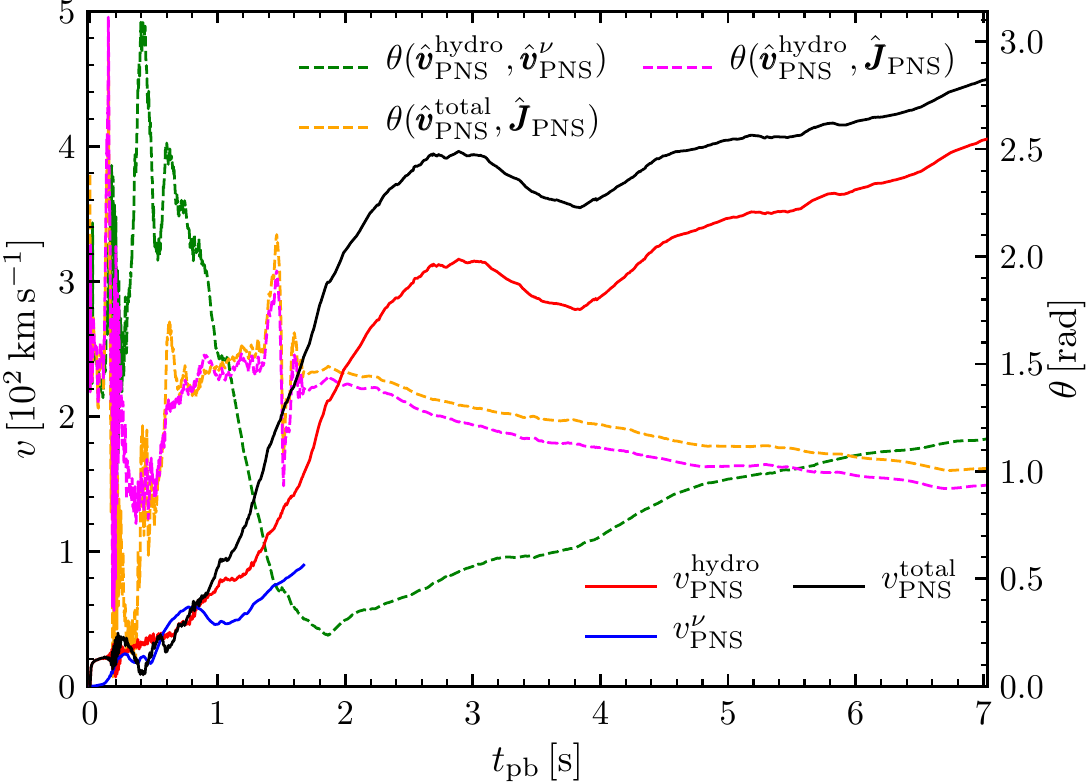}\\\vskip 0pt
        \includegraphics[width=0.48\textwidth]{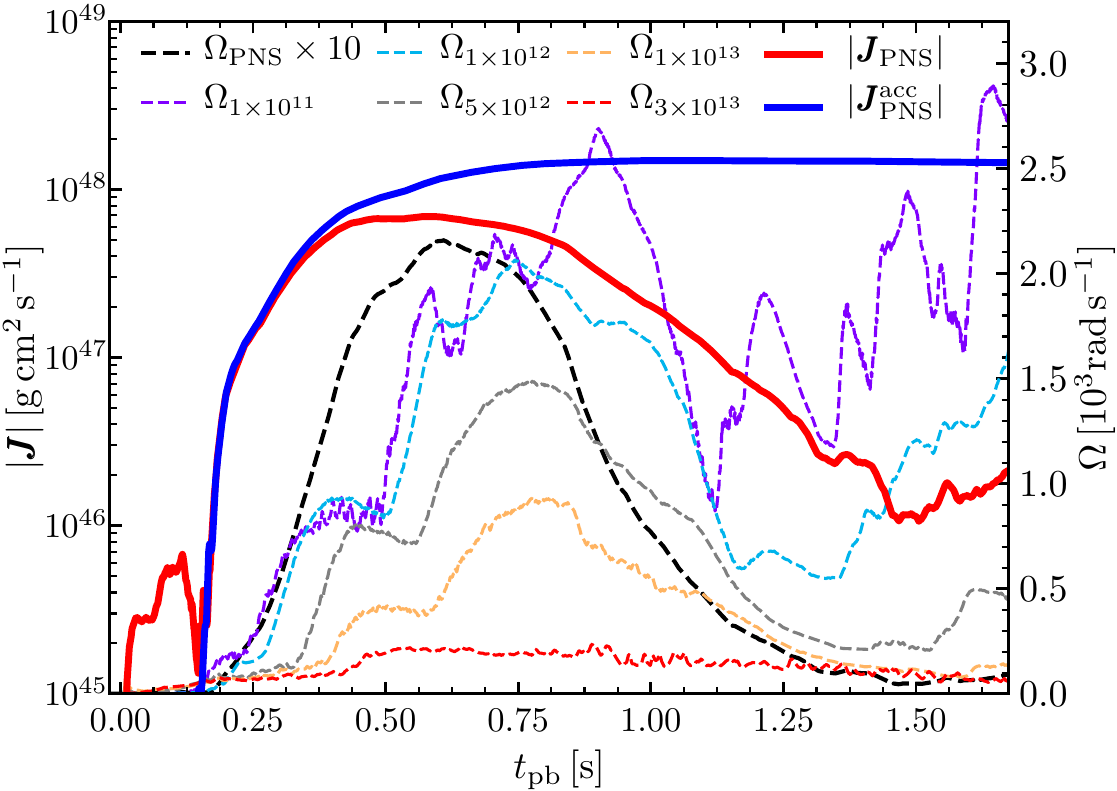}\hskip 15pt
        \includegraphics[width=0.48\textwidth]{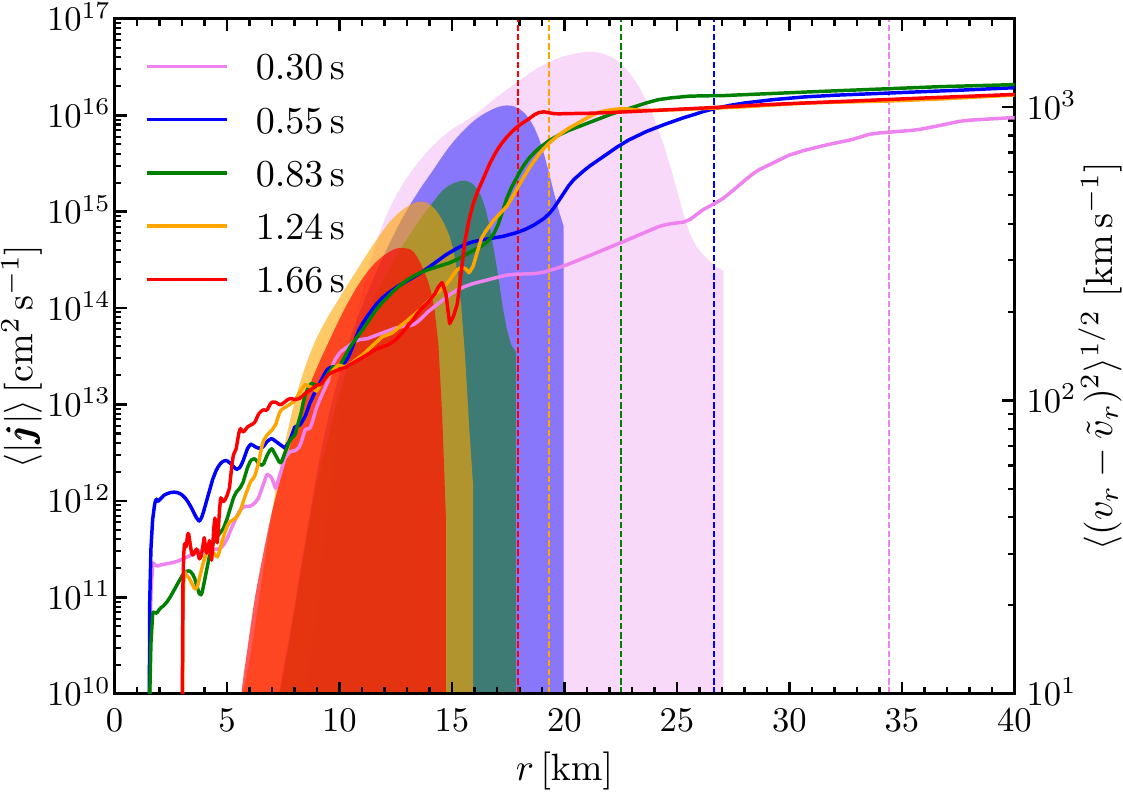}
        \caption{Proto-neutron star (PNS) properties of model \texttt{M\_P3D\_LS220\_m$-$}
        during the simulation with \textsc{Vertex} neutrino transport. The top right panel
        also shows data from the extended calculation of \texttt{M\_P3D\_LS220\_m$-$HC}
        with heating and cooling treatment instead of \textsc{Vertex} transport.
        {\em Top left:} Baryonic mass and radius, both corresponding to the volume where the
        mass-density $\rho \ge 10^{11}$\,g\,cm$^{-3}$.
        {\em Top right:} Neutrino-induced kick velocity ($v_\mathrm{PNS}^\nu$), hydrodynamic kick
        ($v_\mathrm{PNS}^\mathrm{hydro}$), and sum of both (total kick, $v_\mathrm{PNS}^\mathrm{total}$)
        and relative angles between hydrodynamic and neutrino-induced kick velocities, hydrodynamic
        kick and PNS angular momentum vector, and total kick and PNS angular momentum vector. The
        evolution of hydrodynamic and total kick can be tracked also after the end of the \textsc{Vertex}
        calculation, the neutrino-induced kick and the PNS spin direction are unchanged after
        1.675\,s post bounce.
        {\em Bottom left:} Net accretion of angular momentum onto the PNS
        ($|\pmb{J}_\mathrm{PNS}^\mathrm{acc}|$),
        total PNS angular momentum (for $\rho \ge 10^{11}$\,g\,cm$^{-3}$; $|\pmb{J}_\mathrm{PNS}|$),
        average angular velocity $\Omega_\mathrm{PNS}$ if the PNS were rigidly rotating with
        $|\pmb{J}_\mathrm{PNS}|$, and angular velocities $\Omega_\rho$, averaged over isosurfaces of
        density $\rho$ for the values given as subscripts (in g\,cm$^{-3}$).
        {\em Bottom right:} Radial profiles of the angle-averaged (with density-weighting)
        specific angular momentum at different post-bounce times (colored lines). The corresponding
        colored areas show angle-averaged radial turbulent velocities, indicating the convective
        layer inside the PNS. The thin vertical dashed lines mark the radii where the average
        density is $10^{11}$\,g\,cm$^{-3}$. It is obvious that the PNS rotation is highly differential,
        the specific angular momentum is highest in the convectively stable outer PNS layers, and
        inward transport of angular momentum by PNS convection is not efficient.
        }
\label{fig:PNSproperties}
\end{figure*}

Most important in the present context, however, is the fact that in
neutrino-driven explosions there is a tight correlation between the mass ejected in
neutrino-heated freeze-out material, $M_\mathrm{ej}^\mathrm{fo}$, and the diagnostic 
explosion energy, $E_\mathrm{exp}^\mathrm{diag}$. Since on average each ejected nucleon
contributes a net energy of $\bar{e}_\mathrm{out}\sim$\,5\,MeV to the explosion energy
\cite[][see also Section~\ref{sec:explenergy}]{Janka2001,Scheck+2006,Marek+2009,Mueller2015},
about 0.1\,M$_\odot$ of neutrino-heated matter must be expelled to power a SN that
blows up with an energy of 1\,B:\footnote{The diagnostic explosion energy of model
\texttt{L\_P3D\_LS220\_m$-$} is approximately 0.63\,B at 1.884\,s after bounce 
(Table~\ref{tab:simulations}), which is $\sim$0.12\,B higher than
in model \texttt{M\_P3D\_LS220\_m$-$} at roughly the same time. This is fully
compatible with the estimate of Equation~(\ref{eq:freezeoutmass}), because the
baryonic mass of the PNS in model \texttt{L\_P3D\_LS220\_m$-$} at this instant is 
$\sim$$1.2\times 10^{-2}$\,M$_\odot$ lower than in model \texttt{M\_P3D\_LS220\_m$-$},
which suggests a slightly lower accretion efficiency and a correspondingly larger rate 
of mass outflow. The increased ejecta mass carries the additional
supply of explosion energy, because it has absorbed extra energy deposited by neutrinos
near the gain radius or turnaround radius.}
\begin{equation}
	E_\mathrm{exp}^\mathrm{diag} \approx 1\,\mathrm{B}\times 
	\left(\frac{\bar{e}_\mathrm{out}}{5\,\mathrm{MeV/nucleon}}\right)\times
        \left(\frac{M_\mathrm{ej}^\mathrm{fo}}{0.1\,\mathrm{M}_\odot}\right).
        \label{eq:freezeoutmass}
\end{equation}

We can thus, very crudely, bracket the likely yield of $^{56}$Ni by the following
arguments. At 1.7\,s our small network has produced 0.028\,M$_\odot$ of $^{56}$Ni.
At this time the diagnostic explosion energy has just reached half of its final
value. The rest of the explosion energy is provided by continuous neutrino heating
of matter that falls inward with a fairly high rate for seconds and gets re-ejected 
with $\dot M_\mathrm{out}\approx |\dot M_\mathrm{in}|$ after having absorbed energy
from neutrinos (Figure~\ref{fig:explosionprops2}). Since the energy or enthalpy per 
ejected nucleon are nearly constant until about 5\,s, about the same mass of 
neutrino-heated matter shares the explosion energy before and after 1.7\,s.
Assuming the same efficiency for the recombination of neutrino-heated ejecta to
$^{56}$Ni before and after 1.7\,s, an optimistic total yield of $^{56}$Ni can
be estimated roughly as $2\times 0.028\,\mathrm{M}_\odot = 0.056\,\mathrm{M}_\odot$.

However, some of the very early production of $^{56}$Ni might have been made by
explosive burning in the shock-heated ejecta, which takes place only for a few
hundred milliseconds and could account for the steep
rise of the ejected $^{56}$Ni to a short plateau of $\sim$\,0.018\,M$_\odot$ at
about 1\,s (see Figure~\ref{fig:elementmasses}). Assuming that none of this early
$^{56}$Ni comes from freeze-out (which is an unlikely, pessimistic case), freeze-out
until 1.7\,s would only contribute 0.01\,M$_\odot$ of the $^{56}$Ni made until that
time by our small network. Again assuming the same freeze-out production of
$^{56}$Ni before and after 1.7\,s, we thus derive a pessimistic lower estimate of
the total yield as
$0.028\,\mathrm{M}_\odot + 0.01\,\mathrm{M}_\odot = 0.038\,\mathrm{M}_\odot$.

The actual number is probably somewhere between 0.038\,M$_\odot$ and 0.056\,M$_\odot$.
Since some of the $^{44}$Ti, $^{48}$Cr, and $^{52}$Fe might well be $^{56}$Ni instead,
we tend to favor a value towards the optimistic side and consider $\sim$\,0.05\,M$_\odot$
as a reasonable, though crude, estimate of the total nucleosynthesized $^{56}$Ni
mass. This would be well compatible with the yield of SN~1987A, whose
explosion energy and $^{56}$Ni ejecta have been determined from light-curve analysis
to be about 1.5\,B and 0.0765\,M$_\odot$, respectively
\citep[][and references therein]{Utrobin+2011}. We therefore think there are no
reasons to postulate a ``nickel mass problem'' connected to the slow increase of
the explosion energy in neutrino-driven SN blasts~\citep{Suwa+2019,Sawada+2021}.
It is obvious from our 3D simulations that neutrino-driven explosions eject sufficient
freeze-out material to account for the observed $^{56}$Ni masses in Type-II SNe.
Detailed nucleosynthesis calculations will have to reveal the exact fraction of this
radioactive species in the neutrino-heated ejecta.

\subsection{Proto-neutron star properties}
\label{sec:PNSproperties}

\subsubsection{Proto-neutron star spin}
\label{sec:PNSspin}

At $t_\mathrm{pb} \lesssim 0.9$\,s downflows emanating from the convectively perturbed
O-shell (with specific angular momentum up to $j>10^{16}$\,cm$^2$\,s$^{-1}$)
carry considerable angular momentum ($\sim$\,$1.5\times 10^{48}$\,erg\,s;
Figure~\ref{fig:PNSproperties}) 
toward the PNS. Some of this high-$j$ matter ($\sim$50--60\% at 
$t_\mathrm{pb} \lesssim 0.9$\,s) is blown out again by neutrino heating, 
but its partial accretion spins up the outer PNS layers at densities 
$\rho\lesssim 10^{13}\,$g\,cm$^{-3}$ with a dominant negative $y$-component,
which is $\sim$3 times greater than the $x$- and $z$-components. In peaks the
angular velocity reaches (2--$3)\times 10^3$\,rad\,s$^{-1}$ at densities 
$10^{11}\,\mathrm{g\,cm}^{-3}\lesssim\rho\lesssim 10^{12}$\,g\,cm$^{-3}$ 
(Figure~\ref{fig:PNSproperties}) and decreases inward. 
Only about half of the accreted angular momentum 
(i.e., of the excess of inflow compared to outflow of angular momentum), however,
shows up in the PNS mantle at $t_\mathrm{pb} \lesssim 0.6$\,s (compare the blue
and red bold solid lines in the lower left panel of Figure~\ref{fig:PNSproperties}).
After accretion onto the PNS ends at $t_\mathrm{pb} \sim 0.8$\,s, 
the angular momentum stored in the near-surface layers with densities between a 
few times $10^{11}$\,g\,cm$^{-3}$ and some $10^{13}$\,g\,cm$^{-3}$ declines quite 
rapidly with time. Therefore, there is a huge drain of accreted angular momentum
from $t_\mathrm{pb} \sim 0.4$\,s until $t_\mathrm{pb} \sim 1.3$\,s, in course of
which the outer PNS layers above a few $10^{11}$\,g\,cm$^{-3}$ decelerate 
dramatically.

We have not (yet) been able to identify the origin of this angular momentum 
loss unambiguously.
The effect is only moderately dependent on the angular resolution.
The long-time evolution of the angular momentum on the entire computational grid
and in the PNS volume is qualitatively very similar and quantitatively only little
different in model \texttt{M\_P3D\_LS220\_m$-$} and its lower-resolution counterpart
\texttt{L\_P3D\_LS220\_m$-$} (compare the bottom left panel
of Figure~\ref{fig:PNSproperties} with Figure~\ref{figapp:PNSrot4d} and see
Figure~\ref{figapp:angmomevol}).
Despite this modest resolution dependence, we cannot exclude that the deceleration 
of the PNS rotation is caused by imperfect numerical conservation of angular momentum
in the hydrodynamics scheme \citep[as discussed by][]{Mueller2020}, 
possibly enhanced by the use of the Yin-Yang grid and the corresponding 
treatment of the fluxes across the grid interfaces. More information on this aspect
can be found in Appendix~\ref{app:angmom}.

It might also be possible that the angular momentum loss from the PNS surface
layers is, maybe partly, connected to the coupling of hydrodynamics and RbR+ neutrino
transport. This approximation includes nonradial components of the neutrino pressure 
gradients in the fluid equation of motion in the optically thick regime and nonradial 
neutrino advection with the moving medium in the transport equations
\citep{Buras+2006}, but it disregards effects of neutrino viscosity and nonradial
neutrino flux components. Therefore it cannot handle angular momentum transport by 
neutrinos, but it is difficult to track down the exact consequences of the 
nonradial neutrino pressure derivatives for the angular momentum 
evolution of the background medium the neutrinos propagate through.

Interestingly, however, the growing mismatch between the angular momentum accreted
onto the PNS and the angular momentum available in its near-surface layers 
could indeed be caused by neutrinos carrying away this difference. 
Angular momentum is exchanged between 
matter and neutrinos in layers where both are strongly coupled (i.e., at 
$\rho \gtrsim \mathrm{some}~10^{11}$\,g\,cm$^{-3}$), before the neutrinos escape
more easily from lower-density regions. These low-density layers should continue 
to spin rapidly, which indeed is the case in Figure~\ref{fig:PNSproperties}. 
Considering the neutrino-energy loss, $\Delta E_\nu$, during the relevant time
between $\sim$0.4\,s and $\sim$1.3\,s after bounce, the PNS radius,
$R_\mathrm{PNS}$, and the average PNS angular velocity at $R_\mathrm{PNS}$, 
$\overline{\Omega}_\mathrm{PNS}^\mathrm{\,surf}\sim \Omega_{1\times 10^{11}}$, 
one obtains a rough estimate for the angular momentum extracted by neutrinos as 
\begin{eqnarray}
\Delta J_\nu &=& 
\frac{1}{c^2}\,\Delta E_\nu\, R_\mathrm{PNS}^2\,\overline{\Omega}_\mathrm{PNS}^\mathrm{\,surf} \nonumber\\ 
&\approx&
1.4\times 10^{48}\,\left(\frac{\Delta E_\nu}{10^{53}\,\mathrm{erg}}\right) \left(\frac{R_\mathrm{PNS}}{25\,\mathrm{km}}\right)^{\! 2} \nonumber\\
&\phantom{\approx}&
\phantom{1.4\times 10^{48}\,}
\left(\frac{\overline{\Omega}_\mathrm{PNS}^\mathrm{\,surf}}{2000\,\mathrm{rad\,s}^{-1}}\right)\,\,\mathrm{erg\,s}\,.
\label{eq:angmomloss}
\end{eqnarray}
This estimate is comparable to the depletion of accreted angular momentum
visible in Figure~\ref{fig:PNSproperties} between $\sim$0.5\,s and 
$\sim$1\,s. But, of course, the agreement of this estimate 
with the numerical results is not a proof 
that the angular momentum decline in our simulations is indeed 
connected to the neutrino emission. The similar magnitude of the effects 
might also be an astounding coincidence.

In the considered situation the angular momentum is highly concentrated
in near-surface layers of the PNS, which are formed by accreted matter and
where neutrinos undergo their last few interactions before they begin to 
decouple from the stellar background (heavy-lepton neutrinos around 
$\sim$\,$10^{12}$\,g\,cm$^{-3}$, $\nu_e$ and $\bar\nu_e$ between
$\sim$\,$10^{12}$\,g\,cm$^{-3}$ and $\gtrsim$\,$10^{11}$\,g\,cm$^{-3}$). 
Under such conditions the escaping neutrinos can extract angular momentum 
extremely efficiently, thus possibly impeding the spin-up of the
PNSs by SASI spiral modes \citep[e.g.,][]{Blondin+2007,Kazeroni+2017},
(intermittent) accretion disk formation \citep[e.g.,][]{Gilkis+2014},
and magnetic field amplification by angular momentum accretion 
\citep{Soker2020}. Disk formation around the new-born NS is 
a crucial ingredient of the ``jittering-jet mechanism'' 
\citep[e.g.,][]{Papish+2011,Papish+2014}. It was suggested
to be facilitated by the infall of material from the convective 
O-burning and He-layers, where convective motions might possess sufficient 
stochastic angular momentum so that the matter can assemble in orbits
around the PNS \citep{Gilkis+2014}. However, not only angular
momentum loss through neutrino emission (as discussed above)
is a serious handicap for this scenario, but also the intense neutrino
fluxes and the powerful neutrino heating in the vicinity of
the PNS impede disk formation. Instead, the energy input by
neutrinos reverses the infall of high-angular momentum
material from the convective O-burning layer in our 3D simulations 
and thus strengthens the SN blast before 
this material could even start to accumulate in orbits
around the neutrino-emitting PNS (provided it had angular momentum
above the critical threshold of some $10^{16}$\,erg\,s 
needed for disk formation).
Angular momentum extraction by escaping neutrinos
might be of relevance for accretion phenomena, torus formation, and
disk evolution in binary NS mergers, too.

\subsubsection{Proto-neutron star kick}
\label{sec:PNSkick}

Anisotropic mass ejection and neutrino emission also impart a kick to the PNS,
which we evaluate along the lines described by
\citet{Scheck+2006,Wongwathanarat+2013,Stockinger+2020}. Neutrinos escape from
the PNS convection layer anisotropically because of the lepton-emission self-sustained
asymmetry (LESA; \citealt{Tamborra+2014}; 
see also \citealt{Tamborra+2014a,Walk+2018,Walk+2019,Walk+2020,OConnor+2018,Glas+2019a,Powell+2019,Vartanyan+2019,Stockinger+2020,Mueller2020,Nagakura+2021}, and references therein). 
This leads to a kick around 30\,km\,s$^{-1}$ that
saturates with a dominant negative $z$-component at $t_\mathrm{pb} \approx 0.3$\,s, 
because thereafter the LESA emission dipole gets considerably reduced,
presumably by the spin-up of the outer PNS layers \citep[][see also 
Appendix~\ref{app:LESA}, Figure~\ref{figapp:LESA}]{Walk+2019}. 
However, anisotropic neutrino emission, associated with accretion, accelerates the PNS
further (now migrating between the positive $y$- and negative $z$-directions) to 
$v_\mathrm{PNS}^\nu\approx 100$\,km\,s$^{-1}$, when evaluated at the gain radius
(see again Appendix~\ref{app:LESA} for details).
This value is asymptotically reached at $t_\mathrm{pb}\sim 0.8$\,s, because later
accretion downflows hardly penetrate through the gain radius 
(Figure~\ref{fig:explosionprops2}; Section~\ref{sec:explenergy}).
Due to partial absorption of neutrinos in the gain layer, which transfers linear
momentum to the SN ejecta, the relevant neutrino contribution to the total
PNS kick must be evaluated in the free-streaming regime, in order to avoid
that ejecta momentum received from neutrino absorption is erroneously
accounted to the PNS when assuming balance between PNS and ejecta momenta. 
The effective neutrino-induced kick thus grows from
$v_\mathrm{PNS}^\nu\approx 50$--60\,km\,s$^{-1}$ at $t_\mathrm{pb} \sim 0.8$--1\,s to
$\sim$100\,km\,s$^{-1}$ at the end of the \textsc{Prometheus-Vertex} run
(Figure~\ref{fig:PNSproperties}). The growth during this period of time is
again caused by LESA, because the LESA dipole emission becomes stronger again 
after the rotation in the PNS near-surface layers has slowed down. This leads to
a dipole component of the total neutrino luminosity (summed over all neutrino
species) that persists over the entire PNS evolution simulated with the
\textsc{Vertex} neutrino transport, even after the mass accretion by the PNS
has ended. The amplitude of the luminosity dipole amounts to about 5\% of the 
monopole on average 
(see Appendix~\ref{app:LESA}, Figures~\ref{figapp:LESA} and \ref{figapp:LESA4d}).
A considerable fluctuation of the direction of the luminosity dipole, however,
moderates its consequences for the PNS kick.

Asymmetric mass ejection causes
a much larger PNS kick, reaching $v_\mathrm{PNS}^\mathrm{hydro} > 400$\,km\,s$^{-1}$
after 7\,s and drifting considerably over this period of time. This yields a total kick 
of $>$450\,km\,s$^{-1}$, because neutrino-induced and hydrodynamic kicks are in 
the same hemisphere (Figures~\ref{fig:explosionasymmetry} and \ref{fig:PNSproperties}).
The total kick grows further because of the gravitational tug-boat effect.
Note that $v_\mathrm{PNS}^\mathrm{hydro}$ is
computed as the ratio of linear momentum to the (constant) baryonic PNS mass, since
$v_\mathrm{PNS}^\mathrm{hydro}$ should not change in the absence of external forces
when the gravitational mass of the PNS decreases due to isotropic neutrino emission 
in its rest frame. This expresses the fact that neutrinos radiated from a moving
PNS carry linear momentum in the observer frame.

\section{Conclusions}
\label{sec:conclusions}

Model \texttt{M\_P3D\_LS220\_m}$-$ was evolved in 3D with the \textsc{Prometheus-Vertex}
neutrino-hydrodynamics code continuously through the final 7\,min of convective O-shell
burning, core collapse, bounce, and onset of explosion at $t_\mathrm{pb} \sim 0.4$\,s
($\pm \sim$100\,ms depending on resolution and microphysics, i.e., LS220 or SFHo EoS,
and muons)
until $\sim$7\,s after bounce. It constitutes the first case of a self-consistent 3D
simulation with detailed neutrino transport that yields a neutrino-driven explosion with
properties similar to SN~1987A, namely an energy near 1\,B and up to $\sim$0.087\,M$_\odot$
of ejected $^{56}$Ni. Its PNS has 
a baryonic (gravitational) mass of 1.865\,M$_\odot$ ($\lesssim$1.675\,M$_\odot$),
a kick of $\gtrsim$450\,km\,s$^{-1}$, and an average (surface) spin period of roughly
600\,ms (2\,ms) at the end of our simulations (before fallback). Also the kick velocity
is in the ballpark of values inferred for the NS in SN~1987A from measured 
$^{44}$Ti-line redshifts \citep{Boggs2015} and detailed analyses of $^{56}$Co-decay,
$^{44}$Ti-decay, and Fe infrared lines \citep{Jerkstrand+2020}.

The considered non-rotating, solar-metallicity $\sim$19\,M$_\odot$ stellar model was not 
constructed as a SN~1987A progenitor, but our results demonstrate the viability of the
neutrino-driven mechanism to fuel powerful CCSN explosions in principle.
Perturbations in the convective O-burning shell determine the asymmetry of the mass
ejection, which has a pronounced bipolar structure with a large dipole component.
Angular momentum associated with accretion downflows
leads to time-variable spin-up of the surface layers of the PNS. Neutrinos, however, are 
an efficient channel of angular momentum loss because of the highly differential rotation
with the largest velocities in near-surface layers where neutrinos begin to decouple.

We did not see the appearance of a spherical or quasi-spherical neutrino-driven
wind in our long-time 3D simulations. Such a wind is defined as a basically isotropic
outflow of matter expanding away from the PNS surface due to neutrino energy
deposition. It fills the environment of the PNS and was seen as a generic feature of 
1D explosion models and most parametrized multi-D explosion simulations.
However, we could not witness this phenomenon in our long-time 3D explosion 
simulations of the $\sim$19\,M$_\odot$ star discussed in the current paper,
nor did it develop in a pure form during the first $\sim$3\,s after the 
onset of the neutrino-driven explosion in a long-time 3D SN simulation
of a 9.0\,M$_\odot$ iron-core progenitor
presented by \citet{Stockinger+2020}. Instead, accretion downflows and 
neutrino-heated outflows generate turbulent and highly asymmetric mass motions
around the PNS until the end of the simulated evolution of NS birth
and of the beginning SN explosions. In the mentioned 9.0\,M$_\odot$ explosion
the downflows become weaker and less frequent at $\sim$3\,s after bounce, but
some of them achieve to penetrate inward towards the PNS even later,
because a powerful supersonic neutrino-driven wind that could provide a 
strong outward push does not develop.

The possibility of the wind to occur, at least for transient phases, 
is sensitive to the strength of the neutrino heating on the one hand,
and it is likely to depend also on
the rate of mass infall from the progenitor on the other hand. In fact, the
long-lasting downflows ensure a continuous, abundant supply of matter that
can be efficiently heated by neutrinos in the close vicinity of the PNS,
taking up the energy that feeds the blast-wave until its energy
approaches the terminal value several seconds after core bounce.
The presence of these long-lasting downflows is a prerequisite for powerful 
explosions by the neutrino-heating mechanism as in the $\sim$19\,M$_\odot$
SN model reported here.
Therefore ``classical'' neutrino-driven winds do not seem to be a
general feature in 3D explosion simulations, at least not in cases where
strong explosions with energies of more than (0.1--0.2)\,B 
(which is the power-limit of neutrino-driven winds) are obtained.
Instead, isotropic neutrino-driven winds might be a phenomenon that 
occurs predominantly in progenitors with very low core compactness near 
the low-mass end of the SN range. In such progenitors the explosion develops 
quickly, accretion downflows and outflows cease soon after 
shock revival, and the blast-wave energy remains correspondingly low 
\citep[see two of the three 3D simulations of low-mass SNe in][]{Stockinger+2020}.

\software{\textsc{Prometheus-Vertex} \citep{Fryxell1991,Mueller1991,Rampp+2002,Buras+2006},
VisIt \citep{Childs2012}.}

\smallskip
{\em Data availability:} Data of the initial 1D and 3D models, selected 
output times of the 3D SN runs, and of neutrino signals
will be made available upon request in our Core-Collapse Supernova Data Archive
{\tt https://wwwmpa.mpa-garching.mpg.de/ccsnarchive/}.

\acknowledgments
We thank Henk Spruit for useful discussions and Jeanette Plechinger for
her internship work on the 3D data visualization. At Garching, funding by the
European Research Council through Grant ERC-AdG No.~341157-COCO2CASA
and by the Deutsche Forschungsgemeinschaft (DFG, German Research Foundation)
through Sonderforschungsbereich (Collaborative Research Centre)
SFB-1258 ``Neutrinos and Dark Matter in Astro- and Particle Physics
(NDM)'' and under Germany's Excellence Strategy through
Cluster of Excellence ORIGINS (EXC-2094)---390783311 is acknowledged.
This work was also supported by the Australian Research Council through ARC
Future Fellowship FT160100035 (BM) and Future Fellowship FT120100363
(AH), by the Australian Research Council Centre of Excellence for
Gravitational Wave Discovery (OzGrav), through project number
CE170100004 (BM, AH); by the Australian Research Council Centre of
Excellence for All Sky Astrophysics in 3 Dimensions (ASTRO 3D), through
project number CE170100013 (AH), and by the National Science Foundation
under Grant No.~PHY-1430152 (JINA Center for the Evolution of the
Elements; AH).
The authors are grateful to the Gauss Centre for Supercomputing e.V.\
(GCS; www.gauss-centre.eu) for computing time on the GCS Supercomputers
SuperMUC and SuperMUC-NG at Leibniz Supercomputing Centre (LRZ; www.lrz.de)
under GAUSS Call 17 project ID: pr53yi and GAUSS Call 20 project ID: pr53yi,
and to the LRZ under LRZ project ID: pn69ho. They also thank the Max Planck
Computing and Data Facility (MPCDF) for computer resources on the HPC
systems Cobra and Draco.
Parts of this research
were undertaken with the assistance of resources and services from the
National Computational Infrastructure (NCI), which is supported by the
Australian Government. They were also facilitated by resources provided 
by the Pawsey Supercomputing Centre with funding from the Australian
Government and the Government of Western Australia.




\section*{Appendix}
\renewcommand\thesection{\Alph{section}}
\renewcommand\thefigure{\thesection\arabic{figure}}
\renewcommand\theequation{\thesection\arabic{equation}}
\setcounter{section}{0}
\setcounter{figure}{0}
\setcounter{equation}{0}
\renewcommand{\theHsection}{appendixsection.\Alph{section}}
\renewcommand{\theHfigure}{A_fig.\arabic{figure}}


\section{Additional information for model subset with LS220 EoS and
         no muons}
\label{app:addinfos}

For comparison with model \texttt{M\_P3D\_LS220\_m$-$} (and its extension 
\texttt{M\_P3D\_LS220\_m$-$HC}), which we focus our discussion on in the main 
text, we present in this appendix a few selected results for a subset of the
other cases listed in Table~\ref{tab:simulations}. All of these considered models
have been computed with the LS220 EoS and without muons, but possess different 
angular resolutions and are either based on the 1D or 3D progenitor data.

\begin{figure}[tb!]
     \begin{center}
\includegraphics[width=1.0\columnwidth]{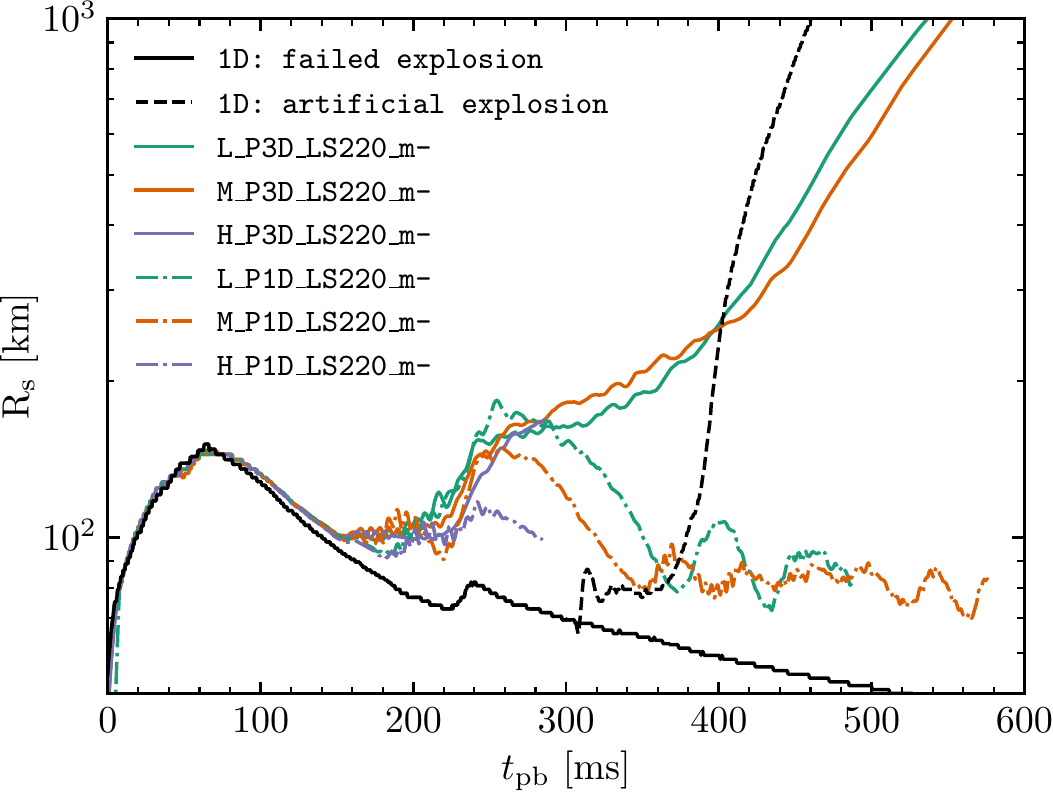}
     \end{center}
     \caption{Average shock radii for all models with LS220 EoS and without muon
     physics. Colored solid lines correspond to the 3D models with different angular
     resolution (\texttt{L}, \texttt{M}, and \texttt{H}) employing 3D pre-collapse
     progenitor data. The colored dash-dotted lines display results for the
     corresponding cases starting from the 1D progenitor data. The black lines
     show the 1D simulation (solid) and the 1D model with artificially triggered
     explosion (dashed), whose neutrino luminosities and mean energies of the PNS
     are employed in the 3D extension run \texttt{M\_P3D\_LS220\_m$-$HC} of model
     \texttt{M\_P3D\_LS220\_m$-$}. In \texttt{M\_P3D\_LS220\_m$-$HC} the neutrino
     treatment with the \textsc{Vertex} code is replaced by a heating and cooling
     description that requires input from the 1D PNS cooling simulation.}
\label{figapp:shockradii}
\end{figure}

\begin{figure*}[tb!]
        \begin{center}
\includegraphics[width=1.0\textwidth]{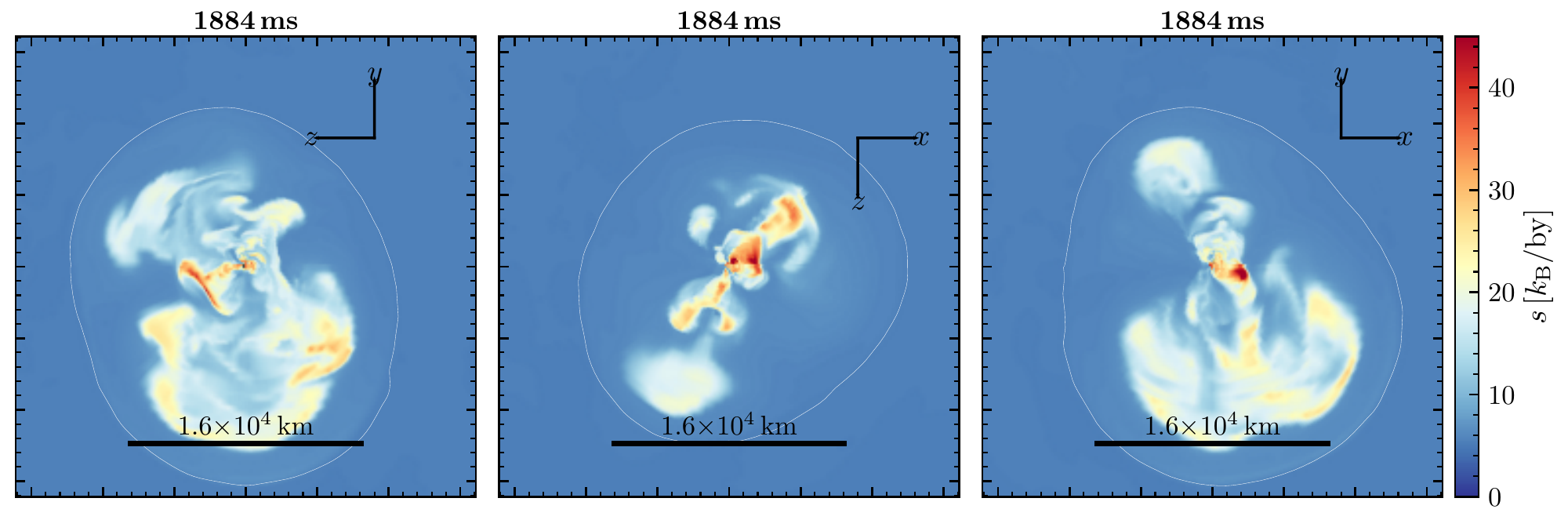}
        \end{center}
        \caption{Entropy distributions in the $y$-$z$, $x$-$z$, and $x$-$y$ planes
        of model \texttt{L\_P3D\_LS220\_m$-$} at the end of the 3D run with
	\textsc{Prometheus-Vertex}. The thin white line indicates the 
	position of the shock.}
\label{figapp:3Ds-cutsL}
\end{figure*}

Figure~\ref{figapp:shockradii} displays the average shock radii of the 
\texttt{L}, \texttt{M}, and \texttt{H} models both for the runs started from 
the 1D progenitor and from the 3D initial conditions. All three 3D core-collapse
simulations initiated with the 1D data do not exhibit an evolution towards 
explosion, whereas the three models based on the 3D progenitor data develop
explosions. The high-resolution case \texttt{H\_P3D\_LS220\_m$-$} is not finally
conclusive, because it had to be stopped at 285\,ms after bounce, but its mean
shock radius follows the two less well resolved and exploding counterparts 
very closely. 

We also plot the non-exploding 1D model (solid black line) as
well as an artificially exploded 1D case, whose explosion was triggered by 
decreasing the density in the infall layer ahead of the shock by a factor of
20, linearly growing from 500\,km to 1500\,km, at 300\,ms after bounce.
When the reduced mass infall rate reaches the
shock, the shock begins to expand and starts running outward at $\sim$400\,ms. 
The PNS left behind has nearly the same mass as the compact remnant 
formed in the exploding 3D models. We therefore used its long-time neutrino
emission to define the time-dependent input for the neutrino HC scheme applied
in model \texttt{M\_P3D\_LS220\_m$-$HC}, which extends the simulation of 
\texttt{M\_P3D\_LS220\_m$-$} to times later than 1.675\,s after bounce.

In all models the shock reaches a maximum radius around 150\,km at $\sim$70\,ms 
post bounce, because it is first pushed outward by the initially high mass
accretion rate, which rapidly adds a mantle layer around the forming NS, and
then it is pulled inward by the settling and shrinking PNS radius when the mass 
accretion rate declines to values that cannot support further expansion.
A second phase of shock expansion sets in when the Si/O composition interface falls
into the shock at about 200\,ms p.b. The strengthening of postshock convection
by the 3D perturbations fosters the explosion of the models with 3D initial
conditions; the corresponding physics was discussed by 
\citet{Mueller+2017}. In all models based on the 1D progenitor data we 
see a brief period of shock expansion, which transitions to a dramatic decrease
after a second maximum of the shock radius. The subsequent
quasi-periodic phases of large-amplitude shock expansion and contraction
indicate the presence of strong SASI activity in the postshock layer. SASI has
an intrinsic oscillation frequency of typically 50--100\,Hz (corresponding roughly
to the advection time scale through the gain layer), but SASI activity pushes
the shock to larger radii until strong convection (a parasitic instability that
taps energy from the SASI) sets in. This leads to cyclic phases of transient
shock expansion and contraction.
A similar behavior was observed in non-exploding 20\,M$_\odot$ simulations
by \citet{Glas+2019} and analysed in detail there.

The evolution of the mean shock radius shows close resemblance in all three exploding 
3D models with no systematic resolution dependence. We interpret this finding
as a consequence of the strong driving of postshock convection due to the large
density and velocity perturbations in the O-shell of the collapsing 3D
progenitor. In contrast, the non-exploding 3D models display a clear ordering
of their mean shock radii, with larger excursions and higher maxima when the
angular resolution is coarser. We explain this fact by stronger SASI activity
in lower-resolution models, where the growth of parasitic Rayleigh-Taylor
and Kelvin-Helmholtz instabilities is suppressed and the saturation amplitude
of SASI is bigger (see also our discussion in Section~\ref{sec:dimresphys}).

Figure~\ref{figapp:3Ds-cutsL} presents the entropy distributions in three cut
planes of the low-resolution model \texttt{L\_P3D\_LS220\_m$-$} at 1.884\,s after
bounce. The explosion is highly asymmetric and exhibits the strongest shock expansion
near the negative $y$-direction, closely resembling model \texttt{M\_P3D\_LS220\_m$-$}
(see Figures~\ref{fig:3D}, \ref{fig:2D}, and \ref{fig:explosionasymmetry}),
despite structural differences in detail, for example an apparent
tilting of the main axis of global deformation by about 50$^\circ$ in anti-clockwise
direction in the $x$-$y$ plane (compare the right panel of Figure~\ref{figapp:3Ds-cutsL}
with the middle and right panels of Figure~\ref{fig:3D}).
The fact that the biggest shock-pushing plume is in the negative $y$-hemisphere confirms
our argument that higher ram pressure of the anisotropically collapsing 3D progenitor
hampers shock revival and subsequent expansion in the opposite hemisphere.

\begin{figure}[tb!]
     \begin{center}
\includegraphics[width=1.0\columnwidth]{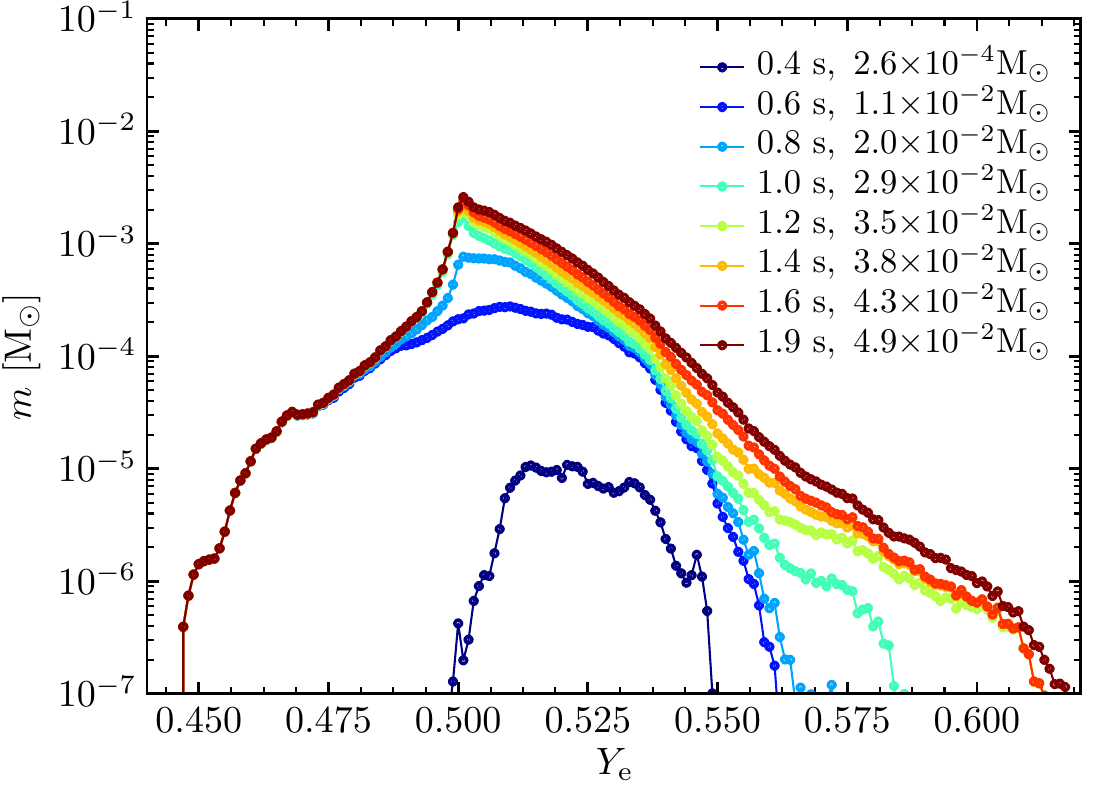}
     \end{center}
     \caption{Same as Figure~\ref{fig:Yemass}, but for model \texttt{L\_P3D\_LS220\_m$-$}.}
\label{figapp:Yemass4d}
\end{figure}

Figure~\ref{figapp:Yemass4d} displays the counterpart of Figure~\ref{fig:Yemass}
for our simulation of \texttt{L\_P3D\_LS220\_m$-$}. The overall similarity
is obvious, and also the low-resolution model fulfills the contraint that less than
roughly $10^{-4}$\,M$_\odot$ of neutrino-heated ejecta with $Y_e\lesssim 0.47$
should be expelled in order to avoid an overproduction of $N = 50$ nuclear species
\citep{Hoffman+1996}. Therefore also this model is compatible with the constraints
deduced from the nucleosynthetic output of typical Type-II SNe.

\setcounter{figure}{0}
\renewcommand{\theHfigure}{B_fig.\arabic{figure}}
\section{Angular momentum evolution}
\label{app:angmom}

\begin{figure}[tb!]
     \begin{center}
\includegraphics[width=1.0\columnwidth]{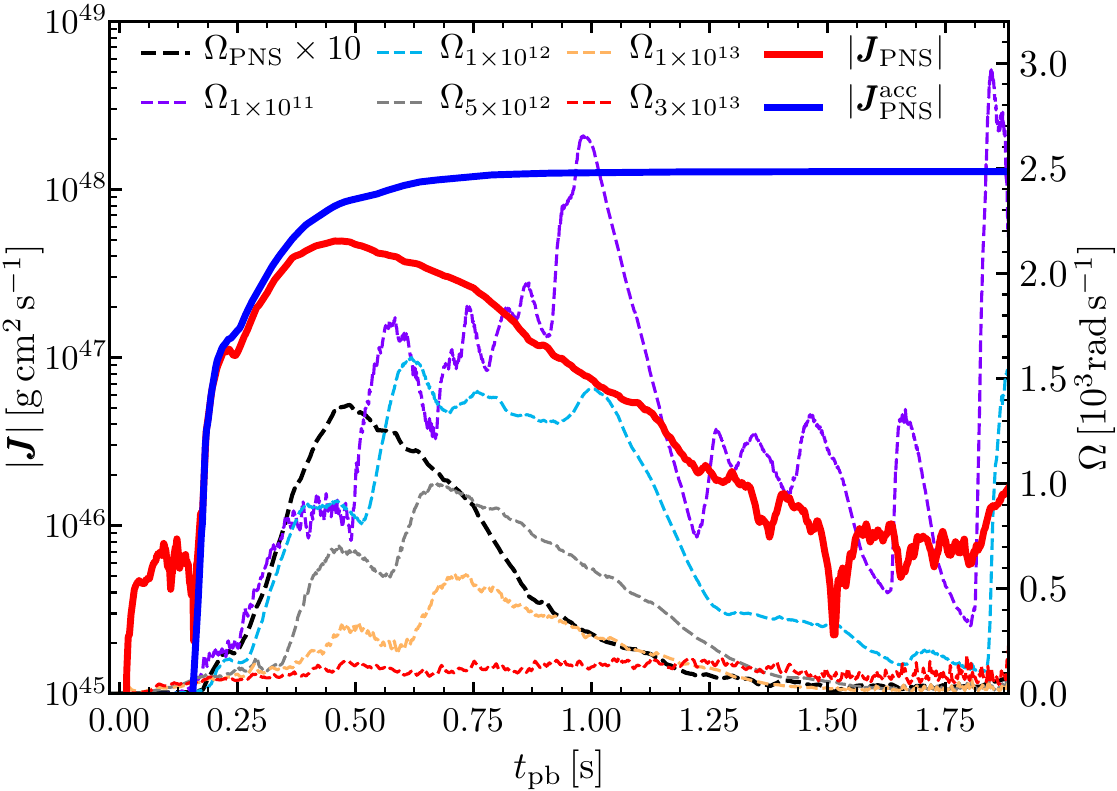}
     \end{center}
     \caption{Same as the lower left panel of Figure~\ref{fig:PNSproperties},
     but for model \texttt{L\_P3D\_LS220\_m$-$}.}
\label{figapp:PNSrot4d}
\end{figure}
\begin{figure*}[!]
        \centering
        \includegraphics[width=0.48\textwidth]{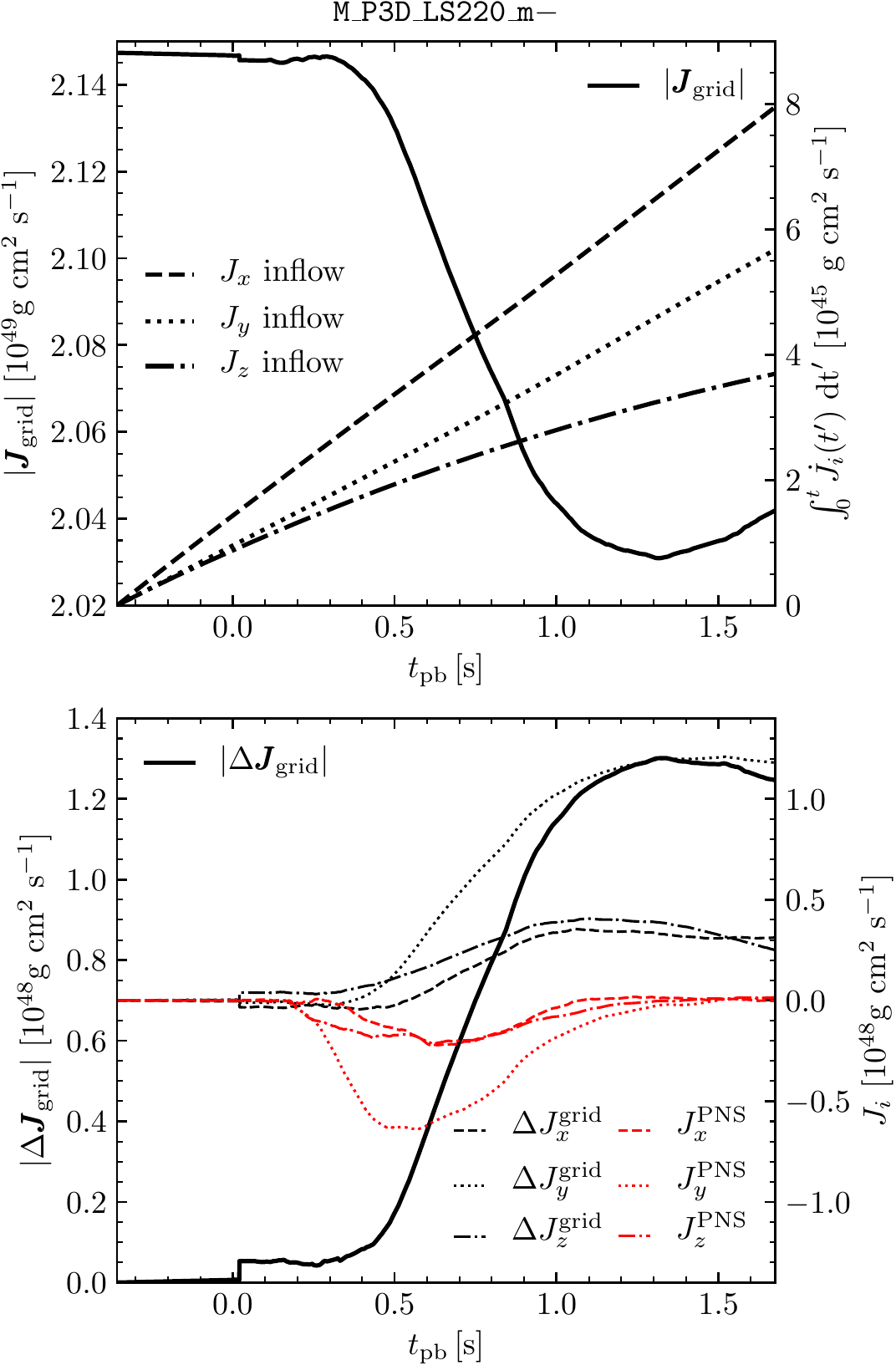}\hskip 15pt
        \includegraphics[width=0.48\textwidth]{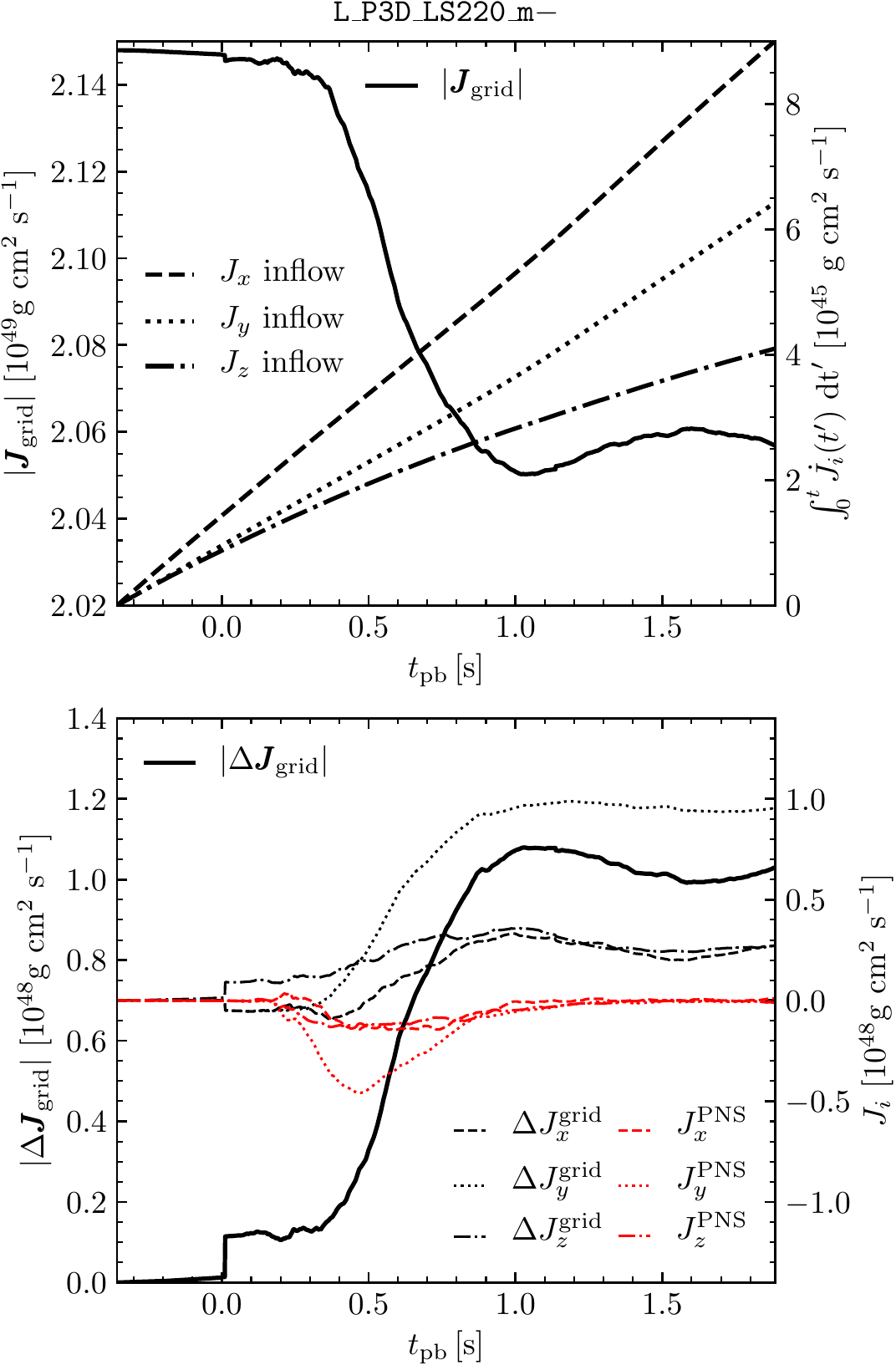}
        \caption{Angular momentum evolution for model \texttt{M\_P3D\_LS220\_m$-$}
        ({\em left}) and model \texttt{L\_P3D\_LS220\_m$-$} ({\em right}).
        {\em Upper panels:} Total angular momentum on the computational
        grid, $|\pmb{J}_\mathrm{grid}|$ (solid line; left $y$-axis scale),
        and the vector components of the time-integrated inflow of angular momentum
        through the open outer radial boundary of the grid (dashed, dotted, and
        dash-dotted lines; right $y$-axis scale). The latter effect is miniscule
	and plays no relevant role in the total angular momentum budget.
        {\em Lower panels:} Change of the total angular momentum on the grid
        compared to the onset of core collapse, $|\Delta \pmb{J}_\mathrm{grid}|$
        (black solid line; left $y$-axis scale), change of the individual components 
        of $\pmb{J}_\mathrm{grid}$ compared to the initial values
	(black dashed, dotted, and dash-dotted lines; right $y$-axis scale),
        and components of the angular momentum contained by the PNS
        (red dashed, dotted, and dash-dotted lines; right $y$-axis scale). The
	discontinuous behavior at $t_\mathrm{pb} = 0$ is connected to a change
	of the numerical grid setup at core bounce and the corresponding mapping
        of the hydrodynamic quantities between the grids.}
\label{figapp:angmomevol}
\end{figure*}

Figure~\ref{figapp:PNSrot4d} shows the analogue of the lower left panel 
of Figure~\ref{fig:PNSproperties} for model \texttt{L\_P3D\_LS220\_m$-$}.
We witness a qualitatively similar behavior of all displayed quantitities
in the \texttt{M} and \texttt{L} models,
independent of the angular resolution, with only smaller quantitative
differences. In the lower-resolution model a deficit in the angular momentum
stored in the PNS, $|\pmb{J}_\mathrm{PNS}|$, compared to the net inflow of 
angular momentum (inflow minus outflow), $|\pmb{J}_\mathrm{PNS}^\mathrm{acc}|$,
becomes visible somewhat earlier, starting at $\sim$0.25\,s, than in model
\texttt{M\_P3D\_LS220\_m$-$}, where the effect shows up only after roughly
0.3--0.35\,s after bounce.
The net angular momentum accreted onto the PNS
is slightly smaller in model \texttt{L\_P3D\_LS220\_m$-$}, the angular
momentum accumulated in the PNS also grows to a slightly lower peak 
with a value around
$5\times 10^{47}$\,erg\,s compared to $\sim$$7\times 10^{47}$\,erg\,s 
in model \texttt{M\_P3D\_LS220\_m$-$}. This maximum is reached at 
$\sim$0.5\,s in the low-resolution model and is sharper than in 
model \texttt{M\_P3D\_LS220\_m$-$}, where a broad plateau of 
$|\pmb{J}_\mathrm{PNS}|$ is visible between $\sim$0.4\,s and $\sim$0.75\,s.
This implies that the decline from the peak sets in earlier in model
\texttt{L\_P3D\_LS220\_m$-$}, but ultimately levels off to about the 
same value of roughly $10^{46}$\,erg\,s at $\sim$1.5\,s post bounce.
Correspondingly, the angular velocities in the accretion layer near
the PNS surface as well as averaged over the PNS (assuming rigid
rotation of the entire PNS mass) stay approximately 30--50\% below 
those in model \texttt{M\_P3D\_LS220\_m$-$}.
These quantitative differences might suggest some influence 
of the angular resolution on the loss of angular momentum in the PNS 
surface layers discussed in Section~\ref{sec:PNSspin}. Alternatively,
they could, however, also be a consequence of small differences in the
downflow and accretion behavior, in course of which less matter with
high angular momentum gets advected through the gain radius to reach the
PNS surface. The low-resolution and higher-resolution models naturally differ
in their detailed flow dynamics and their geometry of inflows and outflows
because of the different resolution and stochastic variations in the 
explosion behavior (cf.\ our discussion around Figure~\ref{figapp:3Ds-cutsL}).

In Figure~\ref{figapp:angmomevol} the evolution of the angular momentum on
the computational grid is displayed as function of time for models 
\texttt{M\_P3D\_LS220\_m$-$} (left) and \texttt{L\_P3D\_LS220\_m$-$} (right).
In both models the total angular momentum decreases roughly in the same way
by approximately 5\% (or $\sim$$10^{48}$\,erg\,s) from an initial 
value of $|\pmb{J}_\mathrm{grid}| \approx 2.147\times 10^{49}$\,erg\,s prior 
to the onset of stellar core 
collapse.\footnote{Although the stellar model is non-rotating and the computational
grid covers the entire convectively perturbed O-shell, the pre-collapse angular
momentum in the simulation volume, $J_\mathrm{grid} = 
|\pmb{J}_\mathrm{grid}| = |\int_{V_\mathrm{grid}} \mathrm{d}V\,\rho\,\pmb{j}\,|$,
is not zero. This angular momentum has 
developed during the 7 minutes of our 3D simulation of convective burning in the
oxygen shell \citep[see][]{Yadav+2020}. Although large by its absolute value,
it accounts for only a few percent of the volume-integrated absolute value 
of the specific angular momentum times density, i.e., of $\tilde{J}_\mathrm{grid} = 
\int_{V_\mathrm{grid}} \mathrm{d}V\,\rho\,|\pmb{j}|$. The much bigger value of
$\tilde{J}_\mathrm{grid}$ means that there is a lot more angular momentum
in the system than $\pmb{J}_\mathrm{grid}$ shows, since
$\tilde{J}_\mathrm{grid}$ measures the angular momentum connected to small-scale
variations and vorticity. Because of these vortex flows on small scales the total
integral performed in $\tilde{J}_\mathrm{grid}$ is much larger than the total
angular momentum added up vectorially, which causes numerical artifacts to grow
more than the total might suggest. Therefore the angular
momentum non-conservation during the pre-collapse evolution stayed within the 
numerical uncertainties that one can expect for 3D simulations of turbulent
convection over several ten convective turnover cycles using a polar Yin-Yang
grid.} This is close to the drain of angular momentum that we discuss
in Section~\ref{sec:PNSspin}.
Interestingly, the decline of the total angular momentum is slightly
smaller in the lower-resolution model ($\sim$$1.04\times 10^{48}$\,erg\,s 
at the end of the simulation compared to $\sim$$1.26\times 10^{48}$\,erg\,s in 
model \texttt{M\_P3D\_LS220\_m$-$}). This fact might be considered to be
in conflict with an explanation of the angular momentum 
loss by numerical grid effects, potentially connected with the 
interfaces of the Yin and Yang patches of the grid. However, such a possibility
cannot be ruled out, because the differences between models 
\texttt{L\_P3D\_LS220\_m$-$} and \texttt{M\_P3D\_LS220\_m$-$} could
simply be a consequence of less accretion of high-angular-momentum matter into
the extremely spinning, highly compressed near-surface layer of the PNS
in the lower-resolution case, as mentioned above.

It is evident that the main reduction of the angular momentum happens in both
models during the period of time when mass accretion leads to the extreme
spin-up of the near-surface layers of the PNS. In both models there is a tight
correlation of the growth of $|\Delta \pmb{J}_\mathrm{grid}|$ with the time
interval when the angular momentum in the PNS exbibits the dramatic decline
(compare the black and red curves in the bottom panels of 
Figure~\ref{figapp:angmomevol}). However, we do not observe any clear 
systematics in the temporal behavior of the three angular momentum components,
neither with respect to angular resolution nor with the coordinate directions.
Since flows near the $y$-$z$ plane cross Yin-Yang grid boundaries
four times per full cycle, whereas flows close to the $x$-$y$ and 
$x$-$z$ planes cross Yin-Yang boundaries only twice per revolution,
one would expect Yin-Yang-associated angular momentum loss to lead to a
stronger damping of $J_x^\mathrm{grid}$ than of the other two 
angular momentum components. Indeed, we find a relative loss of 
$J_x^\mathrm{grid}$ of roughly 15\% in both models, corresponding to a 
change from about $-0.2\times 10^{49}$\,erg\,s to approximately
$-0.17\times 10^{49}$\,erg\,s, whereas $J_y^\mathrm{grid}$ and
$J_z^\mathrm{grid}$ exhibit lower reductions, namely $J_y^\mathrm{grid}$
by $\sim$8\% ($\sim$6\%) from $-1.55\times 10^{49}$\,erg\,s to 
$-1.44\times 10^{49}$\,erg\,s ($-1.46\times 10^{49}$\,erg\,s) in
model \texttt{M\_P3D\_LS220\_m$-$} (\texttt{L\_P3D\_LS220\_m$-$}), and
$J_z^\mathrm{grid}$ by $\sim$2\% from $-1.46\times 10^{49}$\,erg\,s to 
approximately $-1.44\times 10^{49}$\,erg\,s in both cases. The relative
change of $J_x^\mathrm{grid}$ is therefore roughly twice the change of 
$J_y^\mathrm{grid}$ and seven to eight times the change of 
$J_z^\mathrm{grid}$. Moreover, the absolute change of $J_y^\mathrm{grid}$
is several times that of $J_z^\mathrm{grid}$, although both components 
are of similar magnitude initially. All of these facts
seem to be in conflict with the symmetric treatment
of flows on the Yin and Yang patches of the grid and therefore seem to
contradict a grid-connected explanation of the angular momentum loss. Since, 
however, there is
a conservation law only for the total angular momentum but not for its
individual components in 3D, the asymmetric evolution of $J_y^\mathrm{grid}$
and $J_z^\mathrm{grid}$ can be a consequence of torques due to
hydrodynamic forces (and, in principle, also gravitational forces when 3D 
gravity effects play a role), shuffling angular momentum between the
different components. In general, in complex 3D hydrodynamic flows such 
physical processes are always superimposed on possible numerical effects,
for which reason there is no straightforward way to discern grid artifacts 
clearly for conditions as they occur in the turbulent surroundings
of the PNS.  

Effects caused by the Yin-Yang grid can therefore not be diagnosed
unambiguously. The higher relative change and the slighly later increase 
and earlier reduction of $J_x^\mathrm{PNS}$ compared to the other two PNS
angular momentum components (Figure~\ref{figapp:angmomevol}), as well as
the differences in the evolution of $J_y^\mathrm{PNS}$ and $J_z^\mathrm{PNS}$,
may well be
a consequence of the time evolution of the total angular momentum vector in
the accretion flow instead of being connected to the Yin-Yang grid geometry.
Clean test setups are therefore necessary to track down possible grid 
artifacts.

In the described 3D SN simulations
the angular momentum is concentrated in an extremely rapidly
revolving accretion layer near the surface of the PNS. This layer is
geometrically narrow and contains a mass of only several $10^{-2}$\,M$_\odot$.
Presently, we cannot attribute the drain of angular momentum from this region,
which occurs over roughly 100 rotation periods, to any source 
---hydrodynamics solver, Yin-Yang grid, RbR+ transport and its coupling
to the hydro, neutrino-mediated extraction of angular momentum--- unambiguously.
The tests of angular momentum conservation for the Yin-Yang grid in the
literature~\citep[e.g.,][and references therein]{Wongwathanarat+2010} are
inconclusive for such conditions, because they never explored setups that
came anywhere near the extreme situation obtained in the considered
accreting PNSs with respect to densities and rotation rates.
In our previous 3D core-collapse
simulations of rotating progenitors \citep{Summa+2018} we could not witness any
worrisome loss of angular momentum, but the simulated post-bounce evolution
was not as long as in the current models and the angular momentum was also not
concentrated in a narrow near-surface layer of the PNS. 

Further dedicated and specifically designed tests 
of the angular momentum evolution on the Yin-Yang grid for our extreme PNS
conditions are needed with and without neutrino transport. Such tests are 
being carried out with a new implementation of the Yin-Yang grid in the
\textsc{Alcar} code, which also contains a fully multi-dimensional treatment
of the neutrino transport in addition to 
a RbR+ description \citep{Just+2015,Glas+2019}. But this is work in
progress and beyond the scope of the present paper.

\setcounter{figure}{0}
\renewcommand{\theHfigure}{C_fig.\arabic{figure}}
\renewcommand{\theHequation}{C_equ.\arabic{equation}}
\section{Essentials of LESA}
\label{app:LESA}

\begin{figure*}[t!]
        \begin{center}
\includegraphics[width=1.0\textwidth]{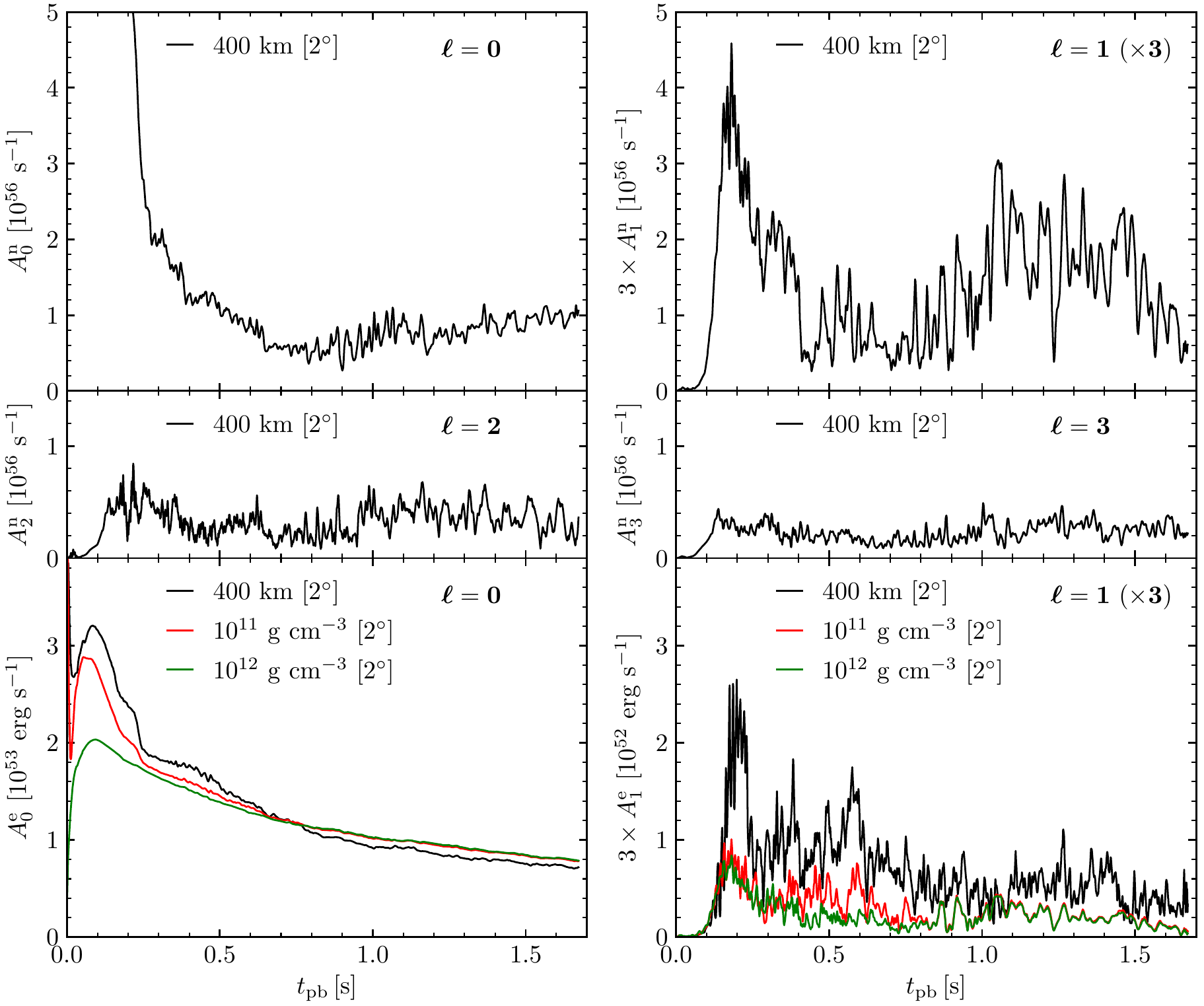}
        \end{center}
	\caption{Multipole components of the neutrino emission in model 
	\texttt{M\_P3D\_LS220\_m$-$} with 2$^\circ$ angular resolution. 
        The plotted data have been smoothed in a running 
	window of 5\,ms to damp short-time scale variations. 
	{\em Upper four panels:}
	Monopole, dipole, quadrupole, and octupole ($\ell = 0,\,1,\,2,\,3$) of the 
	electron-neutrino lepton-number emission. Note that the dipole amplitude as
	defined by Equation~(\ref{eq:shamplitudes2}) is scaled by a factor 3 for better
	comparison with the literature. Even ignoring this scaling, the dipole 
	equals or even dominates the monopole after $\sim$0.2\,s post bounce, whereas
	the quadrupole and octupole are considerably lower. This signals the presence
	of the LESA phenomenon.
	{\em Bottom two panels:} Monopole and dipole of the total neutrino energy 
	emission. The monopole is the (all-sky) total neutrino luminosity given by the 
	sum of the luminosities of all neutrino and antineutrino species. The dipole 
	component (again shown as 3 times the amplitude of Equation~(\ref{eq:shamplitudes2}))
        amounts to roughly 5\% of the monopole on average. 
	All quantities have been transformed to a distant observer in the lab frame.
	The black lines correspond to the analysis being performed in the free-streaming 
	limit (at a radius of 400\,km), the red lines in the bottom two panels show
	the results slightly above the neutrinospheres of $\nu_e$ and $\bar\nu_e$, i.e., 
        at a chosen density of $10^{11}$\,g\,cm$^{-3}$, and the green lines represent 
        the results slightly below the neutrinospheres of $\nu_e$ and $\bar\nu_e$, i.e.,
	at a density of $10^{12}$\,g\,cm$^{-3}$.}
\label{figapp:LESA}
\end{figure*}

\begin{figure*}[t!]
        \begin{center}
\includegraphics[width=1.0\textwidth]{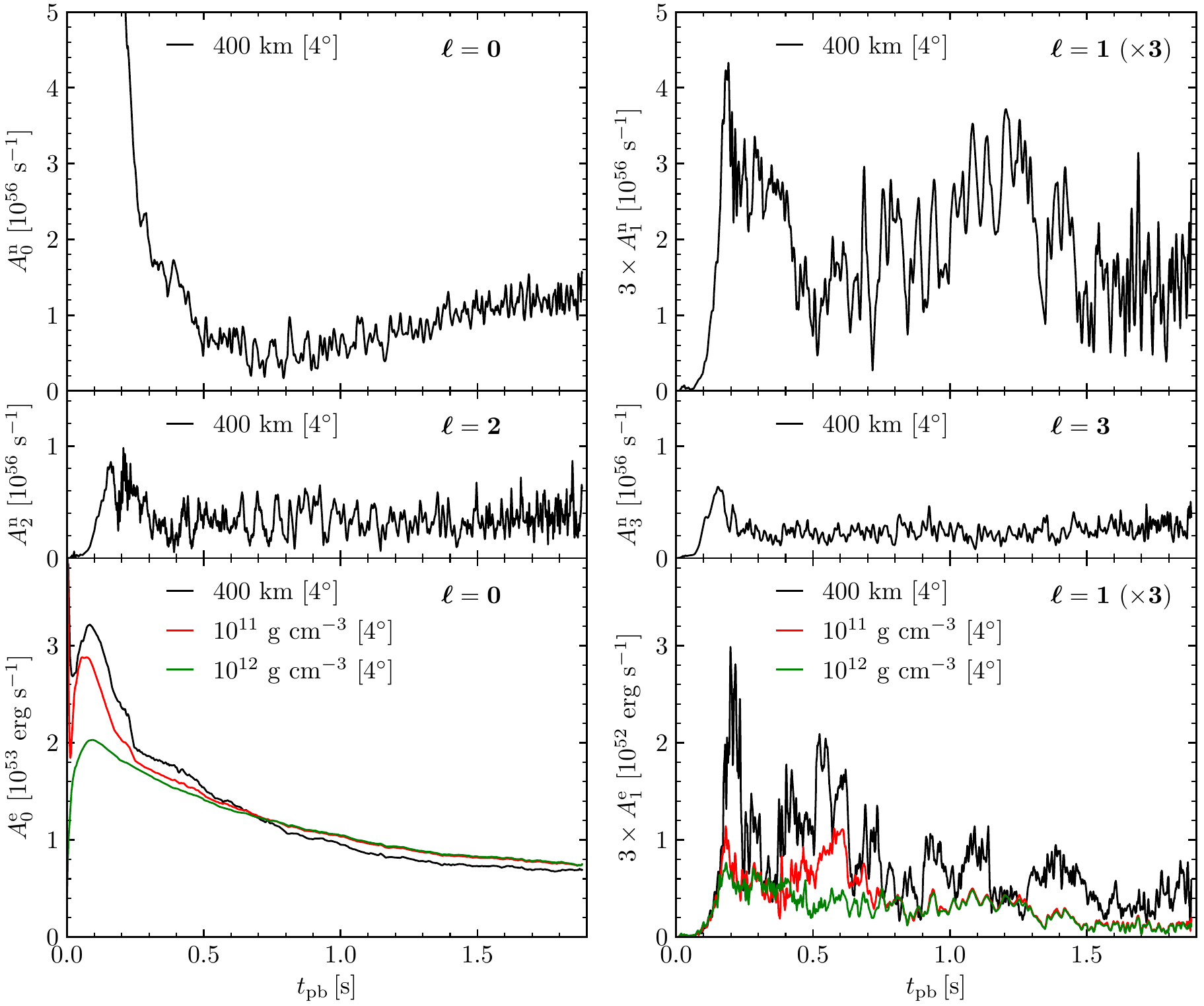}
        \end{center}
        \caption{Same as Figure~\ref{figapp:LESA}, but for model 
        \texttt{L\_P3D\_LS220\_m$-$} with 4$^\circ$ angular resolution.}
\label{figapp:LESA4d}
\end{figure*}

The lepton-number self-sustained asymmetry (LESA) in the neutrino emission of 
new-born NSs was first discovered by \citet{Tamborra+2014} in 3D SN simulations
of the Garching group with the \textsc{Prometheus-Vertex} code.
It is defined by the development of a dominant dipole
emission of the electron-neutrino lepton-number flux (number flux of $\nu_e$
minus $\bar\nu_e$), which can become larger than the corresponding monopole
\citep{Tamborra+2014}. 
This implies that a hemispheric asymmetry of the electron fraction builds up
inside the PNS, associated with the layer where PNS convection takes place 
highly anisotropically, carrying $\nu_e$ out of the PNS core more efficiently
in one hemisphere than in the opposite hemisphere. This can even lead to a 
dominant emission of $\bar\nu_e$ on one side and $\nu_e$ on the other side of 
the PNS. Such a lepton-number emission dipole is accompanied by a dipole of the 
total neutrino luminosity (i.e., of the sum of the luminosities of all species
of neutrinos and antineutrinos) that points in the opposite direction and
reaches an amplitude of only a few percent of the monopole on average 
\citep[i.e., much lower than the LESA dipole relative to the lepton-number 
flux;][]{Tamborra+2014,Tamborra+2014a}.
Because LESA can be found in models after the onset of the explosion, too,
its origin is connected to the anisotropic convection in the interior of the PNS
\citep{Glas+2019a}. Therefore the LESA dipole has to be carefully discriminated
from an accretion dipole of the lepton-number emission. This discrimination
can be achieved
by the criterion that the LESA is characterized by a temporal and spatial
anti-correlation of the $\nu_e$ and $\bar\nu_e$ emission, whereas an accretion
dipole exhibits a correlated behavior of the $\nu_e$ and $\bar\nu_e$ emission
\citep{Tamborra+2014a}. LESA was found to be considerably weaker in 
3D simulations of the collapse of fast-spinning stellar cores \citep{Walk+2019}.
This can be understood from the fact that rapid rotation on the one hand damps the 
convective activity in the PNS due to the constraining influence of specific 
angular momentum conservation, which alters the large-scale structure of 
convective flows \citep{Janka+2001}. On the other hand rapid differential
rotation leads to shear effects in the neutrino advection and thus may
destroy large-scale patterns of the radial neutrino transport in the
optically thick regime.

The characteristic features of LESA, as diagnosed in the 
\textsc{Prome\-theus-Vertex} results of the Garching group, were discussed in a series
of papers by \citet{Tamborra+2014a,Janka+2016,Walk+2018,Walk+2019,Walk+2020}, and 
\citet{Stockinger+2020}. Other groups confirmed the LESA phenomenon in their models
and with different transport treatments 
\citep[e.g.,][]{OConnor+2018,Powell+2019,Vartanyan+2019,Mueller2020,Nagakura+2021}.
In a recent paper by \citet{Glas+2019a} LESA properties were compared in 
simulations with RbR+ as well as fully multi-dimensional neutrino transport
with the \textsc{Alcar} code.

Because of the wealth of the previous literature, 
we will not describe the details of LESA once again here but refer the reader
to the cited papers, also for a comparison of results from different groups and
for suggestions to explain the physical origin of the LESA phenomenon
(see in particular \citealt{Glas+2019a} and \citealt{Mueller2020}). 
Instead of repeating such basic aspects,
we will only present results in the following that are needed in support of our 
discussion of neutrino-induced NS kicks in Section~\ref{sec:PNSkick}.

For a quantitative analysis of the lowest-order multipoles (up to
$\ell =3$) of LESA we follow the definitions by \citet{Glas+2019a}. The (complex) 
amplitude of a spherical harmonics mode $(\ell,m)$ is defined as
\begin{equation}
c_{\ell}^{m}(r) = \sqrt{\frac{4\pi}{2\ell + 1}}\int_{4\pi} \ud{\Omega}\,r^2
	\, F_{\nu,\mathrm{lab}}^\mathrm{n}(\pmb{r})\,Y_l^m\,,
\label{eq:shamplitudes1}
\end{equation}
where $F_{\nu,\mathrm{lab}}^\mathrm{n}(\pmb{r}) = 
F_{\nu_e,\mathrm{lab}}^\mathrm{n}(\pmb{r})-
F_{\bar{\nu}_e,\mathrm{lab}}^\mathrm{n}(\pmb{r})$ 
is the local (energy-integrated) lepton-number 
flux (density) of electron neutrinos minus antineutrinos in the laboratory frame
of the observer, $Y_l^m$ is the (complex) spherical harmonics of degree $l$ and 
order $m$, and the integration is performed over the surface of a closed sphere 
at a chosen radius $r$.
The amplitude of the $\ell$-mode, $A_{\ell}(r)$, is defined by the following formula:
\begin{equation}
	A_{\ell}(r) = \sqrt{\sum_{m=-\ell}^{m=+\ell} c_{\ell}^{m} {c^\ast}_{\ell}^{m}}\,.
\label{eq:shamplitudes2}
\end{equation}
Figures~\ref{figapp:LESA} and \ref{figapp:LESA4d} show the monopole, dipole (multiplied
by a factor of 3 for direct comparison with Figure~3 in \citealt{Tamborra+2014} and
Figure~1 in \citealt{Glas+2019a}),
quadrupole, and octupole of the electron lepton-number emission as functions of time 
(upper four panels) for models \texttt{M\_P3D\_LS220\_m$-$} and \texttt{L\_P3D\_LS220\_m$-$},
respectively. The bottom two panels display the monopole and dipole (again times 3)
of the total energy flux defined as the sum of the energy fluxes of all neutrinos
and antineutrinos: $F_{\nu,\mathrm{lab}}^\mathrm{e}(\pmb{r}) = \sum_{\nu_i}
F_{\nu_i,\mathrm{lab}}^\mathrm{e}(\pmb{r})$ with $\nu_i = \nu_e, \bar\nu_e, \nu_\mu,
\bar\nu_\mu, \nu_\tau, \bar\nu_\tau$. The two monopoles represent, respectively, 
the total loss rate of electron lepton number through neutrino emission and the 
total neutrino luminosity of the PNS (as given, at 400\,km, by the sum of all of 
the luminosities displayed in the right panel of Figure~\ref{fig:explosionprops1}
for model \texttt{M\_P3D\_LS220\_m$-$}). 

The basic features of LESA that are 
visible in Figures~\ref{figapp:LESA} and \ref{figapp:LESA4d}
are in line with the results of the previous \textsc{Prometheus-Vertex} models
discussed in great detail in the literature cited above. However, there are also 
some special aspects due to the fact that the present models develop explosions in a
fairly massive progenitor star and because we were able continue the simulations
to nearly 2\,s after core bounce, thus much longer than before. 

The onset of the explosion reduces the accretion of fresh material by the 
PNS. Since this material fuels the emission of large numbers of electron neutrinos
in the process of its deleptonization, 
the monopole of the lepton-number emission declines much faster to values below
$10^{56}$\,s$^{-1}$ in our present models than in the previous, non-exploding
massive stellar progenitors 
(compare with Figure~3 in \citealt{Tamborra+2014} and Figure~1 in \citealt{Glas+2019a}).
In fact, the decline is similarly fast as in exploding low-mass progenitors
(see Figure~5 of \citealt{Stockinger+2020}), which possess genericly lower 
mass-infall rates and explode relatively fast.
The characteristic LESA dipole of the electron neutrino lepton-number emission
grows to a first maximum within about 0.2\,s after bounce and a peak value
very similar to the 
\textsc{Prometheus-Vertex} simulations discussed by \citet{Tamborra+2014},
\citet{Janka+2016}, and \citet{Stockinger+2020} and roughly compatible with 
results from simulations with the \textsc{Fornax} code \citep{Vartanyan+2019}, but
somewhat faster and higher than in the simulations with \textsc{Alcar} 
\citep{Glas+2019a} and \textsc{Flash} \citep{OConnor+2018}. The origin of these
differences is most probably connected to differences in the development of convection
inside the PNS in these simulations, which in turn can depend on resolution, geometry
of the computational grid, numerical viscosity and order of the computational
method, or on the details of the neutrino transport and neutrino interaction rates.

The dipole of the lepton-number flux displayed in the upper right panels of 
Figures~\ref{figapp:LESA} and \ref{figapp:LESA4d} exceeds the monopole after 
0.2--0.3\,s post bounce. Its value plotted there is $3A_1$ with $A_1$ being the
amplitude of Equation~(\ref{eq:shamplitudes2}). This follows the convention in
\citet{Tamborra+2014,OConnor+2018,Vartanyan+2019},
and \citet{Glas+2019a} and represents the factor weighting the angular variation
when the neutrino number and energy luminosities are written as sum of their
monopole and dipole components as
$L_{\nu,\mathrm{lab}}^\mathrm{n,e}=A_0^\mathrm{n,e}+(3A_1^\mathrm{n,e})\cos\theta$.

After the initial peak the LESA dipole declines and exhibits a wide trough of
lower values before it increases again after $t_\mathrm{pb}\sim 0.9$\,s
in model \texttt{M\_P3D\_LS220\_m$-$} and $t_\mathrm{pb}\sim 0.7$\,s in model 
\texttt{L\_P3D\_LS220\_m$-$}. While the decrease of the dipole after the first 
maximum occurs on a similar time scale in both models, the increase out of the
trough is different and seems to follow the somewhat different shape of the 
time evolution of the PNS angular momentum as visible in the lower left panel
of Figure~\ref{fig:PNSproperties} and in Figure~\ref{figapp:PNSrot4d}.
The trough is deeper and wider in model 
\texttt{M\_P3D\_LS220\_m$-$},\footnote{Some features inside the troughs
of both models, for example the local maximum between $\sim$0.4\,s and 
$\sim$0.8\,s in \texttt{M\_P3D\_LS220\_m$-$}, are connected to anisotropic
accretion downflows onto the PNS, which produce mainly anisotropic $\nu_e$ 
emission (and secondary $\bar\nu_e$ emission) 
between the neutrinosphere and the gain radius. This can be verified by 
comparing the red and green lines in the bottom panels of 
Figures~\ref{figapp:LESA} and \ref{figapp:LESA4d}, because the former
show the luminosities slightly above the neutrinospheres, the latter 
somewhat below the neutrinospheres of $\nu_e$ and $\bar\nu_e$.}
where the PNS angular momentum reaches higher values and declines more slowly
from its plateau. We therefore hypothesize that there might be a causal 
connection, because \citet{Walk+2019} witnessed that LESA is suppressed by 
rapid PNS rotation as mentioned above. Further analysis is needed to test this
hypothesis, but a detailed investigation is quite laborious 
and involved \citep[see][]{Walk+2019} and is therefore deferred to future work.

LESA is present even after accretion onto the PNS has ended, which again 
supports its connection with anisotropic convection inside the PNS, in line
with findings by \citet{Glas+2019a,Powell+2019}, and \citet{Stockinger+2020}.
The late decline of the lepton-number dipole in both of our models after 
about 1.5\,s post bounce may be connected with a weakening of PNS convection
at these late times.

While the dipole of the lepton-number emission dominates the 
corresponding monopole for more than one second, the total energy luminosity
has a much smaller asymmetry with a dipole component 
(again considered to be 3 times the amplitude
$A_1^\mathrm{e}$ from Equation~(\ref{eq:shamplitudes2})) that amounts to only 
$\sim$5\% of the monopole (bottom panels of Figures~\ref{figapp:LESA} and 
\ref{figapp:LESA4d}). The acceleration of the PNS by a dipolar asymmetry of
the neutrino emission scales according to 
$\dot v_\mathrm{PNS}^\nu \propto \frac{2}{3}(3A_1^\mathrm{e})$.
The 5\% neutrino emission asymmetry includes a fluctuating and direction-variable
contribution from anisotropic accretion flows, which produce anisotropic
neutrino emission between neutrinosphere and gain radius, and it also
includes anisotropic neutrino absorption by down- and outflows in the 
gain layer. The corresponding anisotropic neutrino absorption in ejecta 
material adds to the asymmetry of the neutrino energy flux at large radii.
This contribution to the linear momentum of the escaping neutrinos is 
counterbalanced by linear momentum of the ejecta in the opposite direction
(because the momentum transfer to the stellar plasma
by the absorbed neutrinos is taken into account
in our simulations), which implies that it is not transferred to the 
PNS.\footnote{Since the hydrodynamic PNS kick is computed as the negative of 
the linear ejecta momentum, it contains this fluid momentum received from 
neutrinos. The neutrino
kick must therefore be evaluated in the free-streaming limit (and not at the 
neutrinosphere or at the gain radius) in order to account for the partial
compensation of neutrino and gas momenta and thus to correctly compute the 
total PNS kick as the sum of neutrino kick and hydrodynamic kick.}

Once accretion by the PNS has ended and neutrino emission between the 
neutrinosphere and gain radius abates, the true neutrino-induced
kick of the PNS depends on the anisotropic neutrino radiation that leaves the
neutrinosphere. It is considerably smaller then the neutrino kick evaluated
in the free-streaming limit (due to the mentioned anisotropic neutrino 
absorption in the gain layer), namely typically only about half of the total
effect. This can be concluded from comparing the dipole amplitudes
of the total neutrino luminosity in the free-streaming regime (e.g., at 400\,km)
and near the neutrinosphere (e.g., at $10^{11}$\,g\,cm$^{-3}$) in the lower
right panels of Figures~\ref{figapp:LESA} and \ref{figapp:LESA4d}.
The neutrino-induced PNS kick velocity shown in 
the upper right panel of Figure~\ref{fig:PNSproperties} corresponds to the 
total effect deduced from the neutrino luminosities in the free-streaming
limit (as briefly mentioned in Section~\ref{sec:PNSkick}). The persistent
dipole neutrino emission until the end of our \textsc{Vertex} simulations 
leads to a fairly constant PNS acceleration after $\sim$1\,s post bounce,
which however is dwarfed by the hydrodynamic kick
(see Figure~\ref{fig:PNSproperties}).

\bibliography{references}
\bibliographystyle{aasjournal}
\end{document}